\documentclass[aps,prb,preprint,showpacs]{revtex4}
\usepackage{natbib}
\usepackage{epsfig}

\begin{document}

\title{Exact Ground States of the Periodic Anderson Model
\\ in $D=3$ Dimensions}
\author{ Zsolt~Gul\'acsi$^{a,b}$ and Dieter~Vollhardt$^{a}$}
\affiliation{$^{(a)}$ Theoretical Physics III, Center for
Electronic Correlations and Magnetism, Institute for Physics,
University of Augsburg, D-86135 Augsburg,
Germany \\
$^{(b)}$ Department of Theoretical Physics, University of
Debrecen, H-4010 Debrecen, Hungary}
\date{\today }

\begin{abstract}
We construct a class of exact ground states of three-dimensional
periodic Anderson models (PAMs)
--- including the conventional PAM
--- on regular Bravais lattices at and above $3/4$
filling, and discuss their physical properties. In general, the
$f$ electrons can have a (weak) dispersion, and the hopping and
the non-local hybridization of the $d$ and $f$ electrons extend
over the unit cell. The construction is performed in two steps.
First the Hamiltonian is cast into positive semi-definite form
using composite operators in combination with coupled non-linear
matching conditions. This may be achieved in several ways, thus
leading to solutions in different regions of the phase diagram. In
a second step, a non-local product wave function in position space
is constructed which allows one to identify various stability
regions corresponding to insulating and conducting states.
The compressibility of the insulating state is shown to diverge at
the boundary of its stability regime. The metallic phase  is a
non-Fermi liquid
with one dispersing and one flat band. This state is also an exact
ground state of the conventional PAM and has the following
properties: (i) it is non-magnetic with spin-spin correlations
disappearing in the thermodynamic limit, (ii) density-density
correlations are short-ranged, and (iii) the momentum
distributions of the interacting electrons are analytic functions,
i.e., have no discontinuities even in their derivatives. The
stability regions of the ground states extend through a large
region of parameter space, e.g., from weak to strong on-site
interaction $U$. Exact itinerant, \emph{ferromagnetic}  ground
states are found at and below $1/4$ filling.
\end{abstract}

\maketitle


\section{Introduction}


The periodic Anderson model (PAM) is the basic microscopic model
for the investigation of compounds with heavy-fermion or
intermediate valence properties such as cerium or uranium
\cite{one}. The model describes $f$ electrons which interact via a
strong on-site Coulomb repulsion $U$ and hybridize with
noninteracting $d$ electrons. In the simplest case the $f$
electrons are assumed dispersionless, the hybridization purely
local, and the $d$ electron hopping nonzero only between nearest
neighbor sites. However, for real systems this is an
oversimplification since there is experimental evidence for (i) a
weak, but finite dispersion of the $f$ electrons, especially in
uranium compounds \cite{two,fast,schm}, (ii) nonlocal
contributions to the hybridization, and (iii) hopping of the $d$
electrons beyond nearest neighbors \cite{three}.

Recently the PAM was employed to study the dramatic volume
collapse at the $\alpha \to \gamma$ transition in cerium
compounds.\cite{negy,ot,hat} These investigations called attention
to the possibility of a Mott metal-insulator transition in the
PAM. In fact, a remarkable similarity between the Hubbard model
with nearest-neighbor hopping \cite{het} and the PAM with nearest
neighbor hybridization and $d$ hopping \cite{het,nyo,ki,tiz} was
found. These results show that the range of the hopping and
hybridization in the PAM are quite important but still poorly
understood.

In this situation exact results on the existence of insulating and
metallic phases in the PAM and their dependence on the hopping,
hybridization, and interaction parameters are particularly
desirable -- especially in three dimensions. So far exact results
for the PAM were mostly limited to special regions of
parameter space, namely, for infinite repulsion of the $f$ electrons \cite%
{tizegy,tizket,tizhar}, and for finite repulsion in low dimensions
$D=1,2$ \cite{tine,tineu,tiot,tiot1}

In this paper we not only present details of the construction and
the
physical properties of a class of exact ground states of three-dimensional ($%
D=3$) periodic Anderson models reported in Ref.\cite{prl}, but
extend the range of applicability of our approach substantially.
In particular, we (i) demonstrate the uniqueness of the metallic
and insulating solutions discovered at $3/4$ filling \cite{prl},
(ii) explicitly present and analyze the non--linear matching
conditions connecting the starting Hamiltonian to the transformed
Hamiltonian, (iii) deduce the current operator and the sum rule
for the charge conductivity, (iv) derive several local and global
expectation values such as the magnetization of the system, (v)
calculate correlation functions, (vi) extend the solutions to the
conventional PAM case, and (vii) show that, by employing different
procedures to cast the Hamiltonian into positive semi-definite
form, one arrives at exact ground states in different regions of
the parameter space.

The paper is structured as follows. In Section II we present the
Hamiltonian, discuss its transformation into positive
semi-definite form, and construct a class of exact ground states.
Section III describes the localized solution, and Section IV
characterizes the metallic non-Fermi liquid state. In Section V
the approach is generalized, leading to solutions in other regions
of parameter space, and in Section VI the results are summarized.
Technical details are discussed in Appendices A - E.


\section{Transformation of the Hamiltonian and Construction of Exact Ground States}


\subsection{General form of the periodic Anderson model}


We consider a general form of the periodic Anderson model (PAM)
describing non-interacting $d$ electrons which hybridize with
interacting $f$ electrons. In contrast to the conventional PAM we
do not assume the $f$ electrons to be localized, i.e., the
Hamiltonian is given by
\newcounter{equ}
\setcounter{equ}{1}
\def\theequation{\arabic{equation}.\alph{equ}}
\begin{eqnarray}
\hat{H} &=&\hat{H}_{0}+\hat{H}_{U},
\label{Eq1} \\
\stepcounter{equ} \setcounter{equation}{1}
\hat{H}_{0} &=&\sum_{\mathbf{k},\sigma }[(\epsilon _{%
\mathbf{k}}^{d}\hat{n}_{\mathbf{k}\sigma }^{d}+\epsilon _{\mathbf{k}}^{f}%
\hat{n}_{\mathbf{k}\sigma
}^{f})+(V_{\mathbf{k}}\hat{d}_{\mathbf{k}\sigma
}^{\dagger }\hat{f}_{\mathbf{k}\sigma }+V_{\mathbf{k}}^{\ast }\hat{f}_{%
\mathbf{k}\sigma }^{\dagger }\hat{d}_{\mathbf{k}\sigma })],
\label{Eq2} \\
\stepcounter{equ} \setcounter{equation}{1}
\hat{H}_{U} &=&U\sum_{\mathbf{i}}\hat{n}_{\mathbf{i},\uparrow }^{f}\hat{n}_{%
\mathbf{i},\downarrow }^{f}. \label{Eq2i}
\end{eqnarray}
\def\theequation{\arabic{equation}}

We denote the two types of electrons by $b=d,f$, i.e., $\hat{b}_{\mathbf{k}%
\sigma }^{\dagger }$ creates a $b(=d,f)$ electron with momentum
$\mathbf{k}$ and spin $\sigma $. The corresponding particle number
operators are $\hat{n}_{\mathbf{k}\sigma
}^{b}=\hat{b}_{\mathbf{k}\sigma }^{\dagger
}\hat{b}_{\mathbf{k}\sigma }$,
and the dispersion relations of the $b$-electrons are given by $\epsilon _{%
\mathbf{k}}^{b}$. Furthermore, the hybridization amplitude and the
local (Hubbard) interaction are denoted by $V_{\mathbf{k}}$ and
$U$, respectively.

In real space the model is defined on a general Bravais lattice in
$D=3$
dimensions, with a unit cell $I$ defined by the primitive vectors $\{\mathbf{%
x}_{\tau }\}$, $\tau =1,2,3$. The non-interacting part of the Hamiltonian, $%
\hat{H}_{0}$, reads%
\begin{eqnarray}
\hat{H}_{0} &=&\sum_{\mathbf{i},\sigma }\{\sum_{\mathbf{r}}[(t_{\mathbf{r}%
}^{d}\hat{d}_{\mathbf{i},\sigma }^{\dagger }\hat{d}_{\mathbf{i}+\mathbf{r}%
,\sigma }+t_{\mathbf{r}}^{f}\hat{f}_{\mathbf{i},\sigma }^{\dagger }\hat{f}_{%
\mathbf{i}+\mathbf{r},\sigma })+(V_{\mathbf{r}}^{d,f}\hat{d}_{\mathbf{i}%
,\sigma }^{\dagger }\hat{f}_{\mathbf{i}+\mathbf{r},\sigma }+V_{\mathbf{r}%
}^{f,d}\hat{f}_{\mathbf{i},\sigma }^{\dagger }\hat{d}_{\mathbf{i}+\mathbf{r}%
,\sigma })+H.c.]  \nonumber \\
&+&(V_{0}\hat{d}_{\mathbf{i},\sigma }^{\dagger
}\hat{f}_{\mathbf{i},\sigma
}+H.c.)+E_{f}\hat{n}_{\mathbf{i},\sigma }^{f}\},  \label{Eq3}
\end{eqnarray}%
where $t_{\mathbf{r}}^{b}$ characterizes the hopping of $b$
electrons
between sites $\mathbf{i}$ and $\mathbf{i}+\mathbf{r}$, $V_{\mathbf{r}%
}^{b,b^{\prime }}$ is the hybridization of $b$ and $b^{\prime }$
electrons at sites $\mathbf{i}$ and $\mathbf{i}+\mathbf{r}$,
$V_{0}$ is the on-site hybridization, and $E_{f}$ is the local
on-site $f$-electron energy. The separation between a site
$\mathbf{i}$ and its neighbors in the unit cell is
denoted by $\mathbf{r}$, with $\mathbf{r}\neq 0$. While the amplitudes $t_{%
\mathbf{r}}^{b}$ are real, $V_{0},V_{\mathbf{r}}^{b,b^{\prime }}$
can, in principle, be complex (whether
$V_{\mathbf{r}}^{b,b^{\prime }}$ is real, imaginary or complex,
depends on the linear combination of the corresponding
electronic orbitals and hence on the lattice symmetry \cite%
{het,tiot,im1,im2,im3}) and obey the relations
\begin{eqnarray}
\epsilon _{\mathbf{k}}^{b}=E_{f}\delta _{b,f}+\sum_{\mathbf{r}}(t_{\mathbf{r}%
}^{b}e^{-i\mathbf{k}\mathbf{r}}+{t_{\mathbf{r}}^{b}}^{\ast }e^{+i\mathbf{k}%
\mathbf{r}}),\quad V_{\mathbf{k}}=V_{0}+\sum_{\mathbf{r}}(V_{\mathbf{r}%
}^{d,f}e^{-i\mathbf{k}\mathbf{r}}+{V_{\mathbf{r}}^{f,d}}^{\ast }e^{+i\mathbf{k}%
\mathbf{r}}). \label{Eq4}
\end{eqnarray}

In particular, for $t_{\mathbf{r}}^{f}=0$, the $f$-electrons are
localized, and the model reduces to the conventional PAM.


\subsection{Transformation of the Hamiltonian.}


\subsubsection{Representation of sites in a unit cell}


The separation from a site $\mathbf{i}$ in (\ref{Eq3}) is
indicated by the vector $\mathbf{r}$ which corresponds to
neighboring sites located in different coordination spheres. In
our investigation $\mathbf{r}$ may extend over a \emph{unit cell}
$I$ of a general Bravais lattice in $D=3$. This implies 26
different inter-site hopping and non-local hybridization
amplitudes. To avoid multiple counting of contributions by the
$H.c.$ term in (\ref{Eq3}) the vector $\mathbf{r}$ must be
properly defined. To this end the sites within $I_{\mathbf{i}}$,
the unit cell defined at site $\mathbf{i}$,
are denoted by $\mathbf{r}_{I_{\mathbf{i}}}=\mathbf{i}+\mathbf{r%
}_{\alpha \beta \gamma }$, with $\mathbf{r}_{\alpha \beta \gamma
}=\alpha \mathbf{x}_{1}+\beta \mathbf{x}_{2}+\gamma
\mathbf{x}_{3}$; $\alpha ,\beta
,\gamma =0,1$. As shown in Fig. 1\ the eight sites $\mathbf{r}_{I_{\mathbf{i}%
}}$ can be numbered by the indices $n(\alpha ,\beta ,\gamma
)=1+\alpha +3\beta +4\gamma -2\alpha \beta $ \emph{without}
reference to $I_{\mathbf{i}}$. Then $\mathbf{r}=\mathbf{r}_{\alpha
^{\prime }\beta ^{\prime }\gamma ^{\prime }}-\mathbf{r}_{\alpha
\beta \gamma }$, with $n(\alpha \prime ,\beta \prime ,\gamma
\prime )>n(\alpha ,\beta ,\gamma )$, connects any two sites within
a unit cell. It corresponds to half of the 26 possibilities, i.e.,
to the 13 possibilities $\mathbf{x}_{1},\mathbf{x}_{2},\mathbf{x}%
_{3},\mathbf{x}_{2}\pm \mathbf{x}_{1},\mathbf{x}_{3}\pm \mathbf{x}_{1},%
\mathbf{x}_{3}\pm \mathbf{x}_{2},\mathbf{x}_{3}\pm
\mathbf{x}_{2}\pm \mathbf{x}_{1}$. The remaining (negative) values
of $\mathbf{r}$\ are taken into account in (\ref{Eq3}) by the
$H.c.$ contributions.


\subsubsection{Transformation of $\hat{H}$ into positive semi-definite form}


To construct exact ground states the Hamiltonian $\hat{H}$ needs
to be
rewritten in terms of positive semi-definite operators. This is made possible%
\cite{mi4} by the construction of two new operators -- one for the
transformation of $\hat{H}_{0}$ and one for $\hat{H}_{U}.$

The first of these operators is the "unit cell operator" $\hat{A}_{I_{%
\mathbf{i}},\sigma }^{\dagger }$ which represents a superposition
of fermionic operators creating $d$ or $f$ electrons with spin
$\sigma $ inside every unit cell $I_{\mathbf{i}}$ as
\begin{eqnarray}
&&\hat{A}_{I_{\mathbf{i}},\sigma }^{\dagger }=\sum_{n(\alpha
,\beta ,\gamma )=1}^{8}[a_{n,d}^{\ast
}\hat{d}_{\mathbf{i}+\mathbf{r}_{\alpha ,\beta
,\gamma },\sigma }^{\dagger }+a_{n,f}^{\ast }\hat{f}_{\mathbf{i}+\mathbf{r}%
_{\alpha ,\beta ,\gamma },\sigma }^{\dagger }]  \nonumber \\
&=&(a_{1,d}^{\ast }\hat{d}_{\mathbf{i},\sigma }^{\dagger }+a_{2,d}^{\ast }%
\hat{d}_{\mathbf{i}+\mathbf{x}_{1},\sigma }^{\dagger }+a_{3,d}^{\ast }\hat{d}%
_{\mathbf{i}+\mathbf{x}_{1}+\mathbf{x}_{2},\sigma }^{\dagger
}+a_{4,d}^{\ast
}\hat{d}_{\mathbf{i}+\mathbf{x}_{2},\sigma }^{\dagger }+....+a_{8,d}^{\ast }%
\hat{d}_{\mathbf{i}+\mathbf{x}_{2}+\mathbf{x}_{3},\sigma
}^{\dagger })
\nonumber \\
&&+(a_{1,f}^{\ast }\hat{f}_{\mathbf{i},\sigma }^{\dagger }+a_{2,f}^{\ast }%
\hat{f}_{\mathbf{i}+\mathbf{x}_{1},\sigma }^{\dagger }+a_{3,f}^{\ast }\hat{f}%
_{\mathbf{i}+\mathbf{x}_{1}+\mathbf{x}_{2},\sigma }^{\dagger
}+a_{4,f}^{\ast
}\hat{f}_{\mathbf{i}+\mathbf{x}_{2},\sigma }^{\dagger }+....+a_{8,f}^{\ast }%
\hat{f}_{\mathbf{i}+\mathbf{x}_{2}+\mathbf{x}_{3},\sigma
}^{\dagger }). \label{Eq7}
\end{eqnarray}%
Because of the translational symmetry of the lattice, the
numerical prefactors $a_{n,b}^{\ast }$, $n=n(\alpha ,\beta ,\gamma
)$ are \emph{the same} in every unit cell. The composite operators $\hat{A}_{I_{%
\mathbf{i}},\sigma }^{\dagger }$ do not obey canonical
anticommutation rules, since $\{\hat{A}_{I,\sigma
},\hat{A}_{I^{\prime },\sigma ^{\prime
}}^{\dagger }\}\neq 0$ for all $I\neq I^{\prime }$. This is because $\hat{A}%
_{I_{\mathbf{i}},\sigma }^{\dagger }$ creates electrons also at
the boundaries of the unit cell $I_{\mathbf{i}}$ with neighboring
unit cells. It should be stressed that for this reason
$\hat{A}_{I_{\mathbf{i}},\sigma }^{\dagger }$ has a genuine
dependence on the lattice structure and
thereby on the spatial dimension. Furthermore the relations $\{\hat{A}%
_{I,\sigma }^{\dagger },\hat{A}_{I^{\prime },\sigma ^{\prime
}}^{\dagger
}\}=\{\hat{A}_{I,\sigma },\hat{A}_{I^{\prime },\sigma ^{\prime }}\}=0$ and $%
\{\hat{A}_{I,\sigma },\hat{A}_{I,\sigma }^{\dagger }\}=K_{d}+K_{f}$, $%
K_{b}=\sum_{n=1}^{8}|a_{n,b}|^{2}$ imply \setcounter{equ}{1}
\def\theequation{\arabic{equation}.\alph{equ}}
\begin{eqnarray}
&&(\hat{A}_{I,\sigma }^{\dagger })^2=0,
\label{Eq8} \\
\stepcounter{equ}\setcounter{equation}{5} &&-\hat{A}_{I,\sigma }^{\dagger }%
\hat{A}_{I,\sigma }=\hat{A}_{I,\sigma }\hat{A}_{I,\sigma
}^{\dagger }-(K_{d}+K_{f}).  \label{Eq9}
\end{eqnarray}
\def\theequation{\arabic{equation}}

The second operator, $\hat{P}$, originates from the relation $\sum_{\mathbf{i%
}}\hat{n}_{\mathbf{i},\uparrow }^{f}\hat{n}_{\mathbf{i},\downarrow }^{f}=%
\hat{P}+\sum_{\mathbf{i},\sigma }\hat{n}_{\mathbf{i},\sigma
}^{f}-N_{\Lambda }$, where
\begin{eqnarray}
\hat{P}=\sum_{\mathbf{i}}\hat{P}_{\mathbf{i}},\quad \hat{P}_{\mathbf{i}}=%
\hat{n}_{\mathbf{i},\uparrow }^{f}\hat{n}_{\mathbf{i},\downarrow }^{f}-(\hat{%
n}_{\mathbf{i},\uparrow }^{f}+\hat{n}_{\mathbf{i},\downarrow
}^{f})+1, \label{pp}
\end{eqnarray}%
and $N_{\Lambda }$ is the number of lattice sites. The local operators $\hat{%
P}_{\mathbf{i}}$ are positive semi-definite and assume their
lowest eigenvalue (=0) whenever there is \emph{at least} one $f$
electron on site $\mathbf{i}$. By contrast, the interaction
operator $\hat{H}_{U}$ itself, which is also positive
semi-definite for $U>0$, assumes the eigenvalue 0 only if the
 double occupancy is \emph{exactly zero}.
For this reason $\hat{P}$ is more useful for our investigation
than $\hat{H}_{U}$.



Taking into account periodic boundary conditions and allowing
$\mathbf{r}$ to take only the values discussed above,
(\ref{Eq1}, \ref{Eq3}) may be written as
\begin{eqnarray}
\hat{H} =  \hat{P}_{A}+U\hat{P}+E_{g}, \label{ham}
\end{eqnarray}%
where $\hat{P}_{A}=\sum_{\mathbf{i},\sigma }\hat{A}_{I_{\mathbf{i}},\sigma }%
\hat{A}_{I_{\mathbf{i}},\sigma }^{\dagger }$,
$E_{g}=K_{d}N+UN_{\Lambda }-2N_{\Lambda }(2K_{d}-E_{f})$, and $N$
is the total number of particles which is assumed to be fixed.

For (\ref{ham}) to reproduce (\ref{Eq1}) the prefactors
$a_{n,b}^{\ast }$ in $\hat{A}%
_{I_{\mathbf{i}}\sigma }^{\dagger }$ must be expressed in terms of
the
microscopic parameters $t_{\mathbf{r}}^{d}$, $t_{\mathbf{r}}^{f}$, $V_{%
\mathbf{r}}$, $V_{\mathbf{r}}^{\ast }$, $V_{0}$, $V_{0}^{\ast }$, $E_{f}$, $%
U $, for all $\mathbf{r}\in I_{\mathbf{i}}$, taking into account
periodic boundary conditions. This leads to 55 coupled, non-linear
matching conditions which can be written in compact notation as
\cite{prl}
\[
\sum_{\beta _{1},\beta _{2},\beta
_{3}=-1}^{1}(\prod_{i=1}^{3}D_{\beta
_{i},\alpha _{i}})a_{n^{+},b}^{\ast }a_{n^{-},b^{\prime }}=T_{\bar{\mathbf{r}%
},\nu }^{b,b^{\prime }}.
\]
These matching conditions have the explicit form
\begin{eqnarray}
&&J_{\mathbf{x}_{1}}^{b,b^{\prime }}=a_{1,b}^{\ast }a_{2,b^{\prime
}}+a_{4,b}^{\ast }a_{3,b^{\prime }}+a_{5,b}^{\ast }a_{6,b^{\prime
}}+a_{8,b}^{\ast }a_{7,b^{\prime }},\quad J_{\mathbf{x}_{2}+\mathbf{x}%
_{1}}^{b,b^{\prime }}=a_{1,b}^{\ast }a_{3,b^{\prime
}}+a_{5,b}^{\ast
}a_{7,b^{\prime }},  \nonumber \\
&&J_{\mathbf{x}_{2}}^{b,b^{\prime }}=a_{1,b}^{\ast }a_{4,b^{\prime
}}+a_{2,b}^{\ast }a_{3,b^{\prime }}+a_{6,b}^{\ast }a_{7,b^{\prime
}}+a_{5,b}^{\ast }a_{8,b^{\prime }},\quad J_{\mathbf{x}_{3}+\mathbf{x}%
_{1}}^{b,b^{\prime }}=a_{1,b}^{\ast }a_{6,b^{\prime
}}+a_{4,b}^{\ast
}a_{7,b^{\prime }},  \nonumber \\
&&J_{\mathbf{x}_{3}}^{b,b^{\prime }}=a_{1,b}^{\ast }a_{5,b^{\prime
}}+a_{2,b}^{\ast }a_{6,b^{\prime }}+a_{3,b}^{\ast }a_{7,b^{\prime
}}+a_{4,b}^{\ast }a_{8,b^{\prime }},\quad J_{\mathbf{x}_{3}+\mathbf{x}%
_{2}}^{b,b^{\prime }}=a_{1,b}^{\ast }a_{8,b^{\prime
}}+a_{2,b}^{\ast
}a_{7,b^{\prime }},  \nonumber \\
&&J_{\mathbf{x}_{2}-\mathbf{x}_{1}}^{b,b^{\prime }}=a_{2,b}^{\ast
}a_{4,b^{\prime }}+a_{6,b}^{\ast }a_{8,b^{\prime }},\quad J_{\mathbf{x}_{3}-%
\mathbf{x}_{1}}^{b,b^{\prime }}=a_{2,b}^{\ast }a_{5,b^{\prime
}}+a_{3,b}^{\ast }a_{8,b^{\prime }},\quad J_{\mathbf{x}_{3}-\mathbf{x}%
_{2}}^{b,b^{\prime }}=a_{4,b}^{\ast }a_{5,b^{\prime
}}+a_{3,b}^{\ast
}a_{6,b^{\prime }},  \nonumber \\
&&J_{\mathbf{x}_{3}+\mathbf{x}_{2}+\mathbf{x}_{1}}^{b,b^{\prime
}}=a_{1,b}^{\ast }a_{7,b^{\prime }},\quad J_{\mathbf{x}_{3}+\mathbf{x}_{2}-%
\mathbf{x}_{1}}^{b,b^{\prime }}=a_{2,b}^{\ast }a_{8,b^{\prime }},\quad J_{%
\mathbf{x}_{3}-\mathbf{x}_{2}+\mathbf{x}_{1}}^{b,b^{\prime
}}=a_{4,b}^{\ast
}a_{6,b^{\prime }},  \nonumber \\
&&J_{\mathbf{x}_{3}-\mathbf{x}_{2}-\mathbf{x}_{1}}^{b,b^{\prime
}}=a_{3,b}^{\ast }a_{5,b^{\prime }},\quad
V_{0}=-\sum_{n=1}^{8}a_{n,d}^{\ast }a_{n,f},\quad
U+E_{f}=K_{d}-K_{f}, \label{Eq12}
\end{eqnarray}%
where $J_{\mathbf{r}}^{b,b^{\prime }}=-[\delta _{b,b^{\prime }}t_{\mathbf{r}%
}^{b}+(1-\delta _{b,b^{\prime }})V_{\mathbf{r}}^{b,b^{\prime }}]$, $%
b,b^{\prime }=d,f$.

Details of this transformation are presented in Appendix A.


\subsection{Construction of exact ground states}


Apart from the constant term $E_g$ in (\ref{ham}) $\hat H$ is a
positive semi-definite operator. A state $|\Psi_g\rangle$ which
fulfills the conditions \setcounter{equ}{1}
\def\theequation{\arabic{equation}.\alph{equ}}
\begin{eqnarray}
&&\hat P_{\mathbf{i}} |\Psi_g\rangle = 0, 
\label{Eq15} \\
\stepcounter{equ} \setcounter{equation}{9} &&\hat A^{\dagger}_{I_{\mathbf{i}%
},\sigma} |\Psi_g\rangle = 0 
\label{Eq16}
\end{eqnarray}
\def\theequation{\arabic{equation}}
for all $\mathbf{i}$, and which contains all linearly independent
states with properties (\ref{Eq15},\ref{Eq16}), will then be the
exact ground state of $\hat H$ with energy $E_g$. Since the kernel
of an arbitrary operator $\hat O$, $ker(\hat O)$,
is defined by the linearly independent states $|\phi\rangle$
satisfying $\hat O|\phi\rangle = 0$, the relations (\ref{Eq15})
and (\ref{Eq16}) define the kernel of the operators $\hat P$ and
$\hat P_A$, respectively. Consequently, $|\Psi_g\rangle$ spans the
common part of $ker(\hat P)$ and $ker(\hat P_A)$ denoted by the
Hilbert space
\begin{eqnarray}
{\mathcal{H}}_g = ker(\hat P_A) \cap ker(\hat P) . \label{Eq17}
\end{eqnarray}
Using this definition it is ensured that $|\Psi_g\rangle$ is the
complete ground state, and that supplementary degeneracies of $E_g$ do not
occur.

Using (\ref{pp}) and (\ref{Eq15}) it follows that $ker(\hat P)$ is
defined by states
$\hat F^{\dagger}|0\rangle= \prod_{\mathbf{i}%
=1}^{N_{\Lambda}} \hat F^{\dagger}_{ \mathbf{i}}|0\rangle$, where
$\hat F^{\dagger}_{%
\mathbf{i}} = ( \mu_{\mathbf{i},\uparrow} \hat f^{ \dagger}_{\mathbf{i}%
,\uparrow} + \mu_{\mathbf{i},\downarrow} \hat f^{\dagger}_{ \mathbf{i}%
,\downarrow} )$ and $\mu_{\mathbf{i},\sigma}$ are arbitrary
coefficients. Obviously $\hat F^{\dagger}$ creates one $f$
electron on every site $\mathbf{i}$. Furthermore, (\ref{Eq8})
and(\ref{Eq16}) imply that $ker(\hat P_A)$ is defined by states
$\hat G^{\dagger}|0\rangle$, where $\hat
G^{\dagger}=\prod_{\mathbf{i}=1}^{N_{\Lambda}}
(\hat A^{\dagger}_{I_{\mathbf{i}},\uparrow} \hat A^{\dagger}_{I_{\mathbf{i}%
},\downarrow})$ creates at most two ($d$ or $f$) electrons on
$\mathbf{i}$. Since $\hat G^{\dagger}$ also creates contributions
without $f$ electrons, (\ref{Eq17}) implies the (unnormalized)
ground state
\begin{eqnarray}
|\Psi_g\rangle = \hat G^{\dagger}F^{\dagger}|0\rangle=\prod_{\mathbf{i}=1}^{N_{\Lambda}}[ \hat A^{ \dagger}_{I_{%
\mathbf{i}},\uparrow} \hat A^{\dagger}_{I_{\mathbf{i}},\downarrow}
\hat F^{\dagger}_{\mathbf{i}} ] | 0 \rangle   \label{Eq18}
\end{eqnarray}
at $N=3N_{\Lambda}$, e.g., $3/4$ filling. Clearly,
$|\Psi_g\rangle$ has the desired property $\hat H |\Psi_g\rangle =
E_g |\Psi_g\rangle$ and
spans ${\mathcal{H}}_g$ at $3/4$ filling. Thus it is the exact
ground state of $\hat H$ with energy $E_g$.

Since the operator $\hat F^{\dagger}$ was introduced into
$|\Psi_g\rangle$ to take into account the operator $\hat P$ in
(\ref{ham}), i.e., the Hubbard interaction $U$, $|\Psi_g\rangle$
can only be a ground state for $U>0$.

Eq. (\ref{Eq18}) implies that (i) the linearly independent basis
vectors of ${\cal{H}}_g$ have the form $|\Psi_g\{
\sigma_{\mathbf{i}} \} \rangle = \hat G^{\dagger} \hat
F^{\dagger}_{\{ \sigma_{\mathbf{i}} \} } |0\rangle$, where $\hat
F^{\dagger}_{\{ \sigma_{\mathbf{i}} \} } = \prod_{ \mathbf{i}%
=1}^{N_{\Lambda}}\hat
f^{\dagger}_{\mathbf{i},\sigma_{\mathbf{i}}}$,
and (ii) $\hat G^{\dagger}$ does not contribute to the total spin
of the ground state. The overall degeneracy and total spin $S \in
[0, N_{\Lambda}/2]$ of $|\Psi_g\rangle$ are then determined only
by the (arbitrary) set of coefficients ${\mu_{\mathbf{i},\sigma}}$
(see also Sect. III.A.2., Sect.IV.E., and footnote [\cite{mi5}]).
Consequently, the degeneracy of $E_g $ is determined by the (high)
spin degeneracy of $|\Psi_g\rangle$. Since in $|\Psi_g\rangle$
all possible values  $S$ and orientations $\mathbf{S}$ occur the
ground state is globally non-magnetic \cite{aa}.

Exact ground states can also be constructed away from $3/4$
filling. For example, the operator $\hat V^{\dagger}_M = \prod_{j=1}^{M} [ \sum_{\mathbf{i}%
=1}^{N_{\Lambda}} \sum_{b=d,f,\sigma} a^b_{\mathbf{i},j,\sigma}
\hat b^{\dagger}_{\mathbf{i},\sigma}] $ with numerical coefficients $a^b_{\mathbf{i}%
,j,\sigma}$ creates $M < N_{\Lambda}$ additional particles into
the system such that
\begin{eqnarray}
|\Psi_g\rangle = \prod_{\mathbf{i}=1}^{N_{\Lambda}}[ \hat A^{ \dagger}_{I_{%
\mathbf{i}},\uparrow} \hat A^{\dagger}_{I_{\mathbf{i}},\downarrow}
\hat F^{\dagger}_{\mathbf{i}} ] \hat V^{\dagger}_M | 0 \rangle
\label{Eq19}
\end{eqnarray}
is a ground state for $U > 0$ and $N > 3 N_{\Lambda}$.


\section{Exact localized ground state}


The physical properties of $|\Psi_g\rangle$ depend on the values
of the coefficients $a_{n,b}$ in (\ref{Eq7}) which are solutions
of (\ref{Eq12}) for given microscopic parameters. We now identify
different solutions for nonlocal hybridization amplitudes with
$V^{df}_{\mathbf{r}} = V^{fd}_{ \mathbf{r}} = V_{\mathbf{r}}$ and
discuss their physical properties. We start with a localized
ground state at $3/4$ filling.


\subsection{ Derivation of the localized ground state}


From (\ref{Eq7}) it follows that if $a^{*}_{n,d}$ and
$a^{*}_{n,f}$ are proportional, i.e., $a^{*}_{n,d}= p a^{*}_{n,f}$
for all $n$,
the operators $\hat A^{\dagger}_{I_{ \mathbf{i}},\sigma}$ take the
form
\begin{eqnarray}
\hat A^{\dagger}_{I_{\mathbf{i}},\sigma} =
\sum_{n(\alpha,\beta,\gamma)=1}^8
a^{*}_{n(\alpha,\beta,\gamma),f} E^{\dagger}_{\mathbf{i}+\mathbf{r}%
_{\alpha,\beta,\gamma},\sigma} ,  \label{Eq20}
\end{eqnarray}
where $\hat E^{\dagger}_{\mathbf{i},\sigma}= (p\hat d^{\dagger}_{\mathbf{i},%
\sigma} + \hat f^{\dagger}_{\mathbf{i},\sigma})$.
The ground state $|\Psi_g\rangle$, (\ref{Eq18}), then transforms
into $|\Psi_{loc}\rangle = \prod_{\mathbf{i}=1}^{N_{\Lambda}} (
\hat E^{\dagger}_{ \mathbf{i},\uparrow} \hat
E^{\dagger}_{\mathbf{i},\downarrow} \hat F^{\dagger}_{ \mathbf{i}}
) | 0 \rangle$.
Evaluating the product $\hat E^{ \dagger}_{%
\mathbf{i},\uparrow} \hat E^{\dagger}_{\mathbf{i},\downarrow} \hat
F^{ \dagger}_{\mathbf{i}} $, one finds
\begin{eqnarray}
|\Psi_{loc}\rangle = \prod_{\mathbf{i}=1}^{N_{\Lambda}}
[\sum_{\sigma} \mu_{ \mathbf{i},\sigma} ( p \hat
d^{\dagger}_{\mathbf{i},\downarrow} \hat d^{\dagger}_{
\mathbf{i},\uparrow} \hat f^{\dagger}_{\mathbf{i},\sigma} +
\hat f^{\dagger}_{ \mathbf{i},\uparrow} \hat f^{\dagger}_{\mathbf{i}%
,\downarrow} \hat d^{\dagger}_{ \mathbf{i},\sigma} ) ] |0 \rangle
. \label{Eq23}
\end{eqnarray}
Since $\langle \Psi_{loc}| \Psi_{loc} \rangle = (1+
|p|^2)^{ N_{\Lambda}}\prod_{\mathbf{i}=1}^{N_{\Lambda}} (|\mu_{\mathbf{i}%
,\uparrow}|^2 + |\mu_{\mathbf{i},\downarrow}|^2 ) \ne 0$ the
ground state is well-defined.


\subsubsection{The insulating nature of the ground state}


The state $|\Psi_{loc}\rangle$ has
exactly three particles on each site, corresponding to a uniform
electron distribution in the system.
Indeed, for $%
\hat n_{\mathbf{i}} = \sum_{b=d,f} \sum_{\sigma} \hat n^b_{\mathbf{i}%
,\sigma} $ one finds $\hat n_{\mathbf{i}} |\Psi_{loc}\rangle = 3
|\Psi_{loc}\rangle$. Denoting ground state expectation values in
terms of $|\Psi_{loc}\rangle$  by $\langle .... \rangle$, one
obtains
\begin{eqnarray}
\langle \hat d^{\dagger}_{\mathbf{i},\sigma} \hat d_{\mathbf{j}%
,\sigma^{\prime}} \rangle = 0, \quad \langle \hat d^{\dagger}_{\mathbf{i}%
,\sigma} \hat f_{\mathbf{j},\sigma^{\prime}} \rangle = 0, \quad
\langle \hat f^{\dagger}_{\mathbf{i},\sigma} \hat
d_{\mathbf{j},\sigma^{\prime}} \rangle
= 0, \quad \langle \hat f^{\dagger}_{\mathbf{i},\sigma} \hat f_{\mathbf{j}%
,\sigma^{\prime}} \rangle = 0,  \label{Eq24}
\end{eqnarray}
for all $\sigma,\sigma^{\prime}$ and all $\mathbf{i} \ne
\mathbf{j}$. Hence hopping or non-local hybridization does not
occur.

 By separating the Hamiltonian $\hat H$ into an
itinerant part $\hat H_{itin}= \sum_{\mathbf r} \hat
H_{itin}(\mathbf r)$ and a complementary localized part $\hat
H_{loc}=\hat H - \hat H_{itin}$ (see Appendix B), and using
(\ref{Eq24}), one finds $\langle \hat H_{itin}(\mathbf r) \rangle
=0$ for all $\mathbf{r}$, and $\langle \hat H_{loc} \rangle =
E_g$. This clearly demonstrates the localized nature of the ground
state. Furthermore, from (\ref{B7}), the sum rule for the charge
conductivity is obtained as $\int^{\infty}_{0} d \omega Re
\sigma_{\tau,\tau} (\omega) = 0$. Since $Re \sigma_{\tau,\tau}
(\omega)$
is non-negative this relation implies $Re \sigma_{\tau,\tau}
(\omega) = 0$. In particular, $Re \sigma_{\tau,\tau} (0) = 0$, the
DC-conductivity, is also zero. The ground state (\ref{Eq23}) is
therefore insulating \cite{aa1,br}. It should be noted that the
nature of this state is quite nontrivial, since the localization
of the electrons is due to a subtle quantum mechanical
interference
between states with two $d$ and one $f$ electron ($d^{\dagger}_{\mathbf{i},\uparrow} d^{\dagger}_{\mathbf{%
i},\downarrow} f^{\dagger}_{\mathbf{i},\sigma}$) and two $f$ and one $d$ electron ($f^{\dagger}_{\mathbf{i}%
,\uparrow} f^{ \dagger}_{\mathbf{i},\downarrow} d^{\dagger}_{\mathbf{i}%
,\sigma}$) on every site $\mathbf{i}$.

A state with $Re \sigma_{\tau,\tau} (\omega) = 0$ for all $\omega$
appears to be rather unphysical since it implies that not only the
DC-conductivity but even the dynamic conductivity vanishes for all
excitation energies. It should be stressed, however, that the
relation $Re \sigma_{\tau,\tau}(\omega)=0$ was derived in the
framework of the Kubo formula for the charge conductivity, i.e.,
within linear response theory. Consequently, this result is not
valid at high excitation energies $\omega$.


\subsubsection{Global magnetic properties}


The expectation value of the spin \cite{sspin} in terms of the
ground state (\ref{Eq23}) in Cartesian coordinates is found as
\begin{eqnarray}
\langle \hat {\mathbf{S}} \rangle = {\mathbf{x}} \sum_{\mathbf{i}%
=1}^{N_{\Lambda}} \frac{\mu_{\mathbf{i},\downarrow} \mu^{*}_{\mathbf{i}%
,\uparrow} + \mu_{\mathbf{i}, \uparrow}
\mu^{*}_{\mathbf{i},\downarrow}}{2
(|\mu_{\mathbf{i},\uparrow}|^2 + |\mu_{ \mathbf{i},\downarrow}|^2)} + %
{\mathbf{y}} \sum_{\mathbf{i}=1}^{N_{\Lambda}} \frac{(-i)(\mu_{\mathbf{i}%
,\downarrow} \mu^{*}_{\mathbf{i},\uparrow} - \mu_{\mathbf{i},
\uparrow} \mu^{*}_{\mathbf{i},\downarrow})}{2
(|\mu_{\mathbf{i},\uparrow}|^2 + |\mu_{
\mathbf{i},\downarrow}|^2)} + {\mathbf{z}} \sum_{\mathbf{i}%
=1}^{N_{\Lambda}} \frac{|\mu_{\mathbf{i},\uparrow}|^2 - |\mu_{\mathbf{i}%
,\downarrow}|^2}{ 2 (|\mu_{\mathbf{i},\uparrow}|^2 + |\mu_{\mathbf{i}%
,\downarrow}|^2)} ,  \label{Eq26}
\end{eqnarray}
where ${|\mathbf{x}}|=|{\mathbf{y}}|=|{\mathbf{z}}|=1$.
The total spin
is seen to depend on the arbitrary coefficients $\{ \mu_{\mathbf{%
i}, \sigma} \}$. Here the site dependence of the
$\mu_{\mathbf{i},\sigma}$ coefficients should be stressed. Namely,
by choosing $\mu_{\mathbf{i},\sigma} = \mu_{\sigma}$ one obtains
\begin{eqnarray}
\langle {\hat {\mathbf{S}}}^2 \rangle = \frac{N_{\Lambda}}{2} ( \frac{%
N_{\Lambda}}{ 2} + 1 ), \quad \langle \hat S^{z} \rangle = \frac{N_{\Lambda}%
}{2} \frac{ (|\mu_{\uparrow}|^2 -
|\mu_{\downarrow}|^2)}{(|\mu_{\uparrow}|^2 +
|\mu_{\downarrow}|^2)} ,  \label{Eq28}
\end{eqnarray}
which represents a ferromagnetic state with maximal total spin
$S/N_{\Lambda}=1/2$, leading to $\sqrt{\langle {\hat
{\mathbf{S}}}^2 \rangle }/N_{\Lambda} = 1/2$ in the thermodynamic
limit; $\mu_{\sigma}$ is seen to influence only the orientation of
$\mathbf{S}$.

The minimal total spin can be found by considering two distinct
subsystems
of arbitrary shape, both containing the same number of lattice
sites $N_{\Lambda}/2$. Taking $\mu_{\mathbf{i},\sigma} =
\mu_{\sigma},  \mu_{\mathbf{i},-\sigma} = 0$ in each subsystem
one finds
\begin{eqnarray}
\frac{ \sqrt{\langle {\hat {\mathbf{S}}}^2 \rangle }}{N_{\Lambda}} = \frac{1%
}{ \sqrt{2 N_{\Lambda}}} , \quad \langle \hat S^{x} \rangle =
\langle \hat S^{y} \rangle = \langle \hat S^{z} \rangle = 0,
\label{Eq29}
\end{eqnarray}
which implies zero total spin (i.e., a global singlet state) in
the thermodynamic limit. Depending on the choice of the parameters
$\{ \mu_{\mathbf{i},\sigma} \}$ all values of $S$ between these
two extreme values for $S$, and all orientations of ${\mathbf S}$,
can be constructed (see Sec.II.C).


\subsubsection{Local magnetic properties}


Analyzing the local magnetic properties of the ground state one
observes that the expectation value of the double occupancy per
site for both $b=d,f$ electrons is smaller
than unity since $\langle \hat n^b_{\mathbf{i},\uparrow} \hat n^b_{\mathbf{i}%
,\downarrow} \rangle = (\delta_{b,f} + |p|^2
\delta_{b,d})/(1+|p|^2)$. Consequently, each site carries a local
moment. Indeed, irrespective of the values of
$\mu_{\mathbf{i},\sigma}$ and
$p$ one has $\langle {\hat {\mathbf{S}}_{%
\mathbf{i}}}^2 \rangle = 3/4$ on each site $\mathbf{i}$; this is
the result of a quantum mechanical superposition of the
corresponding contributions of $d$ and $f$ electrons. The $f$ and
$d$ moments do not compensate each other locally since
\begin{eqnarray}
(\langle \hat {\mathbf{S}}_{\mathbf{i},f} \rangle + \langle \hat {\mathbf{S}}%
_{\mathbf{i},d} \rangle ) (\sum_{\sigma} |\mu_{\mathbf{i},\sigma}|^2) =%
{\mathbf{x}} Re (\mu_{\mathbf{i},\downarrow} \mu^{*}_{\mathbf{i}%
,\uparrow}) + {\mathbf{y}} Im (\mu_{\mathbf{i},\downarrow}
\mu^{*}_{\mathbf{i}, \uparrow}) + {\mathbf{z}} \frac{|\mu_{\mathbf{%
i},\uparrow}|^2 - |\mu_{\mathbf{i},\downarrow}|^2}{2},
\label{Eq32}
\end{eqnarray} where
 $\langle \hat {\mathbf{S}}_{\mathbf{i},d} \rangle
= |p|^2 \langle \hat {\mathbf{S}}_{\mathbf{i},f} \rangle $ holds.
Furthermore, taking into account fixed (but arbitrary)
$\mu_{\mathbf{i},\sigma}$ the spin-spin correlation function for
$\mathbf{i} \ne \mathbf{j}$ is found as
\begin{eqnarray}
\langle \hat S^z_{\mathbf{i}} \hat S^z_{\mathbf{j}} \rangle =
\frac{1}{4}
\frac{|\mu_{\mathbf{i},\uparrow}|^2-|\mu_{\mathbf{i},\downarrow}|^2}{ |\mu_{%
\mathbf{i},\uparrow}|^2+|\mu_{\mathbf{i},\downarrow}|^2} \frac{|\mu_{\mathbf{%
j},\uparrow}|^2-|\mu_{\mathbf{j},\downarrow}|^2}{ |\mu_{\mathbf{j}%
,\uparrow}|^2+|\mu_{\mathbf{j},\downarrow}|^2} , \quad \langle
\hat
S^{+}_{\mathbf{i}} \hat S^{-}_{\mathbf{j}} \rangle = \frac{\mu_{\mathbf{i}%
,\downarrow} \mu_{\mathbf{i},\uparrow}^{*}}{ |\mu_{\mathbf{i}%
,\uparrow}|^2+|\mu_{\mathbf{i},\downarrow}|^2} \frac{\mu_{\mathbf{j}%
,\uparrow} \mu_{\mathbf{j},\downarrow}^{*}}{ |\mu_{\mathbf{j}%
,\uparrow}|^2+|\mu_{\mathbf{j},\downarrow}|^2}.   \label{Eq33}
\end{eqnarray}
An average over all possible values of $\mu_{\mathbf{i},\sigma},\mu_{\mathbf{j}%
,\sigma} $ therefore implies
$\langle \mathbf{S}_{\mathbf{i}} \mathbf{S}_{%
\mathbf{j}} \rangle =0$. Therefore, in spite of the existence of
local moments the system is globally non-magnetic; this is a
consequence of the large spin degeneracy of the ground state.


\subsection{Solutions of the matching conditions}




The  matching conditions (\ref{Eq12}) for the nonlocal
hybridization amplitudes along the space diagonals of the unit
cell read
\begin{eqnarray}
&&- V_{\mathbf{x}_3 + \mathbf{x}_2 + \mathbf{x}_1} = a^{*}_{1,d}
a_{7,f} = a^{*}_{1,f} a_{7,d}, \quad - V_{\mathbf{x}_3 +
\mathbf{x}_2 - \mathbf{x}_1}
= a^{*}_{2,d} a_{8,f} = a^{*}_{2,f} a_{8,d},  \nonumber \\
&&- V_{\mathbf{x}_3 - \mathbf{x}_2 + \mathbf{x}_1} = a^{*}_{4,d}
a_{6,f} = a^{*}_{4,f} a_{6,d}, \quad - V_{\mathbf{x}_3 -
\mathbf{x}_2 - \mathbf{x}_1} = a^{*}_{3,d} a_{5,f} = a^{*}_{3,f}
a_{5,d}. \label{Eq34}
\end{eqnarray} They hold for all $\{\mathbf{%
x}_{\tau }\}$, $\tau =1,2,3$ and hence imply
$a^{*}_{n,d}/a^{*}_{n,f} = a_{n^{\prime},d}/a_{n^{\prime},f} = p =
p^{*}$.  Since for real $p$, (\ref{Eq12}) leads to the relations
\begin{eqnarray}
t^f_{\mathbf{r}} = \frac{t^d_{\mathbf{r}}}{p^{2}},%
\quad V_{\mathbf{r}} =\frac{%
t^d_{\mathbf{r}}}{p}, \quad \frac{U+E_f}{V_0} = \frac{1-p^{2}}{p} ,%
\label{Eq35}
\end{eqnarray}
\emph{real} $p$ are seen to imply \emph{real} hybridization
amplitudes. In addition to (\ref{Eq35}), (\ref{Eq12}) yields the
following system of 14 coupled nonlinear equations
\begin{eqnarray}
&&-t^d_{\mathbf{x}_1} = a^{*}_{1,d} a_{2,d} + a^{*}_{8,d} a_{7,d}
+
a^{*}_{4,d} a_{3,d} + a^{*}_{5,d} a_{6,d}  ,  \nonumber \\
&&-t^d_{\mathbf{x}_2} = a^{*}_{1,d} a_{4,d} + a^{*}_{6,d} a_{7,d}
+
a^{*}_{2,d} a_{3,d} + a^{*}_{5,d} a_{8,d} ,  \nonumber \\
&&-t^d_{\mathbf{x}_3} = a^{*}_{1,d} a_{5,d} + a^{*}_{3,d} a_{7,d}
+
a^{*}_{2,d} a_{6,d} + a^{*}_{4,d} a_{8,d} ,  \nonumber \\
&&-t^d_{\mathbf{x}_2+\mathbf{x}_1} = a^{*}_{1,d} a_{3,d} +
a^{*}_{5,d} a_{7,d} , \quad -t^d_{\mathbf{x}_2-\mathbf{x}_1} =
a^{*}_{2,d} a_{4,d} +
a^{*}_{6,d} a_{8,d} ,  \nonumber \\
&&-t^d_{\mathbf{x}_3+\mathbf{x}_1} = a^{*}_{1,d} a_{6,d} +
a^{*}_{4,d} a_{7,d} , \quad -t^d_{\mathbf{x}_3-\mathbf{x}_1} =
a^{*}_{2,d} a_{5,d} +
a^{*}_{3,d} a_{8,d} ,  \nonumber \\
&&-t^d_{\mathbf{x}_3+\mathbf{x}_2} = a^{*}_{1,d} a_{8,d} +
a^{*}_{2,d} a_{7,d} , \quad -t^d_{\mathbf{x}_3-\mathbf{x}_2} =
a^{*}_{4,d} a_{5,d} +
a^{*}_{3,d} a_{6,d} ,  \nonumber \\
&&-t^d_{\mathbf{x}_3+\mathbf{x}_2+\mathbf{x}_1} = a^{*}_{1,d}
a_{7,d}, \quad -t^d_{\mathbf{x}_3+\mathbf{x}_2-\mathbf{x}_1} =
a^{*}_{2,d} a_{8,d}, \quad
-t^d_{\mathbf{x}_3-\mathbf{x}_2+\mathbf{x}_1} = a^{*}_{4,d}
a_{6,d},
\nonumber \\
&&-t^d_{\mathbf{x}_3-\mathbf{x}_2-\mathbf{x}_1} = a^{*}_{3,d}
a_{5,d} , \quad p V_0 = - \sum_{n=1}^8 |a_{n,d}|^2.   \label{Eq36}
\end{eqnarray} They determine the unknown complex coefficients $a_{n,d}$
(i.e., 16 unknown real values) from the input parameters
$t^d_{\mathbf{r}}$.  A study of the possible solutions shows that
for $|p| > 1$  the \emph{relative} sizes of the hopping and
hybridization amplitudes are physically very reasonable, e.g.,
$|t^f_{\mathbf{x}_1}| < |t^d_{\mathbf{x}_1}|$, %
$|t^d_{\mathbf{x}_1 + \mathbf{x}_2}| < |t^d_{\mathbf{x}_1}|$, %
$|t^f_{\mathbf{x}_1 + \mathbf{x}_2}| < |t^f_{\mathbf{x}_1}|$.
That is, they decrease with distance, and the magnitude of the
amplitudes of the $d$ electrons is larger than those of the almost
localized $f$ electrons.

Based on (\ref{ham}), the corresponding ground state energy
becomes $E_g/N_{\Lambda} = - U + (1-2/p^{2})\sum_{n=1}^8
|a_{n,d}|^2$. Depending on the solution, $E_g$ has a nontrivial
structure which will be analyzed below.


\subsubsection{Solution for a cubic lattice}


For a simple cubic lattice one has $t^d_{\mathbf{x}_1} =%
t^d_{\mathbf{x}_2} = t^d_{\mathbf{x}_3} = t^d_1$, $t^d_{\mathbf{x}_2 \pm%
\mathbf{x}_1} = t^d_{\mathbf{x}_3 \pm \mathbf{x}_1} = t^d_{\mathbf{x}_3 \pm%
\mathbf{x}_2} = t^d_2$, $t^d_{\mathbf{x}_3 \pm \mathbf{x}_2 \pm \mathbf{x}%
_1} = t^d_3$. Then (\ref{Eq36}) has a solution
\begin{eqnarray}
a_{1,d}=a_1, \quad a_{2,d}=a_{4,d}=a_{5,d}=u a_1, \quad %
a_{3,d}=a_{6,d}= a_{8,d}=u^{2} a_1, \quad a_{7,d}=u^{3} a_1, %
\label{Eq37}
\end{eqnarray}
where $a_1$ is an arbitrary, real quantity which is fixed by the
energy unit. Furthermore, $u$ is real and is determined by the
parameters entering in $\hat H$ through (\ref{Eq36},\ref{Eq37}).
Due to the almost localized nature of the $f$ electrons their
nearest-neighbor hopping amplitude can be expected to be much
smaller than that of the $d$ electrons. Indeed, for
$z\equiv|t^f_1/t^d_1| < 1/2$ the ground state energy is given by
\begin{eqnarray}
\frac{E_g}{UN_{\Lambda}} = - 1 + (1 - 2 \frac{|t^f_1|}{|t^d_1|}) (1 + \frac{%
(1- \sqrt{1-4y^{2}})^{2}}{4y^{2}})^{3} ,  \label{Eq38}
\end{eqnarray}
where $u =(1 - \sqrt{1-4y^{2}})/(2y)$, and $y=|t^d_2/t^d_1| \in
(0,1/2]$ holds.
We see that $|\Psi_{loc}\rangle$ ceases to be the ground state for
$y=|t^d_2/t^d_1| > 1/2=y_c$. This corresponds to rather strong
next-nearest neighbor hopping of $d$ electrons. Apparently, at
$y_c=1/2$ a different --- most probably itinerant --- phase
becomes stable. At $y_c$ the ground state energy $E_g(y)$ has a
finite value, but its derivative diverges due to $\partial
u/\partial y = + \infty$ (see Fig.2). Since the size of the
hopping element may be tuned by pressure, the infinite slope of
$E_g$ at $y=y_c$ is expected to correspond to an \emph{anomaly in
the volume}, or the compressibility, at a critical pressure $P_c$.
Such a feature is indeed observed in some heavy-fermion materials
\cite{hf}.


\subsubsection{Solutions for non-cubic lattices}


Similar results
may be deduced for other lattice structures.
In the most general case, i.e., when all hopping amplitudes
$t^d_{\mathbf{r}}$ are different, the ground state energy for the
localized solution becomes $E_g/N_{\Lambda} = - U + (1-2
|t^f_{\mathbf{x}_1}|/|t^d_{\mathbf{x}_1}|) \sum_{n=1}^8
|a_{n,d}|^2$. When at least one of the terms $|a_{n,d}|^{2}$ (see
(\ref{C8}) in Appendix C) is mathematically no longer defined the
localized solution becomes unstable. Except for accidental
cancellations in the ground state energy an infinite slope of
$E_g$ as a function of the hopping amplitudes is also found in
this more general case (see Appendix C).

\section{Exact Itinerant Ground States}

The localized ground state discussed above has exactly three
electrons per site. In general, however, the inter-site hopping
and hybridization will lead to a variable number of electrons on
each site. In that case the ground state $|\Psi_g\rangle$,
(\ref{Eq18}), becomes conducting. We will now describe solutions
of this kind.

\subsection{Solution for $a_{n,b} \ne 0$ for all $n,b$.}

To solve the matching conditions (\ref{Eq12}) for the case where
$a_{n,b} \ne 0$ for all $n,b$, we define \setcounter{equ}{1}
\def\theequation{\arabic{equation}.\alph{equ}}
\begin{eqnarray}
&& p_n = \frac{a^{*}_{n,d}}{a^{*}_{n,f}},   n=1,2,3,4,
\label{Eq40} \\
\stepcounter{equ} \setcounter{equation}{26} && p_{n'} =
\frac{a_{n',d}}{a_{n',f}},  n'=5,6,7,8,
\label{Eq41}
\end{eqnarray}
\def\theequation{\arabic{equation}}
and consider again
$V^{d,f}_{\mathbf{r}}=V^{f,d}_{\mathbf{r}}=V_{\mathbf{r}}$.
Eq.(\ref{Eq41}) is seen to contain coefficients $a_{n,b}$ instead
of $a_{n,b}^{*}$ since (\ref{Eq34}) holds as well. An itinerant
solution is obtained by choosing $p_n=p=-p^{*}$, i.e., imaginary
$p$. For this choice (\ref{Eq12}) leads to
\begin{eqnarray}
t^f_{\mathbf{r}_1} = \frac{t^d_{\mathbf{r}_1}}{|p|^2}, \quad t^f_{\mathbf{r}%
_2} = - \frac{t^d_{\mathbf{r}_2}}{|p|^2}, \quad V_{\mathbf{r}_2} = p t^f_{%
\mathbf{r}_2}, \quad V_{\mathbf{r}_1} = 0,  \label{Eq42}
\end{eqnarray}
where $\mathbf{r}_1 = \mathbf{x}_1, \mathbf{x}_2, \mathbf{x}_2 \pm \mathbf{x}%
_1$, $\mathbf{r}_2 = \mathbf{x}_3, \mathbf{x}_3 \pm \mathbf{x}_2, \mathbf{x}%
_3 \pm \mathbf{x}_1, \mathbf{x}_3 \pm \mathbf{x}_2 \pm
\mathbf{x}_1$, and $V_{\mathbf{r}_1}=0$ follows from $V_{%
\mathbf{r}_1} = - V_{\mathbf{r}_1}$. As discussed earlier,
imaginary $p$ imply imaginary $V_{\mathbf{r}}$. The local
parameters $U$ and $V_0$
 become
\begin{eqnarray}
U + E_f = \frac{|p|^2 -1}{|p|^2} \sum_{n=1}^8|a_{n,d}|^2  , \quad
p^{*} V_0 = \sum_{n=5}^8 |a_{n,d}|^2 - \sum_{n=1}^4 |a_{n,d}|^2 .
\label{Eq43}
\end{eqnarray}
The remaining relations following from (\ref{Eq12}) are
\begin{eqnarray}
&&-t^d_{\mathbf{x}_1} = a^{*}_{1,d} a_{2,d} + a^{*}_{8,d} a_{7,d}
+
a^{*}_{4,d} a_{3,d} + a^{*}_{5,d} a_{6,d} ,  \nonumber \\
&&-t^d_{\mathbf{x}_2} = a^{*}_{1,d} a_{4,d} + a^{*}_{6,d} a_{7,d}
+
a^{*}_{2,d} a_{3,d} + a^{*}_{5,d} a_{8,d} ,  \nonumber \\
&&-t^d_{\mathbf{x}_3} = a^{*}_{1,d} a_{5,d} + a^{*}_{3,d} a_{7,d}
+
a^{*}_{2,d} a_{6,d} + a^{*}_{4,d} a_{8,d} ,  \nonumber \\
&&-t^d_{\mathbf{x}_2+\mathbf{x}_1} = a^{*}_{1,d} a_{3,d} +
a^{*}_{5,d} a_{7,d} , \quad -t^d_{\mathbf{x}_2-\mathbf{x}_1} =
a^{*}_{2,d} a_{4,d} +
a^{*}_{6,d} a_{8,d}  ,  \nonumber \\
&&-t^d_{\mathbf{x}_3+\mathbf{x}_1} = a^{*}_{1,d} a_{6,d} +
a^{*}_{4,d} a_{7,d} , \quad -t^d_{\mathbf{x}_3-\mathbf{x}_1} =
a^{*}_{2,d} a_{5,d} +
a^{*}_{3,d} a_{8,d} ,  \nonumber \\
&&-t^d_{\mathbf{x}_3+\mathbf{x}_2} = a^{*}_{1,d} a_{8,d} +
a^{*}_{2,d} a_{7,d} , \quad -t^d_{\mathbf{x}_3-\mathbf{x}_2} =
a^{*}_{4,d} a_{5,d} +
a^{*}_{3,d} a_{6,d} ,  \nonumber \\
&&-t^d_{\mathbf{x}_3+\mathbf{x}_2+\mathbf{x}_1} = a^{*}_{1,d}
a_{7,d}, \quad -t^d_{\mathbf{x}_3+\mathbf{x}_2-\mathbf{x}_1} =
a^{*}_{2,d} a_{8,d},
\nonumber \\
&&-t^d_{\mathbf{x}_3-\mathbf{x}_2+\mathbf{x}_1} = a^{*}_{4,d}
a_{6,d}, \quad -t^d_{\mathbf{x}_3-\mathbf{x}_2-\mathbf{x}_1} =
a^{*}_{3,d} a_{5,d} . \label{Eq44}
\end{eqnarray}
Furthermore, $V_{\mathbf{r}_1}=0$ implies
\begin{eqnarray}
&&a^{*}_{1,d} a_{2,d} + a^{*}_{4,d} a_{3,d} = a^{*}_{5,d} a_{6,d}
+ a^{*}_{8,d} a_{7,d} , \quad a^{*}_{1,d} a_{3,d} = a^{*}_{5,d}
a_{7,d} ,
\nonumber \\
&&a^{*}_{1,d} a_{4,d} + a^{*}_{2,d} a_{3,d} = a^{*}_{6,d} a_{7,d}
+ a^{*}_{5,d} a_{8,d} , \quad a^{*}_{2,d} a_{4,d} = a^{*}_{6,d}
a_{8,d} . \label{Eq45}
\end{eqnarray}
From (\ref%
{Eq44}, \ref{Eq45}) it follows that
\begin{eqnarray}
a^{*}_{1,d} = a_1^{*}, \quad a^{*}_{2,d} = u a^{*}_1, \quad
a^{*}_{3,d} = u^{2}
a_1^{*}, \quad a^{*}_{4,d} = u a_1^{*},  \nonumber \\
a^{*}_{5,d} = u a_1^{*}, \quad a^{*}_{6,d} = u^{2} a^{*}_1, \quad
a_{7,d} = u a_1, \quad a_{8,d} = a_1,  \label{Eq46}
\end{eqnarray}
where $u$ is real, $|u| \ne 1$, $V_0=0$, and $\sum_{n=1}^8
|a_{n,d}|^2 = 2 |a_1|^{2} (1+u^{2})^{2}$. These relations can only
be fulfilled for \emph{anisotropic} hopping and hybridization
amplitudes, e.g., for vanishing hybridization in the
basal ($x,y$) plane. Namely, (\ref{Eq44}-\ref{Eq46}) yield $|u|=\sqrt{%
|t^d_{\mathbf{x}_3- \mathbf{x}_2}/t^d_{\mathbf{x}_3 + \mathbf{x}_2}|}$ and $%
|t^b_{\mathbf{x}_{\tau}}|= |t^b_{\mathbf{x}_{\tau^{\prime}}}|$.
The anisotropy in the hopping amplitudes is seen to start at the
level of next-nearest neighbors. The stability region of this
phase is presented in Fig.3.

\subsection{Exact itinerant ground state of the conventional PAM.}

The solutions obtained sofar,
namely (\ref{Eq34}%
-\ref{Eq38}), and (\ref{Eq42}-\ref{Eq46}),
require the $f$ electrons to be itinerant, i.e., $t^f_{\mathbf{r}%
} \ne 0$.
We will now show that exact itinerant ground states can even be
constructed for the \emph{conventional} PAM, i.e., for
$t^f_{\mathbf{r}}=0$. This requires nonzero $d$-electron hopping
up to next-nearest neighbors together with local and
nearest-neighbor hybridizations.

Such a solution can be constructed, for example, for nonzero
coefficients $a_{1,f}, a_{1,d}, a_{2,d}, a_{4,d}, a_{5,d}$, with
the remaining coefficients $a_{n,b}=0$. This is realized in a
tilted unit cell (see Fig.4) where the distances between lattice
sites with indices $(n=1,n=3),(n=1,n=6),(n=1,n=8)$ and corresponding hopping elements $(t^d_{%
\mathbf{x}_2+\mathbf{x}_1}, t^d_{\mathbf{x}_3+\mathbf{x}_1}, t^d_{\mathbf{x}%
_3+\mathbf{x}_2})$ are considerably larger than the distances
between lattice sites with indices $(n=2,n=4),(n=2,n=5),(n=4,n=5)$
and hopping elements $(t^d_{%
\mathbf{x}_2-\mathbf{x}_1}, t^d_{\mathbf{x}_3-\mathbf{x}_1}, t^d_{\mathbf{x}%
_3- \mathbf{x}_2})$. For this reason the former hopping elements
are neglected (i.e., put to zero).
In this case (\ref%
{Eq12}) reduces to the following eleven equations
\begin{eqnarray}
&&t^d_{\mathbf{x}_1}=-a^{*}_{1,d} a_{2,d}, \quad t^d_{\mathbf{x}%
_2}=-a^{*}_{1,d} a_{4,d}, \quad t^d_{\mathbf{x}_3}=-a^{*}_{1,d}
a_{5,d},
\nonumber \\
&&t^d_{\mathbf{x}_2-\mathbf{x}_1}=-a^{*}_{2,d} a_{4,d}, \quad t^d_{\mathbf{x}%
_3-\mathbf{x}_1}=-a^{*}_{2,d} a_{5,d}, \quad t^d_{\mathbf{x}_3-\mathbf{x}%
_2}=-a^{*}_{4,d} a_{5,d},  \nonumber \\
&&V^{f,d}_{\mathbf{x}_1}=-a^{*}_{1,f} a_{2,d}, \quad V^{f,d}_{\mathbf{x}%
_2}=-a^{*}_{1,f} a_{4,d}, \quad
V^{f,d}_{\mathbf{x}_3}=-a^{*}_{1,f} a_{5,d},
\nonumber \\
&&V_0=-a^{*}_{1,d}a_{1,f} , \quad U + E_f = K_d - |a_{1,f}|^2,
\label{Eq47}
\end{eqnarray}
where $K_d=|a_{1,d}|^2 +|a_{2,d}|^2 +|a_{4,d}|^2 +|a_{5,d}|^2 $,
and all other amplitudes are identically zero.

Considering only the simplest case, i.e., $t^d_1=t^d_{\mathbf{x}_{
\tau}}$, $ V_1=
V^{f,d}_{\mathbf{x}_{\tau}}$, $\tau=1,2,3$, and $ t^d_2 = t^d_{\mathbf{x}%
_{\tau}-\mathbf{x}_{\tau^{\prime}}}$, $\tau > \tau^{\prime}$, with
real hopping amplitudes \cite{commentx}, the corresponding
stability region corresponds to the surface in parameter space
(see Fig.5) described by
\begin{eqnarray}
\frac{U + E_f}{|t^d_1|} = x + \frac{1}{x} (1 - y^{2}) ,
\label{Eq49}
\end{eqnarray}
where $x=|t^d_2|/|t^d_1|$ and $y=|V_1|/|t^d_1|$. Further
properties of the solution are discussed in footnotes
[\cite{commentx,commx}]. The fact that $a^{*}_{n,d} = p
a^{*}_{n,f}$ with $p=p^{*}$ does not hold for all $n=1,2,..,8$,
implies a variable number of electrons at each site, i.e., an
itinerant ground state.

\subsection{Diagonalization of the Hamiltonian}

To describe properties of the itinerant state, a
$\mathbf{k}$-representation is more suitable. This can be
introduced without restriction on the $a_{n,b}$ coefficients.
Therefore the two conducting solutions presented in Sections IV.A.
and IV.B. can be treated simultaneously.

Denoting the Fourier transforms of $\hat
A^{\dagger}_{I_{\mathbf{i}}, \sigma} $
by $\hat A^{\dagger}_{\mathbf{k}\sigma} = \sum_{b=d,f} a^{*}_{%
\mathbf{k},b} \hat b^{\dagger}_{\mathbf{k}\sigma}$, with
\begin{eqnarray}
a^{*}_{\mathbf{k},b} & = & a^{*}_{1,b} + a^{*}_{2,b} e^{i \mathbf{k} \mathbf{%
x}_1} + a^{*}_{3,b} e^{i \mathbf{k} (\mathbf{x}_1 +\mathbf{x}_2)}
+ a^{*}_{4,b} e^{i \mathbf{k} \mathbf{x}_2} + a^{*}_{5,b} e^{i
\mathbf{k}
\mathbf{x}_3}  \nonumber \\
& + & a^{*}_{6,b} e^{i \mathbf{k} (\mathbf{x}_1 +\mathbf{x}_3)} +
a^{*}_{7,b} e^{i \mathbf{k} (\mathbf{x}_1 + \mathbf{x}_2 +
\mathbf{x}_3)} + a^{*}_{8,b} e^{i \mathbf{k} (\mathbf{x}_2
+\mathbf{x}_3)} ,  \label{Eq50}
\end{eqnarray}
one may define new canonical Fermi operators describing composite
fermions $\hat C_{\delta,%
\mathbf{k} \sigma}$, $ \delta=1, 2,$ $ \{ \hat C_{\delta,\mathbf{k}%
\sigma}, \hat C^{\dagger}_{\delta^{\prime},\mathbf{k}^{\prime}\sigma^{%
\prime}} \} = \delta_{\delta,\delta^{\prime}} \delta_{\mathbf{k},\mathbf{k}%
^{\prime}} \delta_{\sigma,\sigma^{\prime}}$, $ \{ \hat C_{\delta,\mathbf{k}%
\sigma}, \hat
C_{\delta^{\prime},\mathbf{k}^{\prime}\sigma^{\prime}} \} = 0 $.
Here \cite{commx,commy}
\begin{eqnarray}
\hat C_{1,\mathbf{k}\sigma} = \sqrt{R_{\mathbf{k}}} \hat A_{\mathbf{k}%
\sigma} = \sqrt{R_{\mathbf{k}}} (a_{\mathbf{k},d} \hat d_{\mathbf{k}%
\sigma} + a_{\mathbf{k},f} \hat f_{\mathbf{k}\sigma} )  , \quad \hat C_{2,%
\mathbf{k}\sigma} = \sqrt{R_{\mathbf{k}}} (a^{*}_{\mathbf{k},f}
\hat d_{ \mathbf{k}\sigma} - a^{*}_{\mathbf{k},d} \hat
f_{\mathbf{k}\sigma} ) , \label{Eq51}
\end{eqnarray}
where $R^{-1}_{\mathbf{k}} = \sum_{b=d,f} |a_{\mathbf{k},b}|^2$.
Then the Hamiltonian becomes \cite{commy1}
\begin{eqnarray}
\hat H = \hat H_g + U \hat P - U N_{\Lambda} , \quad \hat H_g = \sum_{%
\mathbf{k},\sigma} [(K_d - R_{\mathbf{k}}^{-1}) \hat C^{\dagger}_{ 1,\mathbf{%
k}\sigma} \hat C_{1,\mathbf{k}\sigma} + K_d \hat C^{\dagger}_{2,\mathbf{k}%
\sigma} \hat C_{2,\mathbf{k}\sigma} ].  \label{Eq55}
\end{eqnarray}
In the ground state where $\hat P |\Psi_g\rangle = 0$ the
Hamiltonian $\hat H$ reduces to $\hat H_g$. Hence the composite
fermion operators introduced above indeed diagonalize $\hat H$.
There are two bands, the lower one having a dispersion
$\xi_{1,\mathbf{k}} = K_d - R^{-1}_{\mathbf{k}}$, while the upper
one is dispersionless, i.e., flat, since $\xi_{2,\mathbf{k}} =
K_d=$const. The lower (upper) diagonalized band contains fermions
created by
$\hat C^{\dagger}_{1,\mathbf{k}\sigma}$ ($\hat C^{\dagger}_{2,\mathbf{k}%
\sigma}$), respectively. Since the total band filling is $3/4$ the
lower band is completely filled, while the upper band is half
filled. Thus the Fermi energy is given by $E_F=K_d$. Band
structures containing a partially filled flat band around $E_F$
have indeed been observed experimentally \cite{ito,xx,xxma}.

\subsection{Conductivity of the itinerant state}

The ground state expectation value of $\hat H_{itin}$, the
itinerant part of $\hat H$, is obtained as
\begin{eqnarray}
\langle \hat H_{itin} \rangle = - \sum_{\bf k} \frac{K_d |a_{{\bf
k},d}|^2 + K_f |a_{{\bf k},f}|^2}{|a_{{\bf k},d}|^2 + |a_{{\bf
k},f}|^2} < 0. \label{ezhitin}
\end{eqnarray}
This is in contrast to the localized case where $\langle \hat
H_{itin} \rangle =0$.\cite{newxx}

Since $\langle \hat H_{itin}(\mathbf{r}) \rangle \ne 0$ the charge
sum rule (\ref{B7}) implies $\int^{\infty}_{0} d \omega Re
\sigma_{\tau,\tau} (\omega) \ne 0$, i.e., the (dynamic)
conductivity is in general non-zero. However, the sum rule does
not allow us to draw conclusions about the DC-conductivity
$\sigma(0)$.
This becomes possible if we calculate the chemical potential of
the system as a function of the particle number. To this end we
observe that in the case of a variable number of particles per
site the itinerant ground state is defined at \emph{and} above
$3/4$ filling, such that $|\Psi_g\rangle$ in (\ref{Eq18}) can be
generalized to fillings beyond $3/4$ via (\ref{Eq19}).
Using (\ref{ham}) this allows one to calculate the energy $E_g$
for different particle
numbers. In particular, one finds $\mu^{+} = E_g(N+2) - E_g(N+1)=K_d$, $%
\mu^{-}=E_g(N+1) - E_g(N) = K_d$, i.e., $\mu^{+} - \mu^{-} =0$.
Therefore, the described itinerant solutions are indeed conducting
\cite{cond}.

\subsection{Magnetic properties.}

Using $\mathbf{k}$-space notation the unnormalized ground state,
 (\ref{Eq18}), can be written as \cite{xxy}
\begin{eqnarray}
|\Psi_g \rangle = [ \prod_{\mathbf{k}}^{N_{\Lambda}} \hat A^{\dagger}_{%
\mathbf{k} \uparrow} \hat A^{\dagger}_{\mathbf{k} \downarrow} ] \{ \prod_{%
\mathbf{i}}^{N_{ \Lambda}} [\sum_{\mathbf{k}}^{N_{\Lambda}} e^{i
\mathbf{k}
\mathbf{i}} (\mu_{\mathbf{i}, \uparrow} \hat f^{\dagger}_{\mathbf{k}%
\uparrow} + \mu_{\mathbf{i},\downarrow} \hat f^{\dagger}_{\mathbf{k}%
\downarrow})] \} |0\rangle ,  \label{Eq56}
\end{eqnarray}
where the sum and product over $\mathbf{k}$ extend over the first
Brillouin zone, and the set $\{ \mu_{\mathbf{i},\sigma} \}$ is
arbitrary. Using (\ref{Eq56}) we will now calculate ground state
 expectation values of the spin for different sets of
$\{\mu_{\mathbf{i},\sigma} \}$.\cite{sspin}

\subsubsection{Maximum total spin}

As in the localized case $\mu_{\mathbf{i},\sigma} = \mu_{\sigma}$
corresponds to maximum total spin. Defining $\hat
F^{\dagger}_{\mathbf{k},b} = (\mu_{\uparrow} \hat
b^{\dagger}_{\mathbf{k} \uparrow} + \mu_{\downarrow} \hat b^{\dagger}_{%
\mathbf{k} \downarrow})$ and employing $(\hat
F^{\dagger}_{\mathbf{k},b})^2 = 0$ the product over sites
$\mathbf{i}$ in  (\ref{Eq56}) may be written as
$\prod_{\mathbf{i}%
}^{N_{\Lambda}} [\sum_{\mathbf{k}}^{ N_{\Lambda}} e^{i \mathbf{k}
\mathbf{i}} \hat F^{\dagger}_{\mathbf{k},f}] =Z \prod_{\mathbf{k}%
}^{N_{\Lambda}} \hat F^{\dagger}_{\mathbf{k},f}$, where $Z$ is
defined in footnote (\cite{xxy}). Then the normalized ground state
becomes
\begin{eqnarray}
|\Psi_g\rangle = \prod_{\mathbf{k}}^{N_{\Lambda}} \frac{ a^{*}_{\mathbf{k}%
,d} \hat d^{\dagger}_{\mathbf{k} \uparrow} \hat d^{\dagger}_{\mathbf{k}%
\downarrow} \hat F^{\dagger}_{\mathbf{k},f} + a^{*}_{\mathbf{k},f}
\hat f^{\dagger}_{\mathbf{k} \downarrow} \hat
f^{\dagger}_{\mathbf{k} \uparrow} \hat F^{\dagger}_{\mathbf{k},d}
}{\sqrt{(|\mu_{\uparrow}|^2 +|\mu_{\downarrow}|^2)
(|a_{\mathbf{k},d}|^2 + |a_{\mathbf{k},f}|^2)}} |0\rangle .
\label{Eq58}
\end{eqnarray}
The spin expectation value then follows as
\begin{eqnarray}
\frac{2 \langle \hat {\mathbf{S}} \rangle}{N_{\Lambda}} =
{\mathbf{x}} \frac{\mu_{\downarrow} \mu^{*}_{\uparrow} +
\mu_{\uparrow} \mu^{*}_{ \downarrow}}{|\mu_{\uparrow}|^2 +
|\mu_{\downarrow}|^2} + {\mathbf{y}} \frac{i(\mu_{\uparrow}
\mu^{*}_{\downarrow} - \mu_{\downarrow} \mu^{*}_{
\uparrow})}{|\mu_{\uparrow}|^2 + |\mu_{\downarrow}|^2} +
{\mathbf{z}} \frac{|\mu_{\uparrow}|^2 -
|\mu_{\downarrow}|^2}{|\mu_{\uparrow}|^2 + |\mu_{ \downarrow}|^2)}
,  \label{Eq59}
\end{eqnarray}
resulting in $|\langle \hat {\mathbf{S}} \rangle / N_{\Lambda}|^2
=1/4$, and $\langle {\hat {\mathbf{S}}}^2 \rangle = (N_{\Lambda}/2)%
(N_{\Lambda}/2 + 1)$. Thus we recover (\ref{Eq28}), the results
for the localized solution. The maximum total spin is again given
by
$\sqrt{\langle {\hat {\mathbf{S}}%
}^2 \rangle }/N_{\Lambda} = 1/2$ in the thermodynamic limit
\cite{mi5}.

\subsubsection{Minimum total spin}

To determine the minimum spin value we proceed as in the localized
case, i.e., divide the system in two sublattices
${\cal D}_{\uparrow}$ and ${\cal D}_{\downarrow}$ containing both
$N_{\Lambda}/2$ lattice sites, such that for all $\mathbf{i}_n \in
{\cal D}_{\uparrow}$ and $\mathbf{j}_n \in {\cal D}_{\downarrow}$
we have $\mathbf{j}_n = \mathbf{i}_n + \mathbf{R}$, where
$\mathbf{R}$ is a fixed Bravais vector \cite{xxy1}.
Choosing $\mu_{\mathbf{i},\sigma}=\mu_{\sigma}$, $\mu_{\mathbf{i}%
,-\sigma}= 0$ for ${\cal D}_{\sigma}$ in (\ref{Eq56}) the
unnormalized ground state becomes
\begin{eqnarray}
|\Psi_g \rangle = [ \prod_{\mathbf{k}}^{N_{\Lambda}} \hat A^{\dagger}_{%
\mathbf{k} \uparrow} \hat A^{\dagger}_{\mathbf{k} \downarrow} ] [\prod_{%
\mathbf{i \in D_1}}^{
\frac{N_{\Lambda}}{2}}(\sum_{\mathbf{k}}^{N_{\Lambda}}
e^{i \mathbf{k} \mathbf{i}} \hat f^{\dagger}_{\mathbf{k} \uparrow})] [\prod_{%
\mathbf{j \in D_2}}^{\frac{
N_{\Lambda}}{2}}(\sum_{\mathbf{k}}^{N_{\Lambda}} e^{i \mathbf{k}
\mathbf{j}} \hat f^{ \dagger}_{\mathbf{k} \downarrow})] |0\rangle
.  \label{Eq60}
\end{eqnarray}
Details of the calculation of ground state expectation values in
terms of (\ref{Eq60}) are presented in Appendix D. Based on
(\ref{E5})
one finds $\langle \hat {\mathbf{S}} \rangle = 0$, and using (%
\ref{E6}), $\sqrt{ \langle {\hat{\mathbf{S}}}^2 \rangle}/N_{
\Lambda} < \sqrt{3}/(2 \sqrt{N_{\Lambda}})$. As a consequence,
this itinerant ground state has zero total spin in the
thermodynamic limit.

Between the two limiting cases of the total spin discussed above
all values of $S$ may exist, depending on the choice of the set
$\{ \mu_{\mathbf{i},\sigma } \}$. The statement regarding the
degeneracy of the ground state presented below (\ref{Eq29}) holds
in the itinerant case as well.

\subsection{Momentum distribution function}

To calculate ground state expectation values of
$\mathbf{k}$-dependent operators involving $\hat C_{\delta,
\mathbf{k} \sigma}$ operators, (\ref{Eq51}), we need to express
$|\Psi_g\rangle$, (\ref{Eq56}), in terms of $\hat
C^{\dagger}_{\delta, \mathbf{k},\sigma}$, too. For the
unnormalized ground state one finds
\begin{eqnarray}
|\Psi_g \rangle = [ \prod_{\mathbf{k}}^{N_{\Lambda}} \hat C^{\dagger}_{1,%
\mathbf{k} \uparrow} \hat C^{\dagger}_{1,\mathbf{k} \downarrow} ] \{ \prod_{%
\mathbf{i}}^{N_{ \Lambda}} [\sum_{\mathbf{k}}^{N_{\Lambda}}
X_{\mathbf{k}}
e^{i \mathbf{k} \mathbf{i}} (\mu_{\mathbf{i},\uparrow} \hat C^{\dagger}_{2,%
\mathbf{k} \uparrow} + \mu_{\mathbf{i}, \downarrow} \hat C^{\dagger}_{2,%
\mathbf{k} \downarrow})] \} |0\rangle ,  \label{Eq68}
\end{eqnarray}
where $\{ \mu_{\mathbf{i},\sigma} \}$ is arbitrary, and
$X_{\mathbf{k}}=-a_{\mathbf{k},d}\sqrt{R_{\mathbf{k}}}$; in the
following we consider $X_{\mathbf{k}} \ne 0$ for all $\mathbf{k}$
\cite{mi6}.
From (\ref%
{Eq68}) it follows that
\begin{eqnarray}
&&\hat C^{\dagger}_{1,\mathbf{k} \sigma} \hat C_{1,\mathbf{k^{\prime}}%
\sigma^{\prime}} |\Psi_g\rangle =
\delta_{\mathbf{k},\mathbf{k}^{\prime}}
\delta_{\sigma,\sigma^{\prime}} |\Psi_g\rangle \: , \quad \hat
C^{\dagger}_{1,\mathbf{k} \sigma} \hat C_{2,\mathbf{k}^{\prime} \sigma^{%
\prime}} |\Psi_g\rangle = 0 .  \label{Eq69}
\end{eqnarray}
Details of the calculation of ground state expectation values of
$\hat C^{\dagger}_{2, \mathbf{k} \sigma} \hat C_{2,\mathbf{k}
\sigma}$ operators are presented in Appendix E. Using
(\ref{Eq69},\ref{D18}), one finds in particular
\begin{eqnarray}
\langle \hat C^{\dagger}_{\delta,\mathbf{k} \sigma} \hat
C_{\delta,\mathbf{k} \sigma} \rangle = \delta_{1,\delta} +
\frac{1}{2} \delta_{2,\delta}  \label{Eq71}
\end{eqnarray}
for all $\mathbf{k}$. This expresses the fact
that the lower band ($\delta=1$) is completely filled, while
the upper band ($\delta=2$) is flat and half filled.

Employing (\ref{Eq71}) the momentum distribution of $b=d,f$
electrons is then obtained as\cite{commy}
\begin{eqnarray}
n^b_{\mathbf{k} \sigma} = \langle \hat b^{\dagger}_{\mathbf{k}
\sigma} \hat
b_{\mathbf{k} \sigma} \rangle = \frac{1}{2} + \frac{1}{2} R_{\mathbf{k}} |a_{%
\mathbf{k},b}|^2 ,  \label{Eq72}
\end{eqnarray}
which implies $n_{\mathbf{k} \sigma} = \sum_{b=d,f}
n^b_{\mathbf{k} \sigma}
= 3/2$. Since the coefficients $R_{\mathbf{k}}|a_{\mathbf{k},b}|^2=|a_{%
\mathbf{k},b}|^2/(|a_{ \mathbf{k},d}|^2+|a_{\mathbf{k},f}|^2)$ are
regular functions of $\mathbf{k}$, this is also the case for
$n^b_{\mathbf{k},\sigma}$ and $n_{\mathbf{k},\sigma}$ (see Fig.6)
\cite{commz}. Consequently, the momentum distributions of the
electrons in the interacting ground state have no discontinuities.
Since the ground state is nonmagnetic and metallic, the system is
therefore a \emph{non-Fermi} liquid. This is a consequence of the
macroscopic degeneracy of the electrons in the upper band. Due to
the flatness of the upper half-filled band all $\mathbf{k}$ states
are equivalent. Hence, even if a Fermi energy $E_F=K_d$ exists the
Fermi \emph{surface}, and the Fermi \emph{momentum} are not
defined. Consequently, Luttinger's theorem  \cite{lut} does not
apply. Non-Fermi liquid properties connected to a flat band have
also been observed in other investigations\cite{NFL1,NFL2,NFL3}.

\subsection{Correlation functions}

To characterize the itinerant ground states further we now
calculate correlation functions of the interacting systems in
$D=3$. This is made possible by the explicit form of the exact
ground states.

\subsubsection{Density-density correlation function}

Using (\ref{Eq69}) the density-density correlation function
\begin{eqnarray}
\rho_{n,n}(\mathbf{r}) = \frac{1}{N_{\Lambda}} \sum_{\mathbf{i}}
(\langle \hat n_{ \mathbf{i}} \hat n_{\mathbf{i}+\mathbf{r}}
\rangle - \langle \hat n_{\mathbf{i}} \rangle \langle \hat
n_{\mathbf{i}+\mathbf{r}} \rangle) , \label{Eq73}
\end{eqnarray}
where $\hat n_{\mathbf{i}} = \sum_{\sigma} \sum_{b=d,f} \hat n^b_{\mathbf{i}%
,\sigma}$, may be written as
\begin{eqnarray}
&&\rho_{n,n}(\mathbf{r}) = \frac{1}{N^2_{\Lambda}} \sum_{\sigma,\sigma^{%
\prime}} \sum_{\mathbf{k}_1,\mathbf{k}_2,\mathbf{k}_3} e^{i (\mathbf{k}_1-%
\mathbf{k}_2)\mathbf{r}} \sqrt{R_{\mathbf{k}_1}R_{\mathbf{k}_2}R_{\mathbf{k}%
_3}R_{\mathbf{k}_4}} [ (a^{*}_{\mathbf{k}_3,d} a_{\mathbf{k}_4,d}
+a^{*}_{\mathbf{k}_3,f} a_{\mathbf{k}_4,f}) \times  \nonumber \\
&&(a^{*}_{\mathbf{k}_1,d} a_{\mathbf{k}_2,d} +a^{*}_{\mathbf{k}_1,f} a_{%
\mathbf{k}_2,f}) (\langle \hat C^{\dagger}_{2,\mathbf{k}_3
\sigma^{\prime}}
\hat C_{2,\mathbf{k}_4 \sigma^{\prime}} \hat C^{\dagger}_{2,\mathbf{k}%
_1 \sigma} \hat C_{2,\mathbf{k}_2 \sigma} \rangle - \langle \hat
C^{\dagger}_{2,\mathbf{k}_3 \sigma^{\prime}} \hat C_{2,\mathbf{k}%
_4 \sigma^{\prime}} \rangle \langle \hat
C^{\dagger}_{2,\mathbf{k}_1 \sigma}
\hat C_{2,\mathbf{k}_2 \sigma} \rangle) +  \nonumber \\
&&(a_{\mathbf{k}_3,d} a_{\mathbf{k}_4,f} -a_{\mathbf{k}_3,f} a_{\mathbf{k}%
_4,d}) (a^{*}_{\mathbf{k}_1,f} a^{*}_{\mathbf{k}_2,d} -a^{*}_{\mathbf{k}%
_1,d} a^{*}_{ \mathbf{k}_2,f}) \langle \hat C^{\dagger}_{1,\mathbf{k}%
_3 \sigma^{\prime}} \hat C_{2, \mathbf{k}_4 \sigma^{\prime}} \hat
C^{\dagger}_{2,\mathbf{k}_1 \sigma} \hat C_{1,\mathbf{k}_2 \sigma}
\rangle ] ,  \label{Eq74}
\end{eqnarray}
where $\mathbf{k}_4=\mathbf{k}_3+\mathbf{k}_1-\mathbf{k}_2$.
Employing (\ref{D14},\ref{D31}) in the first term and
(\ref{Eq69},\ref{D14}) in the second term of (\ref{Eq74}) one
finds
\begin{eqnarray}
&&\sum_{\sigma,\sigma^{\prime}} ( \langle \hat C^{\dagger}_{2,\mathbf{k}%
_3 \sigma^{\prime}} \hat C_{2,\mathbf{k}_4 \sigma^{\prime}} \hat
C^{\dagger}_{2,\mathbf{k}_1 \sigma} \hat C_{2,\mathbf{k}_2 \sigma}
\rangle -
\langle \hat C^{\dagger}_{2,\mathbf{k}_3 \sigma^{\prime}} \hat C_{2,\mathbf{k%
}_4 \sigma^{\prime}} \rangle \langle \hat C^{\dagger}_{2,\mathbf{k}%
_1 \sigma} \hat C_{2,\mathbf{k}_2 \sigma} \rangle ) = \delta_{\mathbf{k}_1,%
\mathbf{k}_4} \delta_{\mathbf{k}_2,\mathbf{k}_3} -  \nonumber \\
&&\frac{1}{N_{\Lambda}}\frac{X_{\mathbf{k}_4\ne\mathbf{k}_2}}{X_{\mathbf{k}%
_3\ne \mathbf{k}_1}} \frac{X_{\mathbf{k}_2}}{X_{\mathbf{k}_1}} \delta_{%
\mathbf{k}_4, \mathbf{k}_3 + \mathbf{k}_1-\mathbf{k}_2} , \quad
\sum_{\sigma,\sigma^{\prime}} \langle \hat C^{\dagger}_{1,\mathbf{k}%
_3 \sigma^{\prime}} \hat C_{2,\mathbf{k}_4 \sigma^{\prime}} \hat
C^{\dagger}_{2,\mathbf{k}_1 \sigma} \hat C_{1,\mathbf{k}_2 \sigma}
\rangle = \delta_{\mathbf{k}_4,\mathbf{k}_1}
\delta_{\mathbf{k}_3,\mathbf{k}_2} . \label{Eq75}
\end{eqnarray}
Then the density-density correlation function becomes
\begin{eqnarray}
&&\rho_{n,n}(\mathbf{r}) = \delta_{\mathbf{r},0} -
\frac{1}{N^3_{\Lambda}}
\sum_{\mathbf{k}_1,\mathbf{k}_2,\mathbf{k}_3,\mathbf{k}_4} e^{i(\mathbf{k}_1-%
\mathbf{k}%
_2)\mathbf{r}} \frac{a_{\mathbf{k}_4\ne \mathbf{k}_2,d} a_{\mathbf{k}_2,d}}{%
a_{\mathbf{k}_3\ne \mathbf{k}_1,d} a_{\mathbf{k}_1,d}} \times  \nonumber \\
&&\frac{ (a^{*}_{\mathbf{k}_3,d} a_{\mathbf{k}_4,d}
+a^{*}_{\mathbf{k}_3,f}
a_{\mathbf{k}_4,f}) (a^{*}_{\mathbf{k}_1,d} a_{\mathbf{k}_2,d} +a^{*}_{%
\mathbf{k}_1,f} a_{\mathbf{k}_2,f})}{ [|a_{\mathbf{k}_2,d}|^2 + |a_{\mathbf{k%
}_2,f}|^2] [|a_{\mathbf{k}_4,d}|^2 + |a_{\mathbf{k}_4,f}|^2] } \delta_{%
\mathbf{k}_4, \mathbf{k}_3 + \mathbf{k}_1-\mathbf{k}_2} .
\label{Eq77}
\end{eqnarray}
In the thermodynamic limit the contributions from $\mathbf{k}_4=\mathbf{k}_2$ and $%
\mathbf{k}_3=\mathbf{k}_1$  in the second term have zero measure.
Hence all $\mathbf{k}$ sums can be calculated without restriction.
$\rho_{n,n}(\mathbf{r})$ is seen to vanish for all $\mathbf{r}$
whenever the $\mathbf{k}$--dependence of the $a_{\mathbf{k},b}$
coefficients is negligible (for example in the case
$|t^d_1/t^d_2|>>1$ for the solution from Sec.IV.B.). It also
vanishes in the limit $|\mathbf{r}| \to \infty$, where the $%
\mathbf{k} \to 0$ limit of the coefficients $a_{\mathbf{k},b}$
gives the dominant contribution to (\ref{Eq77}); this behavior is
indeed seen in Fig.7.

\subsubsection{Spin-spin correlation function}

Similarly, the spin-spin correlation function
\setcounter{equ}{1}
\def\theequation{\arabic{equation}.\alph{equ}}
\begin{eqnarray}
\rho_{\mathbf{S},\mathbf{S}} (\mathbf{r}) &=& \frac{1}{N_{\Lambda}}
\sum_{\mathbf{i%
}} ( \langle \mathbf{\hat S}_{\mathbf{i}} \mathbf{{\hat S}_{i+r}}
\rangle - \langle \mathbf{\hat S}_{\mathbf{i}} \rangle \langle
\mathbf{{\hat S}_{i+r}} \rangle )
\label{corrs}\\
\stepcounter{equ}\setcounter{equation}{50}
&=&
\rho_{S_z,S_z}(\mathbf{r}) +
\rho_{\mathbf{S}_\perp,\mathbf{S}_\perp}(\mathbf{r}) ,
\label{corrs1}
\end{eqnarray}
\def\theequation{\arabic{equation}}
is given by
\setcounter{equ}{1}
\def\theequation{\arabic{equation}.\alph{equ}}
\begin{eqnarray}
\rho_{S_z,S_z}(\mathbf{r}) &=&
\frac{1}{N_{\Lambda}} \sum_{\mathbf{i}} (\langle \hat S^z_{
\mathbf{i}} \hat S^z_{\mathbf{i}+\mathbf{r}} \rangle - \langle
\hat S^z_{\mathbf{i}} \rangle \langle \hat
S^z_{\mathbf{i}+\mathbf{r}} \rangle)
\label{eq51a}\\
\stepcounter{equ}\setcounter{equation}{51}
&=& \frac{1}{4 N^2_{\Lambda}}  \sum_{\sigma}
\sum_{\mathbf{k}_1,\cdots,
\mathbf{k}_4} e^{i (\mathbf{k}_1-\mathbf{k}_2)\mathbf{r}} \sqrt{%
R_{\mathbf{k}_1}R_{\mathbf{k}_2}R_{\mathbf{k}_3}R_{\mathbf{k}_4}}
\Big\{(a^{*}_{\mathbf{k}_3,d} a_{\mathbf{k}_4,d} +a^{*}_{\mathbf{k}_3,f} a_{%
\mathbf{k}_4,f})\times
\nonumber\\
(a^{*}_{\mathbf{k}_1,d} a_{\mathbf{k}_2,d} &+&a^{*}_{\mathbf{k}_1,f} a_{%
\mathbf{k}_2,f})
\Big[(\langle \hat C^{\dagger}_{2,\mathbf{k%
}_3 \sigma} \hat C_{2,\mathbf{k}_4 \sigma} \hat C^{\dagger}_{2,\mathbf{k}%
_1 \sigma} \hat C_{2,\mathbf{k}_2 \sigma} \rangle - \langle \hat
C^{\dagger}_{2,\mathbf{k}_3 \sigma} \hat C_{2,\mathbf{k}_4 \sigma}
\rangle \langle \hat C^{\dagger}_{2,\mathbf{k}_1 \sigma} \hat
C_{2,\mathbf{k}_2 \sigma} \rangle) 
\nonumber\\
- (\langle \hat C^{\dagger}_{2,\mathbf{k}_3 \sigma}%
&\hat C_{2,\mathbf{k}_4 \sigma}& 
\hat C^{\dagger}_{2,\mathbf{k}_1 -\sigma} \hat C_{2,%
\mathbf{k}_2 -\sigma} \rangle - \langle \hat C^{\dagger}_{2,\mathbf{k}%
_3 \sigma} \hat C_{2,\mathbf{k}_4 \sigma} \rangle \langle \hat %
C^{\dagger}_{2,\mathbf{k}_1 -\sigma} \hat C_{2,\mathbf{k}_2 - %
\sigma} \rangle)\Big] + %
\label{eq51b}\\
(a_{\mathbf{k}_3,d} a_{\mathbf{k}_4,f} &-& %
a_{\mathbf{k}_3,f} a_{\mathbf{k}_4,d}) %
(a^{*}_{\mathbf{k}_1,f} a^{*}_{%
\mathbf{k}_2,d} - a^{*}_{\mathbf{k}_1,d} a^{*}_{ \mathbf{k}_2,f}) %
\langle \hat C^{\dagger}_{1,\mathbf{k}_3 \sigma} \hat C_{2,%
\mathbf{k}_4 \sigma} \hat C^{\dagger}_{2,\mathbf{k}_1 \sigma} \hat C_{1,%
\mathbf{k}_2 \sigma} \rangle \Big\} %
\delta_{\mathbf{k}_4, \mathbf{k}_1 - \mathbf{k}_2 +\mathbf{k}_3}
\nonumber
\end{eqnarray}
\def\theequation{\arabic{equation}}
\setcounter{equ}{1}
\def\theequation{\arabic{equation}.\alph{equ}}
\begin{eqnarray}
\rho_{\mathbf{S}_\perp,\mathbf{S}_\perp}(\mathbf{r}) &=&
\frac{1}{N_{\Lambda}} \sum_{\mathbf{i}} [ \langle \hat
S^{x}_{\mathbf{i}} \hat S^x_{\mathbf{i}+\mathbf{r}} + \hat
S^{y}_{\mathbf{i}} \hat S^y_{ \mathbf{i}+\mathbf{r}} \rangle -
(\langle \hat S^{x}_{\mathbf{i}} \rangle \langle \hat
S^x_{\mathbf{i}+\mathbf{r}} \rangle
+ \langle \hat S^{y}_{\mathbf{i}} \rangle \langle \hat S^y_{\mathbf{i}+%
\mathbf{r}} \rangle) ]
\label{eq52a}\\
\stepcounter{equ}\setcounter{equation}{52}
&=& \frac{1}{2
N^2_{\Lambda}} \sum_{\sigma}
\sum_{\mathbf{k}_1,\cdots,
\mathbf{k}_4} e^{i (\mathbf{k}_1-\mathbf{k}%
_2)\mathbf{r}} \sqrt{R_{\mathbf{k}_1}R_{\mathbf{k}_2}R_{\mathbf{k}_3}R_{%
\mathbf{k}_4}} \Big\{ (a^{*}_{\mathbf{k}_3,d} a_{\mathbf{k}_4,d} +a^{*}_{%
\mathbf{k}_3,f} a_{\mathbf{k}_4,f}) \times
\label{eq52b}\\
(a^{*}_{\mathbf{k}_1,d} a_{\mathbf{k}_2,d} &+&a^{*}_{\mathbf{k}_1,f} a_{%
\mathbf{k}_2,f}) ( \langle \hat C^{\dagger}_{2,\mathbf{k}_3
\sigma} \hat C_{2,\mathbf{k}_4 -\sigma} \hat
C^{\dagger}_{2,\mathbf{k}_1 -\sigma}
\hat C_{2,\mathbf{k}_2 \sigma} \rangle - \langle \hat C^{\dagger}_{2,\mathbf{%
k}_3 \sigma} \hat C_{2,\mathbf{k}_4 -\sigma} \rangle \langle \hat
C^{\dagger}_{2,\mathbf{k}_1 -\sigma} \hat C_{2,\mathbf{k}_2
\sigma} \rangle) +
\nonumber\\
(a_{\mathbf{k}_3,d} a_{\mathbf{k}_4,f} &-&a_{\mathbf{k}_3,f} a_{\mathbf{k}%
_4,d}) (a^{*}_{\mathbf{k}_1,f} a^{*}_{\mathbf{k}_2,d} -a^{*}_{\mathbf{k}%
_1,d} a^{*}_{ \mathbf{k}_2,f}) \langle \hat C^{\dagger}_{1,\mathbf{k}%
_3 \sigma} \hat C_{2,\mathbf{k}_4 -\sigma} \hat C^{\dagger}_{2,\mathbf{k}%
_1 -\sigma} \hat C_{1,\mathbf{k}_2 \sigma} \rangle \Big\}
\delta_{\mathbf{k}_4, \mathbf{k}_1 - \mathbf{k}_2 +\mathbf{k}_3}.
\nonumber
\end{eqnarray}
\def\theequation{\arabic{equation}}
Thus
one finds
$\rho_{S_z,S_z}({\mathbf{r}})=\delta_{\mathbf{r},0}/4$ and
$\rho_{\mathbf{S}_\perp,\mathbf{S}_\perp}(\mathbf{r}) =
\delta_{\mathbf{r},0}/2$  in the thermodynamic limit.
Consequently, the spins are uncorrelated at distances
$|\mathbf{r}| \ne 0$. This is a result of the macroscopic spin
degeneracy of the ground state.\\

\section{Alternative Transformations of the Hamiltonian}

\subsection{Generalized cell operators}

The transformation of the PAM Hamiltonian into positive
semi-definite form in terms of composite operators (the unit cell
operators $\hat A_{ I_{\mathbf{i}},\sigma}$ in Sect. II.B.2) is
relatively independent of the form of $\hat A$. Instead of
defining linear combinations of fermionic operators inside a
\emph{unit cell} of the Bravais lattice one may also define this
superposition on an \emph{arbitrary} substructure of the
underlying lattice, e.g., a cell larger than a unit cell (for
example, see Fig. 8). This leads to more general cell operators
which may then be employed to transform the Hamiltonian
into a form similar to (\ref{ham}). The corresponding matching
conditions are similar to (\ref{Eq12}) but are now fulfilled in a
different region of parameter space. In this way it is possible to
construct and investigate various regions of parameter space.

Below we present a different transformation of the PAM Hamiltonian
based on cell operators with octahedral shape, as shown in Fig.8.
Thereby it is possible to make direct contact with the
conventional PAM \cite{mi7}. This clearly shows that a
transformation of the PAM Hamiltonian into positive semi-definite
form is \emph{not} linked to finite $f$-electron hopping
amplitudes in the PAM as one might have suspected from the steps
performed in Sect. II.B. It also shows that the conventional PAM
and the generalized version with extended hopping and
hybridization amplitudes can have ground states with quite similar
properties. This emphasizes the fact that the physical properties
of an interacting electronic system need not depend on the precise
form of the \emph{non}-interacting bands.

We consider an octahedral cell $B_{\mathbf{i}}$ at each lattice
site $\mathbf{i}$. The seven sites within $B_{\mathbf{i}}$ are
denoted by $\mathbf{r}_{B_{\mathbf{i}}} = \mathbf{i} +
\mathbf{r}_{\alpha_1,\alpha_2,\alpha_3}$, where $\mathbf{r}%
_{\alpha_1,\alpha_2,\alpha_3} = \alpha_1(1-|\alpha_2|)(1-
|\alpha_3|) \mathbf{x}_1 + \alpha_2(1-|\alpha_1|)(1-|\alpha_3|)
\mathbf{x}_2 + \alpha_3(1-|\alpha_1|)(1-|\alpha_2|)\mathbf{x}_3$,
$\alpha_{\tau}=-1,0,1$. Here $\mathbf{x}_{\tau}$ are the primitive
vectors of the unit cell of the lattice. As seen from Fig.8., the
seven sites $\mathbf{r}_{B_{\mathbf{i}}}$
can be numbered by the indices $n_{\alpha_1,\alpha_2,\alpha_3}=(1-|%
\alpha_2|)(1- |\alpha_3|)(\alpha_1|\alpha_1|+2)
+(1-|\alpha_1|)(1-|\alpha_3|)(\alpha_2 |\alpha_2| + |\alpha_2| +
2) + (1-|\alpha_1|)(1-|\alpha_2|)(\alpha_3 |\alpha_3| + 4
|\alpha_3| + 2)$ without reference to $B_{\mathbf{i}}$. The
separation vector $\mathbf{r}$ between a site $\mathbf{i}$ and its
neighbors in the cell $B_{\mathbf{i}}$ can take 11 distinct
values ($\mathbf{x}_{\tau}, \mathbf{x}_{ \tau} \pm \mathbf{x}%
_{\tau^{\prime}}, 2 \mathbf{x}_{\tau}$,
$\tau,\tau^{\prime}=1,2,3$, $\tau >
\tau^{\prime}$) over which the summation in (\ref{Eq3},\ref{Eq4}%
) must be performed. Instead of (\ref{Eq7}) the cell operator is
now defined as
\begin{eqnarray}
\hat B^{\dagger}_{B_{\mathbf{i}},\sigma} = a^{*}_{6f} \hat f^{\dagger}_{%
\mathbf{i}, \sigma} + \sum_{n_{\alpha_1,\alpha_2,\alpha_3}=1}^7
a^{*}_{n,d}
\hat d^{ \dagger}_{\mathbf{i}+\mathbf{r}_{\alpha_1,\alpha_2,\alpha_3},%
\sigma} ,  \label{Eq86}
\end{eqnarray}
where $n=n_{\alpha_1,\alpha_2,\alpha_3}$, and the seven vectors
$\mathbf{r}_{ \alpha_1,\alpha_2,\alpha_3} \equiv
\mathbf{r}_{n_{\alpha_1,\alpha_2,\alpha_3}} \equiv
\mathbf{r}_{n}$ in the sum are, in the increasing order of the index $n$, $-\mathbf{x}%
_1, -\mathbf{x}_2, \mathbf{x}_1, \mathbf{x}_2, -\mathbf{x}_3, 0, \mathbf{x}%
_3,$ (see Fig.8.). We note that $a_{n,f}\ne 0$ only on the central
site (e.g., $n=6$) of the cell $B_{\mathbf{i}}$. Consequently
the product $\hat B^{\dagger}_{B_{\mathbf{i}},\sigma} \hat B_{B_{%
\mathbf{i}} \sigma}$ does not introduce $f$-electron hopping terms
into the Hamiltonian, implying that the decomposition discussed
here directly applies to the conventional PAM. In addition, the
Hamiltonian contains only on-site and nearest-neighbor
hybridization amplitudes, and $d$-electron hopping occurs between
nearest and next-nearest neighbor sites.

Instead of (\ref{ham}) the transformed Hamiltonian then becomes
\begin{eqnarray}
\hat H = \sum_{\mathbf{i},\sigma} \hat B_{B_{\mathbf{i}},\sigma}
\hat B^{\dagger}_{ B_{\mathbf{i}},\sigma} + U \hat P + E_{g,b} ,
\label{Eq87}
\end{eqnarray}
where $E_{g,b}= K \hat N - U N_{\Lambda} - 2 N_{\Lambda} (|a_{6,f}|^2+K)$, $%
K=\sum_{n=1}^7 |a_{n,d}|^2$. Furthermore, the matching conditions
from (\ref{Eq12}) transform into the following  non-linear system
of 19 coupled complex algebraic equations
\begin{eqnarray}
&&- t^d_{\mathbf{x}_1} = a^{*}_{6,d} a_{3,d} + a^{*}_{1,d} a_{6,d}
, \quad - t^d_{\mathbf{x}_2} = a^{*}_{6,d} a_{4,d} + a^{*}_{2,d}
a_{6,d} , \quad - t^d_{\mathbf{x}_3} = a^{*}_{6,d} a_{7,d} +
a^{*}_{5,d} a_{6,d} ,  \nonumber
\\
&&- t^d_{\mathbf{x}_2 + \mathbf{x}_1} = a^{*}_{2,d} a_{3,d} +
a^{*}_{1,d} a_{4,d} , \quad - t^d_{\mathbf{x}_2-\mathbf{x}_1} =
a^{*}_{3,d} a_{4,d} + a^{*}_{2,d} a_{1,d} , \quad -
t^d_{\mathbf{x}_3+\mathbf{x}_1} = a^{*}_{5,d}
a_{3,d} + a^{*}_{1,d} a_{7,d} ,  \nonumber \\
&&- t^d_{\mathbf{x}_3-\mathbf{x}_1} = a^{*}_{3,d} a_{7,d} +
a^{*}_{5,d} a_{1,d} , \quad - t^d_{\mathbf{x}_3+\mathbf{x}_2} =
a^{*}_{2,d} a_{7,d} + a^{*}_{5,d} a_{4,d} , \quad -
t^d_{\mathbf{x}_3-\mathbf{x}_2} = a^{*}_{4,d}
a_{7,d} + a^{*}_{5,d} a_{2,d} ,  \nonumber \\
&&- t^d_{2 \mathbf{x}_1} = a^{*}_{1,d} a_{3,d} , \quad - t^d_{2 \mathbf{x}%
_2} = a^{*}_{2,d} a_{4,d} , \quad - t^d_{2 \mathbf{x}_3} =
a^{*}_{5,d}
a_{7,d} , \quad -V_{0} = a^{*}_{6,d} a_{6,f} ,  \nonumber \\
&&-V_{\mathbf{x}_1} = a^{*}_{1,d} a_{6,f} = a^{*}_{6,f} a_{3,d} , \quad -V_{%
\mathbf{x}_2} = a^{*}_{2,d} a_{6,f} = a^{*}_{6,f} a_{4,d} , \quad -V_{%
\mathbf{x}_3} = a^{*}_{5,d} a_{6,f} = a^{*}_{6,f} a_{7,d} ,
\label{Eq88}
\end{eqnarray}
where $V^{f,d}_{\mathbf{r}}=V^{d,f}_{\mathbf{r}}=V_{\mathbf{r}}$,
$E_f = K - U - |a_{6,f}|^2$, and $a_{n,f}=0$ for $n \ne 6$. The
ground state wave function valid for $N \geq 3 N_{\Lambda}$ now
has the form
\begin{eqnarray}
|\Psi_{g,b}\rangle = \prod_{\mathbf{i}=1}^{N_{\Lambda}} [ \hat
B^{\dagger}_{ B_{\mathbf{i}},\uparrow} \hat
B^{\dagger}_{B_{\mathbf{i}},\downarrow} \hat F^{\dagger}_{
\mathbf{i}} ] \hat V^{\dagger}_M | 0 \rangle ,  \label{Eq89}
\end{eqnarray}
where, for $N=3N_{\Lambda}$, $\hat V^{\dagger}_M = 1$ holds. Since $%
a_{n,d}/a_{n,f}$ cannot be constant for all $n=1,2,..7$ (see Sec.
III.) $|\Psi_{g,b}\rangle$ describes an itinerant ground state
with properties similar to those presented in Sec. IV. An
isotropic solution is obtained for \cite{xxx0}
\begin{eqnarray}
a_{1,d}= \sqrt{\frac{|t^d_2|}{2}} e^{i \phi} , \quad a_{6,d}= \frac{|V_0|}{%
V_1} \sqrt{\frac{|t^d_2|}{2}} e^{i(\phi - \phi_{V_0})} , \quad
a_{6,f}= - V_1 \sqrt{\frac{2}{|t^d_2|}} e^{i\phi} .  \label{Eq90}
\end{eqnarray}
where $\phi$ is an arbitrary phase and $\phi_{V_0}$ is the phase
\cite{commentx} of the hybridization
amplitude $V_0$. Introducing the notation $%
t=|t^d_2/t^d_1| $, $v=|V_1/t^d_1|$, this solution emerges in the
parameter region with
\begin{eqnarray}
\frac{U + E_f}{|t^d_1|} = 3 t - \frac{2v^2}{t} + \frac{1}{2 t \cos^2%
\phi_{V_0} }.  \label{Eq91}
\end{eqnarray}
Since $V_0$ and thereby $Im(V_0)$ is in principle arbitrary,
(\ref{Eq91}) defines an entire region in the parameter space where
the solution exists (see Fig.9.). For $Im(V_0) \ne 0$, or real
$V_0$ at $|t^d_1/t^d_2|>6$, the properties presented in Sec.IV.
remain valid. \cite{xxx}

The results presented in this Section show that an itinerant
non-Fermi liquid phase emerges also for real hybridization ---
even in the conventional PAM.

\subsection{Alternative decomposition using unit cell operators}


To emphasize the flexibility of our method for constructing exact
ground states of the PAM we will now show that even if \emph{unit}
cell operators $\hat A_{I_{\bf i},\sigma}$ are used, alternative
decompositions are possible which lead to qualitatively different
ground states in other regions of the parameter space. For
example, instead of (\ref{ham}) the Hamiltonian $\hat H$ defined
by (\ref{Eq1},\ref{Eq2i},\ref{Eq3})
can also be cast into the form
\begin{eqnarray}
\hat H = \sum_{{\bf i},\sigma} \hat A^{\dagger}_{I_{\bf i},\sigma}
\hat A_{I_{\bf i},\sigma} + U \sum_{\bf i} \hat n_{{\bf
i},\uparrow} \hat n_{{\bf i},\downarrow} + E'_g . \label{mag1}
\end{eqnarray}
For periodic boundary conditions the matching conditions can then
be obtained from (\ref{Eq12}) by (i) replacing $J^{b,b'}_{\bf r}$ by ${\bar J}^{b,b'}_{%
\bf r}=-J^{b,b'}_{\bf r}$, and (ii) replacing the last two
equations (for $V_0$ and $U+E_f$) by $V_0 = \sum_{n=1}^8
a^{*}_{n,d}a_{n,f}$ and $E_f=K_d-K_f$, with $E'_g/N=K_d$.

Using the new solutions of the matching conditions for arbitrary
$U
> 0$ we will now identify new ground states of (\ref{mag1})
at and below quarter filling.

\subsubsection{Insulating ground state at quarter filling}

There exist solutions of the form $q_{n,d}/q_{n,f} = p$ for all
$n$, with the additional constraint $p=p^{*}$. In this case, the
unit cell operator $\hat A_{I_{\bf i},\sigma}$ contains only
contributions of the form $\hat f_{{\bf i},\sigma} + p \hat
d_{{\bf i},\sigma}$ for all $\mathbf{i}$ and $\sigma$. Introducing
$\hat I^{\dagger}_{\mathbf{i}}(\mu_{\mathbf{i}}) = [\hat f^{\dagger}_{%
{\bf i},\uparrow} -(1/p) \hat d^{\dagger}_{{\bf i},\uparrow}] +%
\mu_{\mathbf{i}} [\hat f^{\dagger}_{{\bf i},\downarrow} -(1/p) %
\hat d^{\dagger}_{{\bf i},\downarrow}]$, where $\mu_{\mathbf{i}}$
is an arbitrary site dependent constant, one observes that
\begin{eqnarray}
\{\hat f_{{\bf j},\sigma} + p \hat d_{{\bf j},\sigma}, \hat
I^{\dagger}_{\mathbf{j'}}(\mu_{\mathbf{j'}}) \} = 0 , \quad
\forall \mathbf{j}, \mathbf{j'}, \mu_{\mathbf{j'}}, \sigma .
\label{mag1x}
\end{eqnarray}
Since the operator $\hat
I^{\dagger}_{\mathbf{i}}(\mu_{\mathbf{i}})$ acts only on
site $\mathbf{i}$, the operator $\hat I^{\dagger} = \prod^{N_{\Lambda}}_{%
\mathbf{i}=1} \hat I^{\dagger}_{\mathbf{i}}(\mu_{\mathbf{i}})$
does not introduce $f$-electron double occupancies. Consequently,
$\sum_{{\bf i},\sigma} \hat A^{\dagger}_{I_{\bf i},\sigma} %
\hat A_{I_{\bf i},\sigma} \hat I^{\dagger} |0\rangle =0$,
$U \sum_{\bf i} \hat n^f_{{\bf i},\uparrow} \hat n^f_{{\bf i},\downarrow}%
\hat I^{\dagger} |0\rangle = 0$, and the ground state becomes
\begin{eqnarray}
|\Psi_g\rangle = \prod_{{\bf i}=1}^{N_{\Lambda}}[(\hat
f^{\dagger}_{{\bf
i},%
\uparrow} - \frac{1}{p} \hat d^{\dagger}_{{\bf i},\uparrow}) +
\mu_{\bf i}
(\hat f^{\dagger}_{{\bf i},\downarrow} - \frac{1}{p} \hat d^{\dagger}_{%
{\bf i},\downarrow})]|0\rangle. \label{mag5}
\end{eqnarray}
The ground state (\ref{mag5}) has exactly one electron per site,
and is degenerate, localized and globally non-magnetic. One can
directly show that $Re \sigma_{\tau,\tau}(0) = 0$ (see
Sec.III.A.1.), i.e., the state is indeed insulating.

\subsubsection{Conducting ferromagnetic ground state at quarter filling}

If $q_{n,d}/q_{n,f}$ depends on $n$, $\hat
A_{I_{\mathbf{i}},\sigma}$ does not anticommute with $\hat
I^{\dagger}_{\mathbf{i}}(\mu_{\mathbf{i}})$; hence
$\sum_{{\bf i},\sigma} \hat A^{\dagger}_{I_{\bf i},\sigma} %
\hat A_{I_{\bf i},\sigma} \hat I^{\dagger} |0\rangle =0$ does not
hold. In this case one may introduce a complementary unit cell
operator\cite{compop}
\begin{eqnarray}
&&\hat{Q}_{I_{\mathbf{i}},\sigma }^{\dagger }=\sum_{n(\alpha
,\beta ,\gamma )=1}^{8}[q_{n,d}^{\ast
}\hat{d}_{\mathbf{i}+\mathbf{r}_{\alpha ,\beta ,\gamma },\sigma
}^{\dagger }+q_{n,f}^{\ast
}\hat{f}_{\mathbf{i}+\mathbf{r}%
_{\alpha ,\beta ,\gamma },\sigma }^{\dagger }]  \nonumber \\
&=&(q_{1,d}^{\ast }\hat{d}_{\mathbf{i},\sigma }^{\dagger
}+q_{2,d}^{\ast
}%
\hat{d}_{\mathbf{i}+\mathbf{x}_{1},\sigma }^{\dagger
}+q_{3,d}^{\ast
}\hat{d}%
_{\mathbf{i}+\mathbf{x}_{1}+\mathbf{x}_{2},\sigma }^{\dagger
}+q_{4,d}^{\ast }\hat{d}_{\mathbf{i}+\mathbf{x}_{2},\sigma
}^{\dagger }+....+q_{8,d}^{\ast
}%
\hat{d}_{\mathbf{i}+\mathbf{x}_{2}+\mathbf{x}_{3},\sigma
}^{\dagger })
\nonumber \\
&&+(q_{1,f}^{\ast }\hat{f}_{\mathbf{i},\sigma }^{\dagger
}+q_{2,f}^{\ast
}%
\hat{f}_{\mathbf{i}+\mathbf{x}_{1},\sigma }^{\dagger
}+q_{3,f}^{\ast
}\hat{f}%
_{\mathbf{i}+\mathbf{x}_{1}+\mathbf{x}_{2},\sigma }^{\dagger
}+q_{4,f}^{\ast }\hat{f}_{\mathbf{i}+\mathbf{x}_{2},\sigma
}^{\dagger }+....+q_{8,f}^{\ast
}%
\hat{f}_{\mathbf{i}+\mathbf{x}_{2}+\mathbf{x}_{3},\sigma
}^{\dagger }), \label{mag2}
\end{eqnarray}
with the property $\{ \hat A_{I_{\bf i},\sigma}, \hat
Q^{\dagger}_{I_{\bf
j},%
\sigma'} \}=0$ for all ${\bf i},{\bf j}$ and $\sigma,\sigma'$. The
numerical coefficients are given by the relations
\begin{eqnarray}
&&q^{*}_{1,d}=w a_{7,f}, \quad q^{*}_{2,d}=w a_{8,f}, \quad
q^{*}_{3,d}=w a_{5,f}, \quad q^{*}_{4,d}=w a_{6,f},
\nonumber\\
&&q^{*}_{5,d}=w a_{3,f}, \quad q^{*}_{6,d}=w a_{4,f}, \quad
q^{*}_{7,d}=w a_{1,f}, \quad q^{*}_{8,d}=w a_{2,f},
\nonumber\\
&&q^{*}_{1,f}=-w a_{7,d}, \quad q^{*}_{2,f}=-w a_{8,d}, \quad
q^{*}_{3,f}=-w a_{5,d}, \quad q^{*}_{4,f}=-w a_{6,d},
\nonumber\\
&&q^{*}_{5,f}=-w a_{3,f}, \quad q^{*}_{6,f}=-w a_{4,d}, \quad
q^{*}_{7,f}=-w a_{1,d}, \quad q^{*}_{8,f}=-w a_{2,d}, \label{mag3}
\end{eqnarray}
where $w$ is an arbitrary non-zero constant. Introducing
$\hat Q^{\dagger}_{\sigma_1,\sigma_2,...,\sigma_{N_{\Lambda}}} =%
\prod_{{\mathbf{i}}=1}^{N_{\Lambda}} \hat Q^{\dagger}_{I_{\mathbf{i}},%
\sigma_{\mathbf{i}}}$
one has $\sum_{{\bf i},\sigma} \hat A^{\dagger}_{I_{\bf
i},\sigma}\hat A_{I_{\bf
i},%
\sigma} \hat Q^{\dagger}_{\sigma_1,\sigma_2,...,\sigma_{N_{\Lambda}}}%
|0\rangle = 0$ for arbitrary $\sigma_{\mathbf{i}}$. If the sites
$\mathbf{i},\mathbf{i'}$ are adjacent the operator $\hat Q^{\dagger}_{I_{\mathbf{i}},%
\sigma_{\mathbf{i}}} \hat
Q^{\dagger}_{I_{\mathbf{i'}},\sigma_{\mathbf{i'}}}$ introduces two
electrons on the common sites of $I_{\mathbf{i}}$ and
$I_{\mathbf{i'}}$. The state $\hat Q^{\dagger}_{\sigma_1,\sigma_2,...,%
\sigma_{N_{\Lambda}}}|0\rangle$ then provides the minimum possible
eigenvalue (e.g., zero) in the presence of the Hubbard term only
when $\sigma_{\mathbf{i}} = \sigma$ for all $\mathbf{i}$.
Consequently, the ground state becomes
\begin{eqnarray}
|\Psi_g\rangle = \prod_{{\bf i}=1}^{N_{\Lambda}} \hat
Q^{\dagger}_{{\bf
i}%
\sigma} |0\rangle; \sigma=\uparrow,\downarrow  . \label{mag4}
\end{eqnarray}
This corresponds to  a fully saturated, non-degenerate
\emph{ferromagnetic} phase, which is metallic since
in (\ref{mag4}) double occupancies (e.g., $\hat d^{\dagger}_{{\bf i},%
\uparrow} \hat f^{\dagger}_{{\bf i},\uparrow}$), empty sites, and
single occupancies are simultaneously present.\cite{itinq}

Ferromagnetism in the PAM
at and around quarter filling has been investigated rather
extensively in the past already. Following variational results for
Kondo lattices by Fazekas and M\"uller-Hartmann \cite{MM0}
ferromagnetic phases in the PAM itself were found by Dorin and
Schlottmann \cite{MM1A} in the limit $U\to \infty$, and at finite
$U$ by M\"oller and W\"olfle \cite{MM1B} within a slave-boson
approach. Subsequently, ferromagnetic solutions of the PAM at and
near quarter filling were obtained within various other
approximation schemes \cite{MM2,MM3,MM4,MM5,MM6,MM7,MM8,MM9,MM10}.
Exact results were derived in $D=1$ for $U=\infty$ by
Yanagisawa\cite{MM11}.

In the exact solution discussed here ferromagnetism emerges when
the lower diagonalized band of (\ref{Eq3}) becomes non-dispersive
(flat). We note that this can happen even if the \emph{bare} bands
of the Hamiltonian are dispersive.

\subsubsection{Ground states below quarter filling}

Exact ground states of the PAM can even be constructed for $N <
N_{\Lambda}$, i.e., below quarter filling. For the two cases
discussed in Sec.V.B.1-2. they have the same form as (\ref{mag4}) and
(\ref{mag5}). However, the upper limit of the products has to
be replaced by $N$, and the additional geometrical degeneracy of
the electrons needs to be taken into account.

For $n$-dependent $q_{n,d}/q_{n,f}$, the ground state becomes
\begin{eqnarray}
|\Psi_g\rangle = \sum_{{\cal D}_N} \sum_{\{ {\cal D}_{N,m} \} }
\sum_{ \{ \sigma_m \} } \alpha_{{\cal D}_N, \{ {\cal D}_{N,m} \},
\{ \sigma_m \} } \prod_m (\prod_{{\bf i} \in {\cal D}_{N,m}} \hat
Q^{\dagger}_{{\bf i},\sigma_{m} }) |0\rangle \label{mag6}
\end{eqnarray}
where ${\cal D}_N$ denotes an arbitrary domain (a subset of
lattice points) of the full lattice containing $N < N_{\Lambda}$
lattice sites. A given domain ${\cal D}_{N}$ consists of disjoint
subdomains (clusters) denoted by ${\cal D}_{N,m}$ where $m$
enumerates the clusters in ${\cal D}_{N}$; the maximum of $m$ is
denoted by $N_m$. Clearly one has ${\cal D}_N = {\cal D}_{N,1}
\cup {\cal D}_{N,2} \cup ... \cup {\cal D}_{N,N_m}$, and ${\cal
D}_{N,m_1} \cap {\cal D}_{N,m_2} =0$ for $m_1 \ne m_2$.
Furthermore, $\sigma_m = \pm 1/2$ represents a fixed, but
arbitrary, spin-index in the sub-domain ${\cal D}_{N,m}$, and
$\alpha_{{\cal D}_N, \{ {\cal D}_{N,m} \}, \{ \sigma_m \} }$ are
arbitrary coefficients. From a physical point of view (\ref{mag6})
contains disjoint, fully saturated ferromagnetic, conducting
clusters of arbitrary shape and size, whose spin orientation is
arbitrary.

For $q_{n,d}/q_{n,f}=p=p^{*}$ the ground state becomes
\begin{eqnarray}
|\Psi_g\rangle = \sum_{{\cal D}_N} \sum_{\{ {\cal D}_{N,m} \} }
\alpha_{{\cal D}_N, \{ {\cal D}_{N,m} \} } \prod_m
(\prod_{{\bf i} \in {\cal D}_{N,m}} [(\hat f^{\dagger}_{{\bf i},%
\uparrow} - \frac{1}{p} \hat d^{\dagger}_{{\bf i},\uparrow}) +
\mu_{\bf i}
(\hat f^{\dagger}_{{\bf i},\downarrow} - \frac{1}{p} \hat d^{\dagger}_{%
{\bf i},\downarrow})] ) |0\rangle \label{mag7}
\end{eqnarray}
containing again clusters of arbitrary shape and size. But now the
clusters are insulating and non-magnetic, containing strictly one
particle per site with arbitrary spin.

The parameter region corresponding to the ground state solutions
in Sec.V.B. can be obtained from that derived from the matching
conditions (\ref{Eq12}) by the replacements $t^b_{\bf r} \to -
t^b_{\bf r}$, $V_{\bf r} \to -V_{\bf r}$, $U \to 0$. The results
are valid for all $U>0$.

\section{Summary}

We presented details of an analytic scheme which allows one to
construct exact ground states for a general class of
\emph{three}-dimensional periodic Anderson models (PAM), including
the conventional PAM, on regular Bravais lattices.
First the Hamiltonian is cast into positive semi-definite form
with the help of composite fermionic operators in combination with
a set of coupled, non-linear matching conditions for the
microscopic parameters of the Hamiltonian. Then a non-local
product state of these composite operators in position space,
corresponding to 3/4 filling is constructed which yields exact
grounds in various parts of parameter space.

Depending on the choice of the composite operators and the
geometry of the building blocks of the lattice
on which they are defined, the transformation of the Hamiltonian
into positive semi-definite form can be performed in several ways.
Thereby it is possible to construct exact ground states in
different regions of the parameter space of the model.

For real $d,f$ hybridization amplitudes we constructed an
insulating, non-magnetic ground state which is stable on several
different lattice structures. Its ground state energy was shown to
diverge at the boundary of the stability region, implying a
divergence of its compressibility. Such an anomaly is known to
occur in several heavy-fermion materials. Furthermore, we
identified an exact metallic non-Fermi liquid ground state,
characterized by one dispersing band and one upper flat band,
which is stable in different regions of parameter space. This
state is non-magnetic and has vanishing non-local spin-spin
correlations in the thermodynamic limit. Its density-density
correlations are short ranged, and the momentum distributions of
the electrons in the interacting ground state have no
discontinuities. The stability regions of these ground states
extend through a large region of the parameter space -- from weak
to strong on-site interactions $U$.

Exact ground states with conducting and insulating properties,
respectively, were also constructed at and below quarter filling.
In particular, a conducting, fully polarized ferromagnetic state
was found to be the ground state at quarter filling. At lower
fillings a ground state characterized by ferromagnetic clusters of
arbitrary shape was identified.

Our results show that ground states of the conventional PAM and of
generalizations with extended hopping and hybridization amplitudes
can have quite similar properties.

The exact ground states discussed in this paper correspond to
simple solutions of the coupled matching conditions for the
microscopic parameters of the Hamiltonian. In view of their large
number (e.g., 55 conditions in the case of the unit cell operators
in Sec. II.B) and their non-linearity it almost certain that other
solutions exist which then lead to yet other exact ground states
of the three-dimensional PAM and its extensions. In view of the
great relevance of this model for our understanding of correlated
electronic systems, both on the level of models and real
materials, more detailed investigations of the matching conditions
derived here will be worthwhile. Finally it should be stressed
that the concept behind the construction of exact ground states
for the PAM in $D=3$ presented here is quite general, and is also
applicable to other electronic correlation models.


\acknowledgments

This work was supported in part by the Hungarian Scientific
Research Fund through contract OTKA-T-037212, and by the Deutsche
Forschungsgemeinschaft through SFB 484. The numerical calculations
were performed at the Supercomputing Laboratory of the Faculty of
Natural Sciences, University of Debrecen, and supported by
OTKA-M-041537. One of us (ZG) acknowledges valuable discussions
with Ferenc Kun regarding numerical multidimensional integrations.

\appendix
\def\theequation{{\thesection}\arabic{equation}}


\section{ Transformation of the the Periodic Anderson Model Hamiltonian}

In this appendix we present details of the transformation of the
PAM Hamiltonian in its original form with $d$ and $f$ operators
[see
 (1), (2), (6)]
\begin{eqnarray}
\hat H &=& \sum_{{\bf i},\sigma} \{ \sum_{\bf r} [ (t^d_{\bf r}
\hat d^{\dagger}_{{\bf i}, \sigma} \hat d_{{\bf i} +{\bf
r},\sigma} + t^f_{\bf r} \hat f^{\dagger}_{ {\bf i},\sigma} \hat
f_{{\bf i} +{\bf r},\sigma}) + (V_{\bf r}^{df} \hat
d^{\dagger}_{{\bf i},\sigma} \hat f_{{\bf i} +{\bf r}, \sigma} +
V_{\bf r}^{fd} \hat f^{\dagger}_{{\bf i},\sigma} \hat d_{{\bf i} +
{\bf r},\sigma}) + H.c.]
\nonumber\\
&+&  (V_0 \hat d^{\dagger}_{{\bf i},\sigma} \hat f_{ {\bf
i},\sigma} + H.c.)  + (E_f +U) \hat n^f_{{\bf i},\sigma} \} + U
\hat P  - U N_{\Lambda} , \label{A00}
\end{eqnarray}
into the form with
unit cell operators, (\ref{ham}),
\begin{eqnarray}
\hat H = - \sum_{{\bf i},\sigma} \hat A^{\dagger}_{I_{\bf
i},\sigma} \hat A_{ I_{\bf i},\sigma} + K_d \hat N + U \hat P - U
N_{\Lambda} , \label{A0}
\end{eqnarray}
and explain how the matching conditions (\ref{Eq12}) arise in this
process.

One first calculates all terms in the sum $- \sum_{{\bf i},\sigma}
\hat A^{\dagger}_{I_{\bf i},\sigma} \hat A_{ I_{\bf i},\sigma}$
from (\ref{A0}). To identify the contributions generated thereby
with those defined by (\ref{A00}) one needs altogether 55 matching
equations.
This procedure will now be illustrated by four typical examples.
Here we refer to Fig.10 where four neighboring unit cells
$I_{\mathbf{i}_1}, I_{\mathbf{i}_2}, I_{\mathbf{j}_1}$, and
$I_{\mathbf{j}_2}$ -- each denoted according to the notation
introduced in Sec. II.B.1 -- is depicted. Since the precise shape
of the unit cells is unimportance here we use orthorhombic cells
for simplicity.

\underline{Nearest neighbor contributions}: First we analyze a
nearest neighbor amplitude, e.g.,
$t^d_{\mathbf{j}_2,\mathbf{j}_5}$, which appears in the kinetic
energy of
(\ref{A00}) as $t^d_{\mathbf{j}_2,\mathbf{j}_5} \hat d^{\dagger}_{\mathbf{j}%
_2, \sigma} \hat d_{\mathbf{j}_5,\sigma}$. This term acts along
the bond $(\mathbf{j}_2,\mathbf{j}_5)$ represented by the
horizontal, thick dashed line in Fig.10.
In (\ref{A0}) this term originates from $- \sum_{{\bf i},\sigma} \hat A^{\dagger}_{I_{\bf i},\sigma}%
\hat A_{ I_{\bf i},\sigma}$, but only for sites $\mathbf{i}=
\mathbf{i}_1, \mathbf{i}_2,\mathbf{j}_1, \mathbf{j}_2$ in
$\sum_{\mathbf{i}}$.
To calculate $- \sum_{{\mathbf i} = \mathbf{i}_1, \mathbf{i}_2,%
\mathbf{j}_1, \mathbf{j}_2} \hat A^{\dagger}_{I_{\bf i},\sigma}%
\hat A_{ I_{\bf i},\sigma}$ one must write down the unit cell operators $\hat A^{\dagger}_{%
I_{\mathbf{i}}}$, (\ref{Eq7}), corresponding to the aforementioned
four unit cells. For example one finds
\begin{eqnarray}
&&\hat A^{+}_{I_{\mathbf{i}_1},\sigma} = \sum_{b=d,f} [a^{*}_{1,b} %
\hat b^{\dagger}_{\mathbf{i}_1, \sigma} + a^{*}_{2,b} \hat b^{\dagger}_{%
\mathbf{i}_2,\sigma} + a^{*}_{3,b} \hat b^{\dagger}_{\mathbf{i}%
_5,\sigma} + a^{*}_{4,b} \hat b^{\dagger}_{\mathbf{i}_4, \sigma} +
a^{*}_{5,b} \hat b^{\dagger}_{\mathbf{j}_1,\sigma} + a^{*}_{6,b}
\hat b^{\dagger}_{\mathbf{j}_2,\sigma} + a^{*}_{7,b} \hat b^{
\dagger}_{\mathbf{j}_5,\sigma} +
a^{*}_{8,b} \hat b^{\dagger}_{\mathbf{j}_4,\sigma}] ,  \nonumber \\
&&\hat A^{+}_{I_{\mathbf{i}_2},\sigma} = \sum_{b=d,f} [a^{*}_{1,b} %
\hat b^{\dagger}_{\mathbf{i}_2, \sigma} + a^{*}_{2,b} \hat b^{\dagger}_{%
\mathbf{i}_3,\sigma} + a^{*}_{3,b} \hat b^{\dagger}_{\mathbf{i}%
_6,\sigma} + a^{*}_{4,b} \hat b^{\dagger}_{\mathbf{i}_5, \sigma} +
a^{*}_{5,b} \hat b^{\dagger}_{\mathbf{j}_2,\sigma} + a^{*}_{6,b}
\hat b^{\dagger}_{\mathbf{j}_3,\sigma} + a^{*}_{7,b} \hat b^{
\dagger}_{\mathbf{j}_6,\sigma} + a^{*}_{8,b} \hat
b^{\dagger}_{\mathbf{j}_5,\sigma}] .  \nonumber
\end{eqnarray}
Thereby one obtains
contributions of the form
$- a_{6,d}^{*}a_{7,d} \hat d^{\dagger}_{\mathbf{j}_2,\sigma}%
\hat d_{ \mathbf{j}_5,\sigma}$, $- a_{5,d}^{*} a_{8,d} \hat d^{\dagger}_{%
\mathbf{j}_2,\sigma} \hat d_{\mathbf{j}_5,\sigma}$, $- a_{2,d}^{*}
a_{3,d} \hat d^{\dagger}_{\mathbf{j}_2,\sigma} \hat
d_{\mathbf{j}_5,\sigma}$,
and $- a_{1,d}^{ *} a_{4,d} \hat d^{\dagger}_{\mathbf{j}_2,\sigma} \hat d_{%
\mathbf{j}_5,\sigma}$ implying $t^d_{\mathbf{j}_2,\mathbf{j}_5} =
- (a^{*}_{1,d} a_{4,d} + a^{*}_{2,d} a_{3,d} + a^{*}_{5,d}a_{8,d}
+ a^{*}_{6,d} a_{7,d})$.
Since (i) $t^d_{\mathbf{j}_2,\mathbf{j}_5}$ is a specific example of the general hopping amplitude $t^d_{%
\mathbf{i}, \mathbf{i}+\mathbf{x}_2}$, (ii) all amplitudes
$t^d_{\mathbf{r}}$ amplitudes have the same form (i.e.,
$t^d_{\mathbf{x}_2}= t^d_{\mathbf{i},\mathbf{i}+\mathbf{x}_2}$),
and (iii) this also holds for the $f$ electrons, i.e.,
$t^f_{\mathbf{x}_2}= t^f_{\mathbf{i},\mathbf{i}+\mathbf{x}_2}$,
one finds $- t^b_{\mathbf{x}_2} = a^{*}_{1,b} a_{4,b} +
a^{*}_{2,b} a_{3,b} + a^{*}_{5,b} a_{8,b} + a^{*}_{6,b} a_{7,b}$,
where $b=d,f$ .

The same analysis applies to the hybridization amplitudes.
Hence, for the terms $V^{d,f}_{\mathbf{x}_2} \hat d^{\dagger}_{\mathbf{i},%
\sigma} %
\hat f_{\mathbf{i}+{\mathbf{x}_2}, \sigma}$ and $V^{f,d}_{\mathbf{x}_2}%
 \hat f^{\dagger}_{\mathbf{i},\sigma} \hat d_{ \mathbf{i}+\mathbf{x}_2,%
\sigma}$ one finds $- V^{b,b^{\prime}}_{\mathbf{x}_2} =
a^{*}_{1,b} a_{4,b^{\prime}} + a^{*}_{2,b} a_{3,b^{\prime}} +
a^{*}_{ 5,b}a_{8,b^{\prime}} + a^{*}_{6,b} a_{7,b^{\prime}}$.

Thus we have derived the matching condition $J^{b,b^{\prime}}_{\mathbf{x}_2}=%
-[\delta_{%
b,b'} t^b_{\mathbf{x}_2} +
(1-\delta_{b,b'})V^{b,b'}_{\mathbf{x}_2}]$ [the third relation in
(\ref{Eq12})].

It should be noted that the sum $- \sum_{{\bf i},\sigma} \hat
A^{\dagger}_{I_{\bf i},\sigma} \hat A_{ I_{\bf i},\sigma}$ also
produces the Hermitian conjugates of the terms represented by
(\ref{A00}).
Therefore, in addition to the result for $t^d_{\mathbf{j}_2,\mathbf{j%
}_5}$ one also finds $t^d_{\mathbf{j}_5,\mathbf{j}%
_2} = -(a^{*}_{4,d} a_{1,d} + a^{*}_{3,d} a_{2,d} + a^{*}_{8,d}
a_{5,d} + a^{*}_{7,d} a_{6,d})$, where $t^d_{\mathbf{j}_5,\mathbf{j}_2} = %
t^d_{\mathbf{i}, \mathbf{i}-\mathbf{x}_2} = t^d_{-\mathbf{x}_2}= %
{t^d_{\mathbf{x}_2}}^{*}$.

\underline{Second-nearest neighbor contributions}: We now consider
a plaquette-diagonal next-nearest neighbor term, namely
$t^d_{\mathbf{i}_2,\mathbf{j}_5}$, acting on the oblique thick
dashed line in Fig.10. Since the bond
$(\mathbf{i}_2,\mathbf{j}_5)$ appears only in the unit cells
$I_{\mathbf{i}_1}$ and $I_{\mathbf{i}_2}$, only the contributions
from $- \sum_{\mathbf{i}}\hat A^{\dagger}_{ I_{
\mathbf{i}_{\alpha} },\sigma } \hat A_{I_{
\mathbf{i}_{\alpha}},\sigma}$ in (\ref{A0}) with
$\mathbf{i}=\mathbf{i}_1,\mathbf{i}_2$ give rise to
$t^d_{\mathbf{i}_2, \mathbf{j}_5} \hat d^{
\dagger}_{\mathbf{i}_2,\sigma} \hat d_{\mathbf{j}_5,\sigma}$ in
(\ref{A00}).
The sum $- \sum_{\mathbf{i}=\mathbf{i}_1,\mathbf{i}_2} \hat
A^{\dagger}_{ I_{\mathbf{i}},\sigma } \hat
A_{I_{\mathbf{i}},\sigma}$ leads to the expressions  $- a^{*}_{2,d} a_{7,d}%
 \hat d^{\dagger}_{ \mathbf{i}_2,\sigma} \hat d_{\mathbf{j}_5,\sigma}$, and
$- a^{*}_{1,d} a_{8,d} \hat d^{\dagger}_{\mathbf{i}_2,\sigma} \hat d_{%
\mathbf{j}_5,\sigma}$, implying $t^d_{\mathbf{i}_2,\mathbf{j}_5} =%
-(a^{*}_{1,d} a_{8,d} + a^{*}_{2,d}a_{7,d})$. Since (i)
$t^d_{\mathbf{i}_2,\mathbf{j}_5}$ is a specific example of the
general hopping amplitude
$t^d_{\mathbf{i},%
\mathbf{i}+ \mathbf{x}_3+\mathbf{x}_2}$, and (ii) all the amplitudes $t^d_{%
\mathbf{i},\mathbf{i}+ \mathbf{x}_3+\mathbf{x}_2}$ have the same
form (i.e., $t^d_{\mathbf{i},\mathbf{i}+ \mathbf{x}_3+\mathbf{x}_2} =%
t^d_{\mathbf{x}_3+\mathbf{x}_2}$)
one finds $- t^d_{\mathbf{x}_3+\mathbf{x}_2}%
= a^{*}_{1,d} a_{8,d} + a^{*}_{2,d}a_{7,d}$. These relations also
hold for the $f$ electrons. Therefore we arrive at the matching
condition $J^{b,b'}_{\mathbf{x}_3+\mathbf{x}_2} = a^{*}_{1,b}
a_{8,b'} + a^{*}_{2,b}a_{7,b'}$ [the sixth relation in
(\ref{Eq12})].

\underline{Third-nearest neighbor contributions}: The third
neighbor contributions, located along the space diagonals of the
unit cell, are the same in every cell. For example, studying the
bond $(\mathbf{j}_1,\mathbf{j}_5)$ of the cell $I_{\mathbf{i}_1}$,
the
product $- \hat A^{\dagger}_{ I_{ \mathbf{i}_1 },\sigma } \hat A_{I_{%
\mathbf{i}_1},\sigma}$ leads to the term $-a^{*}_{1,d} a_{7,d} %
\hat d^{\dagger}_{\mathbf{i}_1,\sigma} \hat
d_{\mathbf{j}_5,\sigma}$. Hence one finds
$t^d_{\mathbf{i}_1,\mathbf{j}_5} = t^d_{\mathbf{i},\mathbf{i}+%
\mathbf{x}_3+\mathbf{x}_2+ \mathbf{x}_1} = t^d_{\mathbf{x}_3+\mathbf{x}_2+%
\mathbf{x}_1} = - a^{*}_{1,d} a_{7,d}$ and, consequently,
$J^{b,b^{\prime}}_{\mathbf{x}_3 + \mathbf{x}_2 +\mathbf{x}_1} =a^{*}_{1,d} %
a_{7,d}$ [the tenth relation in (\ref{Eq12})].

\underline{Single-site contributions}: Since any site is common to
8 unit cells,  single site contributions require the consideration
of 8 neighboring unit cells. For example, the term $V_0 \sum_{\mathbf{i}%
,\sigma} \hat d^{\dagger}_{\mathbf{i},\sigma} \hat f_{{\hat
i},\sigma}$ is obtained from $- \sum_{\mathbf{i},\sigma} \hat
A^{\dagger}_{I_{\mathbf{i}},\sigma} \hat A_{I_{
\mathbf{i}},\sigma}$ as $-(\sum_{n=1}^8 a^{*}_{n,d} a_{n,f})
\sum_{ \mathbf{i},\sigma} \hat d^{\dagger}_{\mathbf{i},\sigma}
\hat f_{\mathbf{i},\sigma}$, implying $- V_0 = \sum_{n=1}^8
a^{*}_{n,d}a_{n,f}$. In the same way we obtain
the
coefficient of $-\sum_{\mathbf{i},\sigma} \hat b^{\dagger}_{\mathbf{i}%
,\sigma} \hat b_{\mathbf{i},\sigma}$ as $\sum_{n=1}^8 |a_{n,b}|^2%
= K_b$. Therefore the terms $- \sum_{\mathbf{i},\sigma} \hat A^{\dagger}_{%
I_{\mathbf{i}%
},\sigma} \hat A_{I_{\mathbf{i}},\sigma} + K_d \hat N$ in
(\ref{A0}) yield $-(K_d \hat N_d + K_f \hat N_f) + K_d (\hat N_d +
\hat N_f) = (K_d - K_f) \hat N_f$, where $\hat N = \hat N_d + \hat
N_f$. Taking into account the term $(U + E_f) \hat N_f$
 in (\ref{A00}) one obtains $U + E_f = K_d -K_f$ [last relation in
(\ref{Eq12})].

\section{The current operator}

In this appendix we derive the current operator as well as the
charge conductivity sum rule for a general form of the PAM.

The charge density operator at site $\mathbf{i}$ is given
by $\hat \rho(\mathbf{i}) = e \sum_{\sigma} ( \hat d^{\dagger}_{%
\mathbf{i},\sigma} \hat d_{\mathbf{i},\sigma} + \hat f^{\dagger}_{\mathbf{i}%
,\sigma} \hat f_{\mathbf{i},\sigma})$, with $e$ as the electron
charge. The current operator $\hat {\mathbf{j}}$ is defined as
\cite{br} $\hat {\mathbf{j}} = - (i/\hbar) [ \hat
{\mathbf{q}},\hat H ]$, where $V$ is the
volume of the system, $\hat {\mathbf{q}} = (1/V) \sum_{\mathbf{i}} \mathbf{i} \hat \rho(%
\mathbf{i})$ is the charge polarization operator, and $\hat H$ is
the Hamiltonian. With $\hat H$ from (\ref{A00}) the current
operator of the PAM is found as
\begin{eqnarray}
\hat {\mathbf{j}} = \frac{i e}{\hbar V} \sum_{\mathbf{i}}
\sum_{\mathbf{r}} \sum_{\sigma} \sum_{b,b^{\prime}=d,f} (
{J^{b,b^{\prime}}_{\mathbf{r}}}^{*}
\hat {b^{\prime}}^{\dagger}_{\mathbf{i}+\mathbf{r}, \sigma} \hat b_{\mathbf{i%
},\sigma} - J^{b,b^{\prime}}_{\mathbf{r}} \hat
b^{\dagger}_{\mathbf{i}, \sigma} \hat
b^{\prime}_{\mathbf{i}+\mathbf{r},\sigma} ) \mathbf{r} ,
\label{B2}
\end{eqnarray}
where $J^{b,b^{\prime}}_{\mathbf{r}}$ is defined in (\ref{Eq12}), and $%
\mathbf{r}$ is restricted to the values presented in Sec.II.B.1.
We note that only the itinerant part of the Hamiltonian, i.e.,
$\hat H_{itin} = \sum_{\mathbf{r}} \hat H_{itin}(\mathbf{r}) =
\hat H - \hat H_U - \sum_{\mathbf{i},\sigma} [ (V_0 \hat
d^{\dagger}_{\mathbf{i},\sigma} \hat f_{ \mathbf{i},\sigma} +
H.c.) + E_f \hat n^f_{\mathbf{i},\sigma} ]$, where
\begin{eqnarray}
\hat H_{itin}(\mathbf{r}) = \sum_{\mathbf{i},\sigma} [
(t^d_{\mathbf{r}} \hat d^{\dagger}_{\mathbf{i}, \sigma} \hat
d_{\mathbf{i} +\mathbf{r},\sigma}
+ t^f_{\mathbf{r}} \hat f^{\dagger}_{ \mathbf{i},\sigma} \hat f_{\mathbf{i} +%
\mathbf{r},\sigma}) + (V_{\mathbf{r}}^{df} \hat d^{\dagger}_{\mathbf{i}%
,\sigma} \hat f_{\mathbf{i} +\mathbf{r}, \sigma} +
V_{\mathbf{r}}^{fd} \hat f^{\dagger}_{\mathbf{i},\sigma} \hat
d_{\mathbf{i} + \mathbf{r},\sigma}) + H.c.]   \label{B4}
\end{eqnarray}
contributes to (\ref{B2}).

Starting from the current operator in (\ref{B2}), the Kubo formula
for the charge conductivity at zero temperatures becomes
$\sigma_{\tau,\tau} (\omega) = i V \int^{\infty}_{0} d  t e^{-i
\omega t} \langle [ \hat q_{\tau} (t), \hat j_{\tau} ] \rangle$.
The sum rule for the charge conductivity then has the form
\cite{br} $\int^{\infty}_{0} d  \omega  Re \sigma_{\tau,\tau}
(\omega) = - \frac{i \pi V}{2} \langle [ \hat q_{\tau}, \hat
j_{\tau} ] \rangle$,
where $\hat q_{\tau} = \hat {\mathbf{q}}\cdot \mathbf{x}_{\tau}/|\mathbf{x}%
_{\tau}|$ and $\hat j_{\tau} = \hat {\mathbf{j}}\cdot \mathbf{x}_{\tau}/|%
\mathbf{x}_{\tau}|$ represent the $\tau$ components of $\hat
{\mathbf{q}}$ and $\hat {\mathbf{j}}$, respectively; here
$\mathbf{x}_{\tau}$ are the primitive vectors of the unit cell.
For the Hamiltonian under investigation one finds
\begin{eqnarray}
\int^{\infty}_{0} d  \omega  Re  \sigma_{\tau,\tau} (\omega) = -
\frac{\pi e2}{2 \hbar V} \sum_{\mathbf{r}} \frac{(\mathbf{r}\cdot
\mathbf{x}_{\tau})2}{ |\mathbf{x}_{\tau}|^2} \langle \hat
H_{itin}(\mathbf{r}) \rangle  . \label{B7}
\end{eqnarray}


\section{Localized solution for the non-cubic case}


In this appendix the matching conditions (\ref{Eq36}) from section
III.B are solved for localized ground states in the case of
non-cubic systems. We start from the observation that the
coefficients $a_{n,d}$ with $n \geq 5$
 can be expressed by the unit cell diagonal hopping amplitudes
and $a_{n,d}$ with $n < 5$ as
$a_{5,d}= - t^d_{\mathbf{x}_3 - \mathbf{x}_2 - \mathbf{x}%
_1}/a^{*}_{3,d}$, $a_{6,d}= - t^d_{\mathbf{x}_3 - \mathbf{x}_2 + \mathbf{x}%
_1}/a^{*}_{4,d}$, $a_{7,d}= - t^d_{\mathbf{x}_3 + \mathbf{x}_2 + \mathbf{x}%
_1}/a^{*}_{1,d}$, $a_{8,d}= - t^d_{\mathbf{x}_3 + \mathbf{x}_2 - \mathbf{x}%
_1}/a^{*}_{2,d}$. With these relations and using the notations
\begin{eqnarray}
&&\tau_1 \equiv \frac{ 4 t^d_{\mathbf{x}_3 +\mathbf{x}_2 + \mathbf{x}_1} t^d_{%
\mathbf{x}_3 - \mathbf{x}_2- \mathbf{x}_1}}{{t^d_{\mathbf{x}_2 + \mathbf{x}%
_1}}^2}, \quad \tau_2 \equiv \frac{4 t^d_{\mathbf{x}_3 +\mathbf{x}_2 + \mathbf{x}%
_1} t^d_{\mathbf{x}_3 + \mathbf{x}_2-
\mathbf{x}_1}}{{t^d_{\mathbf{x}_3 + \mathbf{x}_2}}^2}, \quad
\tau_3 \equiv \frac{4 t^d_{\mathbf{x}_3 +\mathbf{x}_2 +
\mathbf{x}_1} t^d_{\mathbf{x}_3 - \mathbf{x}_2+ \mathbf{x}_1}}{{t^d_{\mathbf{%
x}_3 + \mathbf{x}_1}}^2},  \nonumber \\
&&\tau_4 \equiv \frac{4 t^d_{\mathbf{x}_3 -\mathbf{x}_2 - \mathbf{x}_1} t^d_{%
\mathbf{x}_3 + \mathbf{x}_2- \mathbf{x}_1}}{{t^d_{\mathbf{x}_3 - \mathbf{x}%
_1}}^2}, \quad \tau_5 \equiv \frac{4 t^d_{\mathbf{x}_3 -\mathbf{x}_2 - \mathbf{x}%
_1} t^d_{\mathbf{x}_3 - \mathbf{x}_2+
\mathbf{x}_1}}{{t^d_{\mathbf{x}_3 - \mathbf{x}_2}}^2}, \quad
\tau_6 \equiv \frac{4 t^d_{\mathbf{x}_3 -\mathbf{x}_2 +
\mathbf{x}_1} t^d_{\mathbf{x}_3 + \mathbf{x}_2- \mathbf{x}_1}}{{t^d_{\mathbf{%
x}_2 - \mathbf{x}_1}}^2},  \label{C7}
\end{eqnarray}
(\ref{Eq36}) leads to
\begin{eqnarray}
&&|a_{1,d}|^2 = - \frac{t^d_{\mathbf{x}_2+\mathbf{x}_1} t^d_{\mathbf{x}_3+%
\mathbf{x}_2}}{ 2 t^d_{\mathbf{x}_3-\mathbf{x}_1} \tau_4} F(\tau_1)F(%
\tau_2)F(\tau_4), \quad |a_{2,d}|^2 = - \frac{t^d_{\mathbf{x}_2-\mathbf{x}%
_1} t^d_{\mathbf{x}_3+\mathbf{x}_2} \tau_2}{2 t^d_{\mathbf{x}_3+\mathbf{x}_1}%
} \frac{F(\tau_3)F(\tau_6)}{ \tau_3 F(\tau_2)} ,  \nonumber \\
&&|a_{3,d}|^2 = - \frac{t^d_{\mathbf{x}_2 + \mathbf{x}_1}
t^d_{\mathbf{x}_3 - \mathbf{x}_1}}{ 2 t^d_{\mathbf{x}_3 +
\mathbf{x}_2}} \frac{ \tau_4
F(\tau_1)}{F(\tau_2) F(\tau_4)} , \quad |a_{4,d}|^2 = - \frac{ t^d_{\mathbf{x%
}_2 - \mathbf{x}_1} t^d_{\mathbf{x}_3 + \mathbf{x}_1} \tau_3}{ 2 t^d_{%
\mathbf{x}_3 + \mathbf{x}_2}} \frac{F(\tau_2)F(\tau_6)}{\tau_2
F(\tau_3)} , \label{C8}
\end{eqnarray}
where $F(x)=1 \pm \sqrt{1-x}$. The function required for the
expression of $E_g$ becomes
\begin{eqnarray}
\sum_{n=1}^8 |a_{n,d}|^2 &=& (|a_{1,d}|^2 +
\frac{{t^d}^2_{\mathbf{x}_3 + \mathbf{x}_2 +
\mathbf{x}_1}}{|a_{1,d}|^2}) + (|a_{2,d}|^2 + \frac{{t^d}^2_{
\mathbf{x}_3 + \mathbf{x}_2 - \mathbf{x}_1}}{|a_{2,d}|^2})  \nonumber \\
&+& (|a_{3,d}|^2 + \frac{{t^d}^2_{\mathbf{x}_3 - \mathbf{x}_2 - \mathbf{x}_1}%
}{ |a_{3,d}|^2}) + (|a_{4,d}|^2 + \frac{{t^d}^2_{\mathbf{x}_3 -
\mathbf{x}_2 + \mathbf{x}_1}}{|a_{4,d}|^2}) .  \label{C9}
\end{eqnarray}
For the localized solution to be stable the quantities
$|a_{n,d}|^2$ in (\ref{C8}), and hence the function $F(\tau_i)$,
(\ref{C9}), need to be real.
When $F(\tau_i)$ becomes complex the localized solution becomes
unstable. At these points the derivative $\partial F(x)/\partial
x$ diverges. Except for accidental cancellations in (\ref{C9}),
this implies an infinite slope of $E_g$ as a function of hopping
amplitudes.


\section{Sector of minimal spin of the itinerant solution}


Here we present details of the calculation of spin expectation
values in the itinerant case for minimum total spin (see
Sec.IV.E.2).

In (\ref{Eq60}) the terms
\begin{eqnarray}
[\prod_{\mathbf{i \in
D_{\sigma}}}^{N_{\Lambda}/2}(\sum_{\mathbf{k}}^{N_{ \Lambda}} e^{i
\mathbf{k} \mathbf{i}} \hat f^{\dagger}_{\mathbf{k} \sigma})] =
\sum_{ \{ \mathbf{k}_i \} } \beta_{ \{ \mathbf{k}_i \},\sigma}
\hat F^{\dagger}_{f, \{ \mathbf{k}_i\},\sigma} \: , \quad \beta_{
\{ \mathbf{k}_i \},\sigma}= \sum_{P_{ \{ \mathbf{k}_i \} }}
(-1)^{\bar p} e^{i (\mathbf{i}_1
\mathbf{k}_{i_1} + ... + \mathbf{i}_{ N_{\Lambda}/2} \mathbf{k}%
_{i_{N_{\Lambda}/2}})} ,  \label{E1}
\end{eqnarray}
appear where $\: \hat F^{\dagger}_{f,\{ \mathbf{k}_i\},\sigma }= (
\hat
f^{\dagger}_{ \mathbf{k}_{i_1} \sigma}\hat f^{\dagger}_{\mathbf{k}%
_{i_2} \sigma}... \hat f^{
\dagger}_{\mathbf{k}_{i_{N_{\Lambda}/2}} \sigma})$. Here the sum
over $\{ \mathbf{k}_i \}$ extends over all sets of momentum with
$N_{\Lambda}/2$ elements chosen from the first Brillouin zone, the
sum $\sum_{P_{ \{ \mathbf{k}_i \} }}$ goes over all permutations
$P_{ \{\mathbf{k}_i \}}$ of the momenta in each set $\{ \mathbf{k}_i \} = (\mathbf{k}%
_1,... , \mathbf{k}_{N_{ \Lambda}/2})$, and $\bar p$ represents
the
parity of $P_{ \{ \mathbf{k}_i \} }$. Since for $\mathbf{i}%
_{n,\sigma} \in {\cal D}_{\sigma}$ we have $\mathbf{i}_{n,\downarrow}= \mathbf{i}%
_{n,\uparrow} + \mathbf{R}$, one finds $\beta_{\{ \mathbf{k}_i \},
\downarrow}= e^{i\phi_{\{ \mathbf{k}_i \}}} \beta_{\{ \mathbf{k}_i
\},\uparrow}$, where $\phi_{\{ \mathbf{k}_i \}}= \mathbf{R}
\sum_{\mathbf{k} \in \{ \mathbf{k}_i \}} \mathbf{k}$. Thus,
(\ref{Eq60}) becomes
\begin{eqnarray}
|\Psi_g\rangle = \sum_{\{ \mathbf{k}_i \}} \sum_{\{ \mathbf{k}_j
\}} \alpha_{ \{ \mathbf{k}_i \}, \{ \mathbf{k}_j \}} |v_{ \{
\mathbf{k}_i \}, \{ \mathbf{k}_j \}}\rangle \: , \quad |v_{ \{
\mathbf{k}_i \}, \{ \mathbf{k}_j
\}} \rangle = [ \prod_{\mathbf{k}}^{ N_{\Lambda}} \hat A^{\dagger}_{\mathbf{k%
},\uparrow} \hat A^{\dagger}_{\mathbf{k}, \downarrow} ] \hat
F^{\dagger}_{f,\{ \mathbf{k}_i\},\uparrow } \hat F^{\dagger}_{ f,\{ \mathbf{k%
}_i\},\downarrow } | 0 \rangle ,  \label{E2}
\end{eqnarray}
where $\alpha_{ \{ \mathbf{k}_i \}, \{ \mathbf{k}_j \}}= \beta_{\{ \mathbf{k}%
_i \}, \uparrow} \beta_{\{ \mathbf{k}_i \},\downarrow}$,
$|\alpha_{ \{ \mathbf{k}_i \}, \{ \mathbf{k}_j \}}|= |\alpha_{ \{
\mathbf{k}_j \}, \{ \mathbf{k}_i \}}|$, and $|v_{ \{ \mathbf{k}_i
\}, \{ \mathbf{k}_j \}} \rangle$ are orthogonal states. Using
(\ref{E2}) one finds\cite{sspin}
\begin{eqnarray}
\langle \hat S^x \rangle = \langle \sum_{\mathbf{k}} \hat
S^x_{\mathbf{k}} \rangle = 0  , \quad \langle \hat S^y \rangle =
\langle \sum_{\mathbf{k}} \hat S^y_{\mathbf{k}} \rangle = 0  ,
\quad
\langle \hat S^z \rangle = \langle \sum_{\mathbf{k}}^{N_{\Lambda}} \hat S^z_{%
\mathbf{k}} \rangle = 0 , \quad \langle \hat {\mathbf{S}} \rangle
= 0 . \label{E5}
\end{eqnarray}
To calculate expectation values of the square of the spin we use
normalized wave functions $ |w_{ \{ \mathbf{k}_j \}, \{
\mathbf{k}_i \}} \rangle = |v_{ \{ \mathbf{k}_j \}, \{
\mathbf{k}_i \}} \rangle / ( \langle v_{ \{ \mathbf{k}_j \}, \{
\mathbf{k}_i \}} | v_{ \{ \mathbf{k}_j \}, \{ \mathbf{k}_i \}}
\rangle)^{1/2} $, in terms of which the ground state can be
written as $|\Psi_g\rangle = \sum_{\{ \mathbf{k}_i\}} \sum_{\{
\mathbf{k}_j \}} \alpha^{\prime}_{\{
\mathbf{k}_i \}, \{ \mathbf{k}_j \}} |w_{ \{ \mathbf{k}_i \}, \{ \mathbf{k}%
_j \}}\rangle$, where $\alpha^{\prime}_{\{ \mathbf{k}_i \}, \{
\mathbf{k}_j \}}$ are new numerical coefficients. Thereby one
finds
\begin{eqnarray}
\langle (\hat S^z)2 \rangle = \frac{1}{4} \langle (\hat S^{+} \hat
S^{-} + \hat S^{-} \hat S^{+}) \rangle = \frac{ \sum_{\{
\mathbf{k}_i\}}
\sum_{\{ \mathbf{k}_j \}} |\alpha^{\prime}_{\{ \mathbf{k}_i \}, \{ \mathbf{k}%
_j \}}|^2 (N_{\Lambda} - d_{\{ \mathbf{k}_i \}, \{ \mathbf{k}_j
\}})}{ 4 \sum_{\{ \mathbf{k}_i\}} \sum_{\{ \mathbf{k}_j \}}
|\alpha^{\prime}_{\{ \mathbf{k}_i \}, \{ \mathbf{k}_j \}}|^2} <
\frac{ N_{\Lambda}}{4} , \label{E6}
\end{eqnarray}
where $d_{\{ \mathbf{k}_i \}, \{ \mathbf{k}_j \}}$ is the number
of common elements of the sets $\{ {\bf k}_i \}$ and $\{ {\bf k}_j
\}$.

\section{Expectation values for the flat band}

In this appendix we present details of the calculation of ground
state expectation values
for the itinerant solution in Sec.IV. The ground state wave vector
(\ref{Eq68}) is a superposition of states
\begin{eqnarray}
|\Psi_{g, \{ \sigma \} } \rangle = [
\prod_{\mathbf{k}}^{N_{\Lambda}} \hat
C^{ \dagger}_{1,\mathbf{k} \uparrow} \hat C^{\dagger}_{1,\mathbf{k}%
\downarrow} ] [\prod_{n=1}^{N_{\Lambda}} ( \frac{1}{\sqrt{N_{\Lambda}}} \sum_{%
\mathbf{k}_n}^{N_{ \Lambda}} X_{\mathbf{k}_n} e^{i \mathbf{k}_n
\mathbf{i}_n} \hat C^{\dagger}_{2,\mathbf{k}_n
\sigma_{\mathbf{i}_n}})]|0\rangle,  \label{D1}
\end{eqnarray}
where $\{ \sigma \} = (\sigma_{\mathbf{i}_1},
\sigma_{ \mathbf{i}_2}, \sigma_{\mathbf{i}_3},..., \sigma_{\mathbf{i}%
_{N_{\Lambda}}})$. By modifying the $\{ \sigma \}$ sets in
(\ref{D1}) one obtains $2^{N_{\Lambda}}$ states, which obey
$\langle \Psi_{g,\{ \sigma \}} | \Psi_{g,\{ \sigma^{\prime}\} }\rangle =%
Det[ x_{\mathbf{i},\mathbf{j}}(\{X^{*}_{\mathbf{k}}\},\{X_{\mathbf{k}%
}\},\sigma_{\mathbf{i}}, \sigma^{\prime}_{\mathbf{j}})]$,
where $x_{\mathbf{i},\mathbf{j}}(\{V^{*}_{\mathbf{k}}\},\{W_{\mathbf{k}%
}\},\sigma_{\mathbf{i}}, \sigma^{\prime}_{\mathbf{j}}) = \delta_{ \sigma_{%
\mathbf{i}},\sigma^{\prime}_{\mathbf{j}}} (1/N_{\Lambda})\sum_{\mathbf{k}%
}^{N_{\Lambda}} V^{*}_{\mathbf{k}} W_{\mathbf{k}} e^{i \mathbf{k}(\mathbf{j}-%
\mathbf{i})}$. We see that for $\{\sigma^{\prime}\} \ne
\{\sigma\}$ the states
$| \Psi_{g,\{ \sigma\} }\rangle$, and $| \Psi_{g,\{ \sigma^{\prime}\} }%
\rangle$ are not necessarily orthogonal.


\subsection{The case $X_k \neq 0$}

We first consider $X_{\mathbf{k}} \ne 0$ for all $\mathbf{k}$.
Introducing the states
\begin{eqnarray}
|\Psi_{g}^{ \{ \sigma \} } \rangle = [
\prod_{\mathbf{k}}^{N_{\Lambda}} \hat
C^{ \dagger}_{1,\mathbf{k} \uparrow} \hat C^{\dagger}_{1,\mathbf{k}%
 \downarrow} ] [\prod_{n=1}^{N_{\Lambda}} (\frac{1}{\sqrt{N_{\Lambda}}} \sum_{%
\mathbf{k}_n}^{N_{ \Lambda}} \frac{1}{X^{*}_{\mathbf{k}_n}} e^{i
\mathbf{k}_n \mathbf{i}_n} \hat C^{\dagger}_{2,\mathbf{k}_n
\sigma_{\mathbf{i}_n}})]|0\rangle ,  \label{D4}
\end{eqnarray}
one finds $\langle
\Psi_{g}^{\{ \sigma \}} | \Psi_{g,\{ \sigma^{\prime}\} } \rangle = Det[ z_{%
\mathbf{i},\mathbf{j}}(\sigma_{\mathbf{i}},\sigma^{\prime}_{\mathbf{j}})]$,
where $z_{\mathbf{i},\mathbf{j}}(\sigma_{\mathbf{i}},\sigma^{\prime}_{%
\mathbf{j}}) = x_{\mathbf{i},\mathbf{j}}(\{1/X_{\mathbf{k}}\},\{X_{\mathbf{k}%
}\},\sigma_{\mathbf{i}}, \sigma^{\prime}_{\mathbf{j}}) =
\delta_{\sigma_{\mathbf{i}},\sigma^{\prime}_{\mathbf{i}}} \delta_{ \mathbf{i}%
,\mathbf{j}}$, such that $Det[z_{\mathbf{i},\mathbf{j}}(\sigma_{\mathbf{i%
}},\sigma^{\prime}_{\mathbf{j}})] = \delta_{ \{ \sigma \}, \{
\sigma^{\prime}\} }$. This yields $\langle \Psi_{g}^{\{ \sigma \}}
| \Psi_{g,\{ \sigma^{\prime}\} } \rangle = \delta_{ \{ \sigma \},
\{ \sigma^{\prime}\} }$, i.e., the set $\{| \Psi_{g,\{ \sigma\} }
\rangle\}$ is linearly independent
and for $N=3N_{\Lambda}$ provides a basis for ${\mathcal{H}}_g$
(see (\ref{Eq17}%
)). An arbitrary state of the form of (\ref{Eq68}) can then be
written as $|\Psi_g\rangle = \sum_{ \{ \sigma \} } \alpha_{\{
\sigma \} } |\Psi_{g,\{ \sigma \}} \rangle$, where $\alpha_{\{
\sigma \} }$ are numerical coefficients \cite{tiot}. Furthermore,
one finds
\begin{eqnarray}
&&\hat C^{\dagger}_{2,\mathbf{k}_1 \sigma_1} \hat C_{2,\mathbf{k}%
_2 \sigma_2}| \Psi_{g, \{ \sigma^{\prime}\} } \rangle =
\sum_{n=1}^{N_{\Lambda}} |\Psi_{g,\{\sigma^{\prime}\},n}^{(\mathbf{k}%
_1 \sigma_1),(\mathbf{k}_2 \sigma_2)} \rangle , \quad
|\Psi_{g,\{\sigma^{\prime}\},n}^{(\mathbf{k}_1 \sigma_1),(\mathbf{k}%
_2 \sigma_2)} \rangle = [
\prod_{\mathbf{k}^{\prime}}^{N_{\Lambda}} \hat
C^{\dagger}_{1,\mathbf{k}^{\prime} \uparrow} \hat C^{\dagger}_{1,\mathbf{k}%
^{\prime} \downarrow} ] \times  \nonumber \\
&&[ (\frac{1}{\sqrt{N_{\Lambda}}}
\sum_{\mathbf{k}^{\prime}}^{N_{\Lambda}} X_{\mathbf{k}^{\prime}}
e^{ i \mathbf{k}^{\prime}\mathbf{i}_1} \hat
C^{\dagger}_{2,\mathbf{k}^{\prime} \sigma^{\prime}_{\mathbf{i}_1}}) .... (%
\frac{1}{\sqrt{N_{\Lambda}}} \sum_{\mathbf{k}^{\prime}}^{N_{\Lambda}} X_{%
\mathbf{k}^{\prime}} e^{ i \mathbf{k}^{\prime}\mathbf{i}_{n-1}}
\hat
C^{\dagger}_{2,\mathbf{k}^{\prime} \sigma^{\prime}_{\mathbf{i}_{n-1}} })(%
\frac{1}{\sqrt{N_{\Lambda}}}X_{\mathbf{k}_2} e^{ i
\mathbf{k}_2\mathbf{i}_n}
\hat C^{ \dagger}_{2,\mathbf{k}_1 \sigma_1} \delta_{\sigma^{\prime}_{\mathbf{%
i}_n},\sigma_2})  \nonumber \\
&&(\frac{1}{\sqrt{N_{\Lambda}}} \sum_{\mathbf{k}^{\prime}}^{N_{\Lambda}} X_{%
\mathbf{k}^{\prime}} e^{ i \mathbf{k}^{\prime}\mathbf{i}_{n+1}}
\hat
C^{\dagger}_{2,\mathbf{k}^{\prime} \sigma^{\prime}_{\mathbf{i}_{n+1}} })....(%
\frac{1}{\sqrt{N_{\Lambda}}} \sum_{\mathbf{k}^{\prime}}^{N_{\Lambda}} X_{%
\mathbf{k}^{\prime}} e^{i
\mathbf{k}^{\prime}\mathbf{i}_{N_{\Lambda}}} \hat
C^{\dagger}_{2,\mathbf{k}^{\prime} \sigma^{\prime}_{ \mathbf{i}%
_{N_{\Lambda}}}}) ] | 0 \rangle ,  \label{D8}
\end{eqnarray}
from which
\begin{eqnarray}
\langle \Psi_g^{ \{\sigma^{\prime\prime}\}} | \hat C^{\dagger}_{2,\mathbf{k}%
_1 \sigma_1} \hat C_{2,\mathbf{k}_2 \sigma_2}| \Psi_{g, \{
\sigma^{\prime}\}
} \rangle = \sum_{n=1}^{N_{\Lambda}} Det[\bar z_{ \mathbf{i},\mathbf{j}%
}(n,\sigma^{\prime\prime}_{\mathbf{i}},\sigma^{\prime}_{\mathbf{j}})]
 \label{D9}
\end{eqnarray}
follows. Here, the matrix $\bar z_{\mathbf{i},\mathbf{j}}(n,\sigma^{%
\prime\prime}_{%
\mathbf{i}},\sigma^{\prime}_{\mathbf{j}})$
is the same as
$z_{\mathbf{i},\mathbf{j}}(\sigma^{\prime%
\prime}_{\mathbf{i}},\sigma^{\prime}_{\mathbf{j}})$, except for
the matrix elements in the $n$-th column
which are given by
$\bar z_{\mathbf{i},%
\mathbf{n}} = (1/N_{\Lambda})(X_{\mathbf{k}_2}/X_{\mathbf{k}_1}) e^{i (%
\mathbf{k}_2 \mathbf{i}_n- \mathbf{k}_1\mathbf{i})}
\delta_{\sigma^{\prime\prime}_{\mathbf{i}},\sigma_1}\delta_{\sigma^{\prime}_{%
\mathbf{i}_n},\sigma_2}$.
From (\ref{D9}) one finds that
$\langle \Psi_g^{ \{\sigma^{\prime\prime}\}}| \hat C^{\dagger}_{2,\mathbf{k}%
_1 \sigma} \hat C_{2,\mathbf{k}_2 -\sigma}| \Psi_{g, \{%
\sigma^{\prime}\} } \rangle$ vanishes in the thermodynamic limit
as $1/N_{\Lambda}$.  In the case of
$\sigma_1=\sigma_2=\sigma$, only $\{\sigma^{\prime\prime}\}=\{%
\sigma^{\prime}\}$ components remain in (\ref{D9}) yielding
\begin{eqnarray}
\langle \Psi_g^{ \{\sigma^{\prime\prime}\}}| \hat C^{\dagger}_{2,\mathbf{k}%
_1 \sigma} \hat C_{2,\mathbf{k}_2 \sigma}| \Psi_{g, \{
\sigma^{\prime}\} }
\rangle = \delta_{\{ \sigma^{\prime\prime}\},\{ \sigma^{\prime}\}} \frac{X_{%
\mathbf{k}_2}}{X_{\mathbf{k}_1}} \frac{1}{N_{\Lambda}} \sum_{n=1}^{N_{%
\Lambda}}\delta_{\sigma,\sigma^{\prime}_{\mathbf{i}_n}} e^{i (\mathbf{k}%
_2-\mathbf{k}_1) \mathbf{i}_n} ,  \label{D12}
\end{eqnarray}
where $\sum_{n=1}^{ N_{\Lambda}}\delta_{\sigma,\sigma^{\prime}_{\mathbf{i}%
_n}}$ quantifies the number of $\sigma$ spins from
$\{\sigma^{\prime}\}$. Since $\sum_{ \sigma}
\delta_{\sigma,\sigma^{\prime}_{\mathbf{i}_n}} = 1$ for arbitrary
$\sigma^{\prime}_{ \mathbf{i}_n}$, it follows that $\langle
\Psi_g^{ \{\sigma^{\prime\prime}\}}|\sum_{\sigma} \hat C^{\dagger}_{2,%
\mathbf{k}_1 \sigma} \hat C_{2,\mathbf{k}_2 \sigma}|\Psi_{g, \{
\sigma^{\prime}\} } \rangle = \delta_{\mathbf{k}_2, \mathbf{k}_1}
\delta_{\{ \sigma^{\prime\prime}\},\{ \sigma^{\prime}\}}$. Using
the notation $ \langle ...\rangle = \langle
\Psi_g|...|\Psi_g\rangle/ \langle \Psi_g|\Psi_g\rangle$, one
finally obtains
\begin{eqnarray}
&&\langle \sum_{\sigma} \hat C^{\dagger}_{2,\mathbf{k}_1 \sigma}
\hat C_{2, \mathbf{k}_2 \sigma} \rangle =
\delta_{\mathbf{k}_2,\mathbf{k}_1} . \label{D14}
\end{eqnarray}
The spin dependent expectation value $\langle \hat C^{\dagger}_{2,\mathbf{k}%
\sigma} \hat C_{2,%
\mathbf{k} \sigma} \rangle$
may be calculated, for example,
by taking the $T\to 0$ limit of
$\langle \hat C^{\dagger}_{2,\mathbf{k} \sigma} \hat
C_{2,\mathbf{k} \sigma} \rangle$ as
\begin{eqnarray}
\langle \hat A \rangle = \lim_{T\to 0} \frac{Tr (\hat A e^{-\beta
\hat H})}{
Tr (e^{-\beta \hat H})} = \lim_{T\to 0} \frac{Tr (\hat A e^{-\beta \hat H_g})%
}{Tr(e^{-\beta \hat H_g})} = \frac{\sum_{\{ \sigma \} } \langle
\Psi_{g}^{ \{ \sigma \} }|\hat A |\Psi_{g, \{ \sigma \} }\rangle}{
\sum_{\{ \sigma \} } \langle \Psi_{g}^{\{ \sigma \} } |\Psi_{g,\{
\sigma \} }\rangle} , \label{D16}
\end{eqnarray}
where $\hat A$ is an arbitrary operator and $\hat H_g$ is defined
in (\ref{Eq55}). The second equality in (\ref{D16}) holds since
for $N_\Lambda<\infty$ excited states of $\hat H$ are always
separated by a finite energy from the ground state and thus give
only exponentially small corrections to the contribution of $\hat
H_g$. Eq.(\ref{D16}) therefore yields
\begin{eqnarray}
\langle \hat C^{\dagger}_{2,\mathbf{k}_1 \sigma_1} \hat C_{2,\mathbf{k}%
_2 \sigma_2} \rangle = \frac{1}{2}
\delta_{\mathbf{k}_2,\mathbf{k}_1} \delta_{\sigma_2,\sigma_1}  .
\label{D18}
\end{eqnarray}


\subsection{The case $X_k=0$}

Here we consider the case  $a_{\mathbf{k}%
^{*},d}=0$ for a given vector $\mathbf{k}=\mathbf{k}^{*}$ which
implies $X_{\mathbf{k}^{*}}=0$ (we note that the $\hat
C_{\delta,\mathbf{k} \sigma}$ operators are then still
well-defined and $R^{-1}_{\mathbf{k}^{*}} \ne 0$), with $X_{%
\mathbf{k} \ne \mathbf{k}^{*}} \ne 0$. Such a situation arises,
for example, in the case of the itinerant solution (see Sec.IV.B)
for large next-nearest neighbor hopping amplitude $|t^d_1/t^d_2|
\leq 3$. Except for $\mathbf{k}^{*}$ (\ref{D14},\ref{D18}) remain
valid the thermodynamic limit, and the ground state expectation
values of the momentum occupation becomes
\begin{eqnarray}
\langle \hat C^{\dagger}_{1,\mathbf{k} \sigma} \hat
C_{1,\mathbf{k} \sigma} \rangle = 1 , \quad \langle \hat
C^{\dagger}_{2,\mathbf{k}^{*} \sigma} \hat C_{2,\mathbf{k}^{*}
\sigma} \rangle = 0 , \quad \langle \hat
C^{\dagger}_{2,\mathbf{k}\ne\mathbf{k}^{*} \sigma} \hat
C_{2,\mathbf{k}\ne \mathbf{k}^{*} \sigma}\rangle = \frac{1}{2} .
\label{D25}
\end{eqnarray}
We note that for one vector ${\mathbf k}^{*}$ the maximum total
spin decreases from $N_{\Lambda}/2$ to $N_{\Lambda}/2-1$ since in
the sum over ${\bf k}_n$ in  (E1, E2) the term with ${\bf k}^{*}$
is missing.

\subsection{Expectation values for correlation functions}

A key problem in the calculation of correlation functions is the
evaluation of expectation values of products of four $\hat C_{2,
\mathbf{k} \sigma}$ operators. Using (\ref{D8}) one finds
\begin{eqnarray}
&&\hat C^{\dagger}_{2,\mathbf{k}_3 \sigma_3} \hat C_{2,\mathbf{k}%
_4 \sigma_4} \hat C^{\dagger}_{2,\mathbf{k}_1 \sigma_1} \hat C_{2,\mathbf{k}%
_2 \sigma_2}| \Psi_{g, \{ \sigma^{\prime}\} } \rangle = \delta_{\mathbf{k}_1,%
\mathbf{k}_4} \delta_{ \sigma_1,\sigma_4}\sum_{n=1}^{N_{\Lambda}}
|\Psi_{g,\{\sigma^{\prime}\},n}^{(\mathbf{k}_3 \sigma_3),(\mathbf{k}%
_2 \sigma_2)} \rangle +  \nonumber \\
&&(1-\delta_{\mathbf{k}_1,\mathbf{k}_3}
\delta_{\sigma_1,\sigma_3}) (1-\delta_{\mathbf{k}_2,\mathbf{k}_4}
\delta_{\sigma_2,\sigma_4}) \sum_{n=1}^{N_{\Lambda}}
\sum_{m=1,m\ne n}^{N_{\Lambda}}
|\Psi_{g,\{\sigma^{\prime}\},m,n}^{[(\mathbf{k}_3 \sigma_3),(\mathbf{k}%
_4 \sigma_4)];[(\mathbf{k}_1 \sigma_1), (\mathbf{k}_2 \sigma_2)]}
\rangle ,  \label{D26}
\end{eqnarray}
where $\mathbf{k}_4
=\mathbf{k}_3+\mathbf{k}_1-\mathbf{k}_2$. Here $|\Psi_{g,\{\sigma^{\prime}\},n}^{(\mathbf{k}_3 \sigma_3),(\mathbf{k}%
_2 \sigma_2)} \rangle$ is defined in (\ref{D8}),  and
\begin{eqnarray}
&&|\Psi_{g,\{\sigma^{\prime}\},m,n}^{[(\mathbf{k}_3 \sigma_3),(\mathbf{k}%
_4 \sigma_4)];[ (\mathbf{k}_1 \sigma_1),(\mathbf{k}_2 \sigma_2)]}
\rangle =
[ \prod_{\mathbf{k}^{\prime}}^{N_{\Lambda}} \hat C^{ \dagger}_{1,\mathbf{k}%
^{\prime} \uparrow} \hat C^{\dagger}_{1,\mathbf{k}^{\prime}
\downarrow} ] \:
[ (\frac{1}{\sqrt{N_{\Lambda}}} \sum_{\mathbf{k}^{\prime}}^{N_{\Lambda}} X_{%
\mathbf{k}^{\prime}} e^{ i \mathbf{k}^{\prime}\mathbf{i}_1} \hat
C^{\dagger}_{2,\mathbf{k}^{\prime}
\sigma^{\prime}_{\mathbf{i}_1}}) ...
\nonumber \\
&&(\frac{1}{\sqrt{N_{\Lambda}}} \sum_{\mathbf{k}^{\prime}}^{N_{\Lambda}} X_{%
\mathbf{k}^{\prime}} e^{ i \mathbf{k}^{\prime}\mathbf{i}_{m-1}}
\hat
C^{\dagger}_{2,\mathbf{k}^{\prime} \sigma^{\prime}_{\mathbf{i}_{m-1}} })(%
\frac{1}{\sqrt{N_{\Lambda}}}X_{\mathbf{k}_4} e^{ i
\mathbf{k}_4\mathbf{i}_m}
\hat C^{ \dagger}_{2,\mathbf{k}_3 \sigma_3} \delta_{\sigma^{\prime}_{\mathbf{%
i}_m},\sigma_4}) (\frac{1}{\sqrt{N_{\Lambda}}} \sum_{\mathbf{k}%
^{\prime}}^{N_{\Lambda}} X_{\mathbf{k}^{\prime}} e^{ i \mathbf{k}^{\prime}%
\mathbf{i}_{m+1}} \hat C^{\dagger}_{2,\mathbf{k}^{\prime} \sigma^{\prime}_{%
\mathbf{i}_{m+1}} })...  \nonumber \\
&&(\frac{1}{\sqrt{N_{\Lambda}}} \sum_{\mathbf{k}^{\prime}}^{N_{\Lambda}} X_{%
\mathbf{k}^{\prime}} e^{ i \mathbf{k}^{\prime}\mathbf{i}_{n-1}}
\hat
C^{\dagger}_{2,\mathbf{k}^{\prime} \sigma^{\prime}_{\mathbf{i}_{n-1}} })(%
\frac{1}{\sqrt{N_{\Lambda}}}X_{\mathbf{k}_2} e^{ i
\mathbf{k}_2\mathbf{i}_n}
\hat C^{ \dagger}_{2,\mathbf{k}_1 \sigma_1} \delta_{\sigma^{\prime}_{\mathbf{%
i}_n},\sigma_2}) (\frac{1}{\sqrt{N_{\Lambda}}} \sum_{\mathbf{k}%
^{\prime}}^{N_{\Lambda}} X_{\mathbf{k}^{\prime}} e^{ i \mathbf{k}^{\prime}%
\mathbf{i}_{n+1}} \hat C^{\dagger}_{2,\mathbf{k}^{\prime} \sigma^{\prime}_{%
\mathbf{i}_{n+1}} })  \nonumber \\
&&....(\frac{1}{\sqrt{N_{\Lambda}}}
\sum_{\mathbf{k}^{\prime}}^{N_{\Lambda}} X_{\mathbf{k}^{\prime}}
e^{i \mathbf{k}^{\prime}\mathbf{i}_{N_{\Lambda}}}
\hat C^{\dagger}_{2,\mathbf{k}^{\prime} \sigma^{\prime}_{ \mathbf{i}%
_{N_{\Lambda}}}}) ] | 0 \rangle   \label{D27}
\end{eqnarray}
contains only contribution with $n\ne m$. The expectation value of
the density--density correlation function can be calculated from
(\ref{D26}), with $\sigma_1=\sigma_2= \sigma_a$,
$\sigma_3=\sigma_4=\sigma_b$, where a summation $\sum_{\sigma_a}
\sum_{\sigma_b}$ has to be included. Introducing the notations
$\bar Z^{\mathbf{k}_4,\mathbf{k}_2}_{\mathbf{k}_3,\mathbf{k}_1}
= Z^{\mathbf{k}_4 \ne \mathbf{k}_2,\mathbf{k}_2}_{\mathbf{k}_3 \ne \mathbf{k}%
_1,\mathbf{k}_1}$, where
$Z^{\mathbf{k}_4,\mathbf{k}_2}_{\mathbf{k}_3,\mathbf{k}_1}= X_{\mathbf{%
k}_4} X_{ \mathbf{k}_2}/(X_{\mathbf{k}_3} X_{\mathbf{k}_1}
N_{\Lambda})$, and using the procedure leading to (\ref{D14}) one
then finds
\begin{eqnarray}
\langle \sum_{\sigma_a,\sigma_b} \hat C^{\dagger}_{2,\mathbf{k}_3
\sigma_b} \hat C_{2,\mathbf{k}_4 \sigma_b} \hat
C^{\dagger}_{2,\mathbf{k}_1 \sigma_a}
\hat C_{2,\mathbf{k}_2 \sigma_a} \rangle = \delta_{\mathbf{k}_1,\mathbf{k}%
_4} \delta_{\mathbf{k}_2,\mathbf{k}_3} +
\delta_{\mathbf{k}_1,\mathbf{k}_2}
\delta_{\mathbf{k}_3,\mathbf{k}_4} - \bar Z^{\mathbf{k}_4,\mathbf{k}_2}_{%
\mathbf{k}_3,\mathbf{k}_1} \delta_{\mathbf{k}_4, \mathbf{k}_3 + \mathbf{k}_1-%
\mathbf{k}_2} .  \label{D31}
\end{eqnarray}
The same expression is obtained if we calculate the expectation
value in the $T\to 0$ limit, as described in connection with
(\ref{D16} - \ref{D18}).

In the case of the spin-spin correlation function for the $S^z$
components one again has $ \sigma_1=\sigma_2 =\sigma_a$ and
$\sigma_3=\sigma_4=\sigma_b$, but the sum over the spin indices
must be separately performed for $ \sigma_b=\sigma_a=\sigma$ and
$\sigma_b=- \sigma_a=\sigma$. Using
(\ref{D16}-\ref{D18},\ref{D26}), one finds in the $T \to 0$ limit
\begin{eqnarray}
&&\langle \sum_{\sigma} \hat C^{\dagger}_{2,\mathbf{k}_3 \sigma} \hat C_{2,%
\mathbf{k}_4 \sigma} \hat C^{\dagger}_{2,\mathbf{k}_1 \sigma} \hat C_{2,%
\mathbf{k}_2 \sigma} \rangle = \delta_{\mathbf{k}_1,\mathbf{k}_4} \delta_{%
\mathbf{k}_2,\mathbf{k}_3} + \frac{1}{2}
\delta_{\mathbf{k}_1,\mathbf{k}_2}
\delta_{\mathbf{k}_3,\mathbf{k}_4} - \frac{1}{2} \bar Z^{\mathbf{k}_4,%
\mathbf{k}_2}_{\mathbf{k}_3,\mathbf{k}_1} \delta_{\mathbf{k}_4,
\mathbf{k}_3
+ \mathbf{k}_1-\mathbf{k}_2} ,  \nonumber \\
&&\langle \sum_{\sigma} \hat C^{\dagger}_{2,\mathbf{k}_3 \sigma} \hat C_{2,%
\mathbf{k}_4 \sigma} \hat C^{\dagger}_{2,\mathbf{k}_1 -\sigma} \hat C_{2,%
\mathbf{k}_2 -\sigma} \rangle = \frac{1}{2}\delta_{\mathbf{k}_1,\mathbf{k}%
_2} \delta_{\mathbf{k}_3,\mathbf{k}_4} - \frac{1}{2} Z^{\mathbf{k}_4,\mathbf{%
k}_2}_{\mathbf{k}_3,\mathbf{k}_1} \delta_{\mathbf{k}_4,
\mathbf{k}_3 + \mathbf{k}_1-\mathbf{k}_2} .  \label{D35}
\end{eqnarray}

To calculate the spin-spin correlation functions for the $S^x$,
$S^y$ components, (\ref{D26}) must be evaluated for $\sigma_3 =
\sigma_2 = -\sigma_1 = -\sigma_4 = \sigma$, and the summation
$\sum_{\sigma}$ must be performed. In the $T \to 0$ limit one
finds
\begin{eqnarray}
&&\langle \sum_{\sigma} \hat C^{\dagger}_{2,\mathbf{k}_3  \sigma} \hat C_{2,%
\mathbf{k}_4 -\sigma} \hat C^{\dagger}_{2,\mathbf{k}_1 -\sigma} \hat C_{2,%
\mathbf{k}_2 \sigma} \rangle = \delta_{\mathbf{k}_1,\mathbf{k}_4} \delta_{%
\mathbf{k}_2,\mathbf{k}_3} ,
\nonumber\\
&&\langle \sum_{\sigma} \hat C^{\dagger}_{1,\mathbf{k}_3 \sigma}
\hat C_{2, \mathbf{k}_4 -\sigma} \hat C^{\dagger}_{2,\mathbf{k}_1
-\sigma} \hat C_{1,
\mathbf{k}_2 \sigma} \rangle \delta_{\mathbf{k}_4,\mathbf{k}_3+\mathbf{k}_1-%
\mathbf{k}_2} = \delta_{\mathbf{k}_2,\mathbf{k}_3} \delta_{\mathbf{k}_1,%
\mathbf{k}_4} .  \label{D39}
\end{eqnarray}



\newpage

\begin{figure}[h]
\centerline{\epsfbox{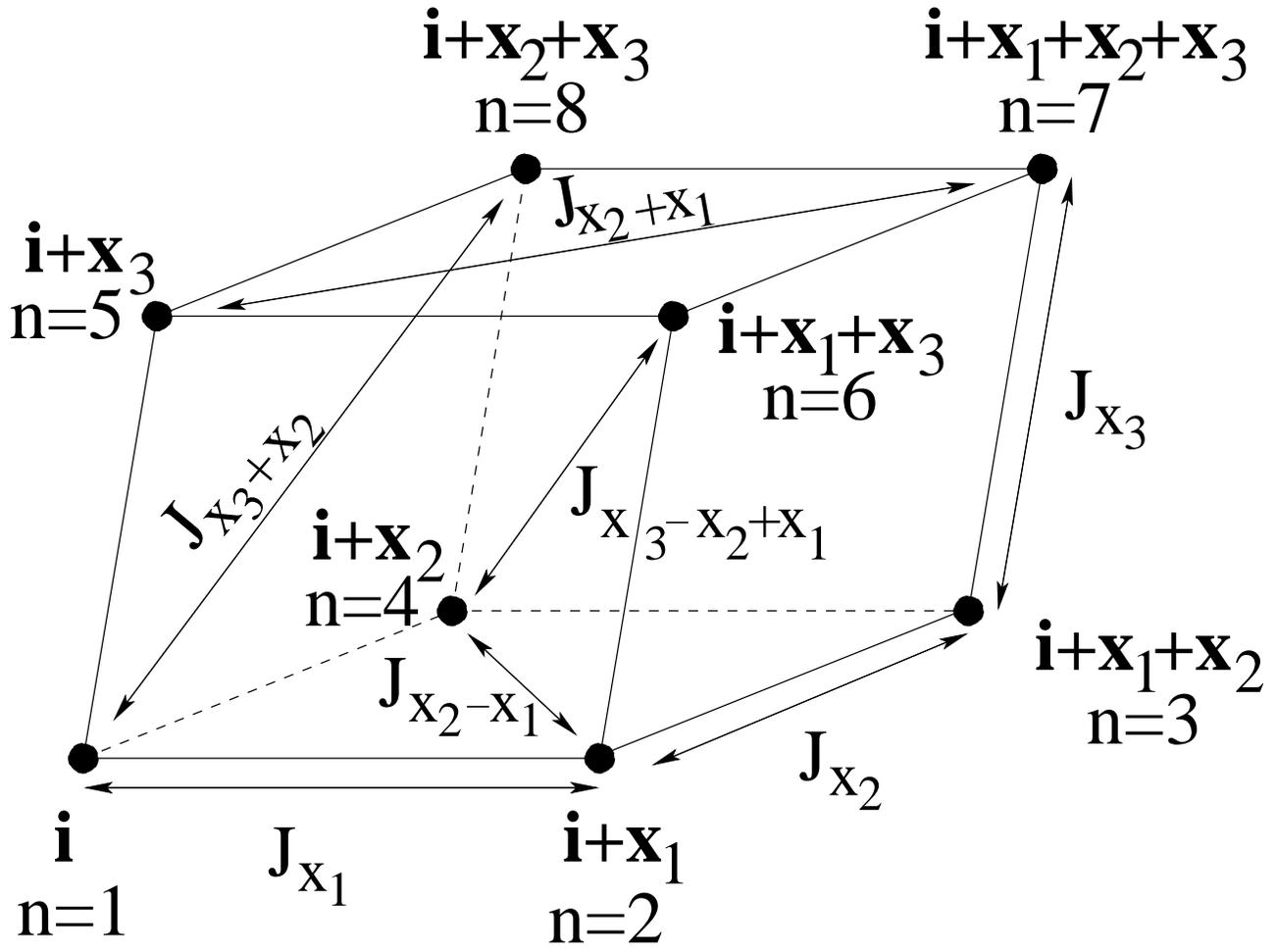}} \caption{ A unit cell
$I_{\mathbf{i}}$ connected to an arbitrary site $\mathbf{i}$
showing the primitive vectors $\mathbf{x}_{\protect\tau}$ and
indices $n$ of the sites
in $I$. Arrows depict some of the hopping and hybridization amplitudes $%
(J=t,V)$ defined within $I_{\mathbf{i}}$.} \label{fig1}
\end{figure}

\newpage

\begin{figure}[h]
\centerline{\epsfbox{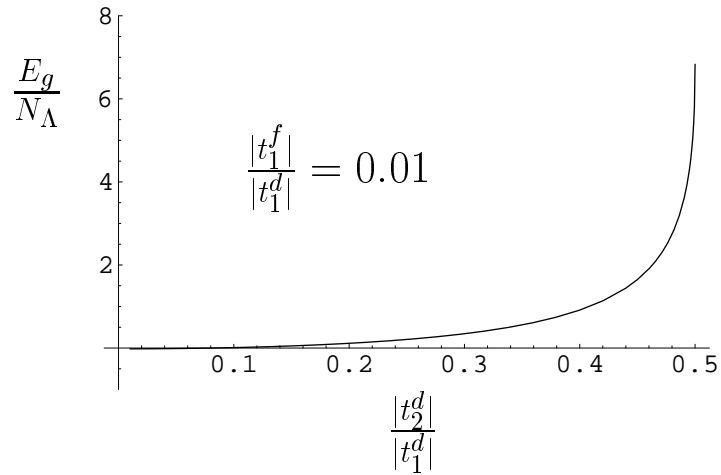}} \caption{Ground state energy
per lattice site,(\protect\ref{Eq38}), in units of $U$ as a
function
of $y = |t^d_2/t^d_1|$ for the localized solution in a simple cubic crystal
for $|t^f_1/t^d_1|=0.01$.
As seen from the plot, $E_g/N_{\Lambda}$ is finite at $y_c=1/2$ but has
infinite slope.} \label{fig2}
\end{figure}

\newpage

\begin{figure}[h]
\centerline{\epsfbox{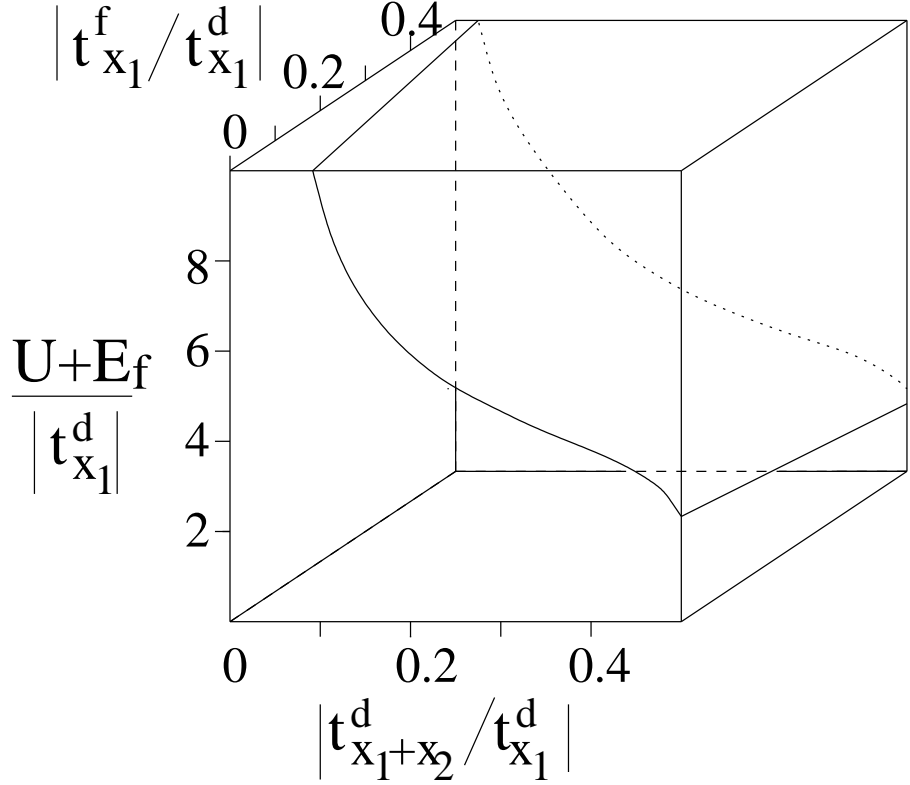}} \caption{Surface in parameter
space representing the stability region of the conducting ground
state discussed in Sec.IV.A.} \label{fig3}
\end{figure}

\begin{figure}[h]
\centerline{\epsfbox{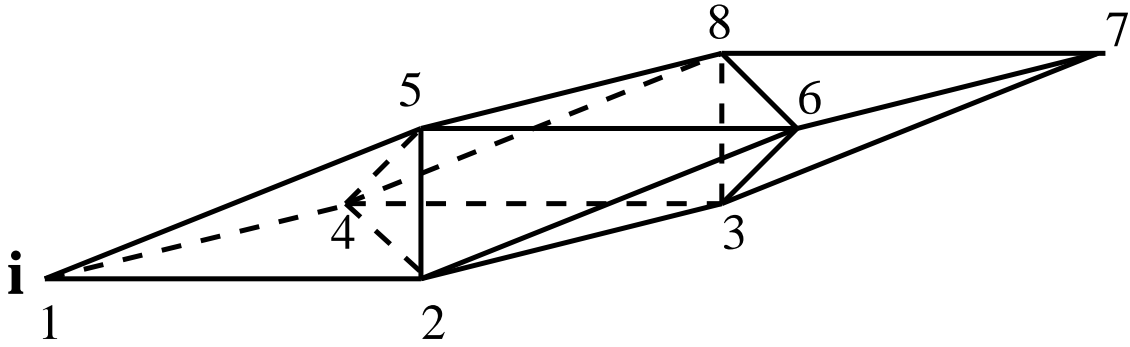}} \caption{Tilted unit cell located
at site $\mathbf{i}$ discussed in Sec.IV.B. Numbers represent the
intra-cell numbering of lattice sites. Only those bonds are
presented along which hopping of $d$ electrons occurs.}
\label{fig4}
\end{figure}

\newpage

\begin{figure}[h]
\centerline{\epsfbox{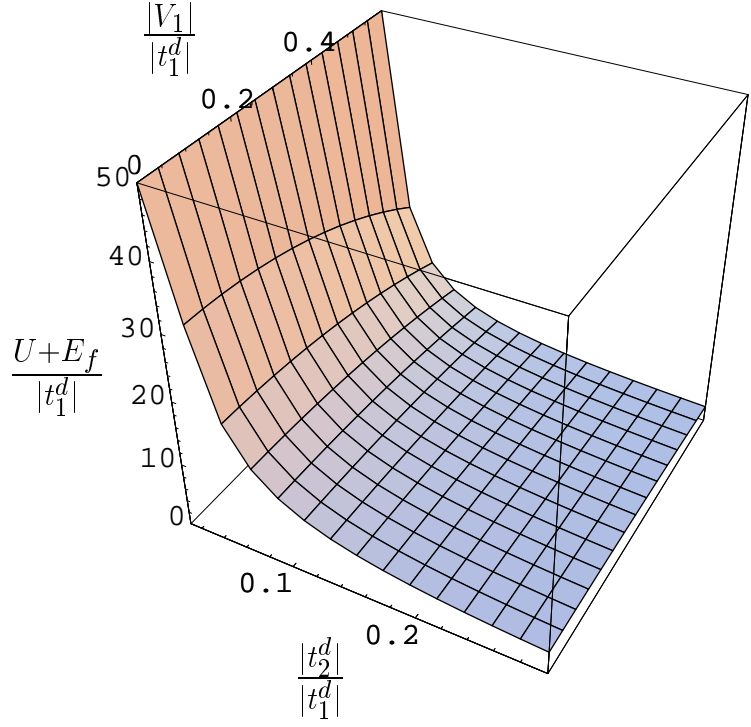}} \caption{Surface in parameter
space representing the stability region of
the itinerant solution derived in the case of the conventional PAM, (%
\protect\ref{Eq49}). For $|t^d_2/t^d_1| \to 0$, the surface
asymptotically approaches the $(|V_1/t^d_1|, (U+E_f)/|t^d_1|)$
plane. } \label{fig5}
\end{figure}

\newpage

\begin{figure}[h]
\centerline{\epsfbox{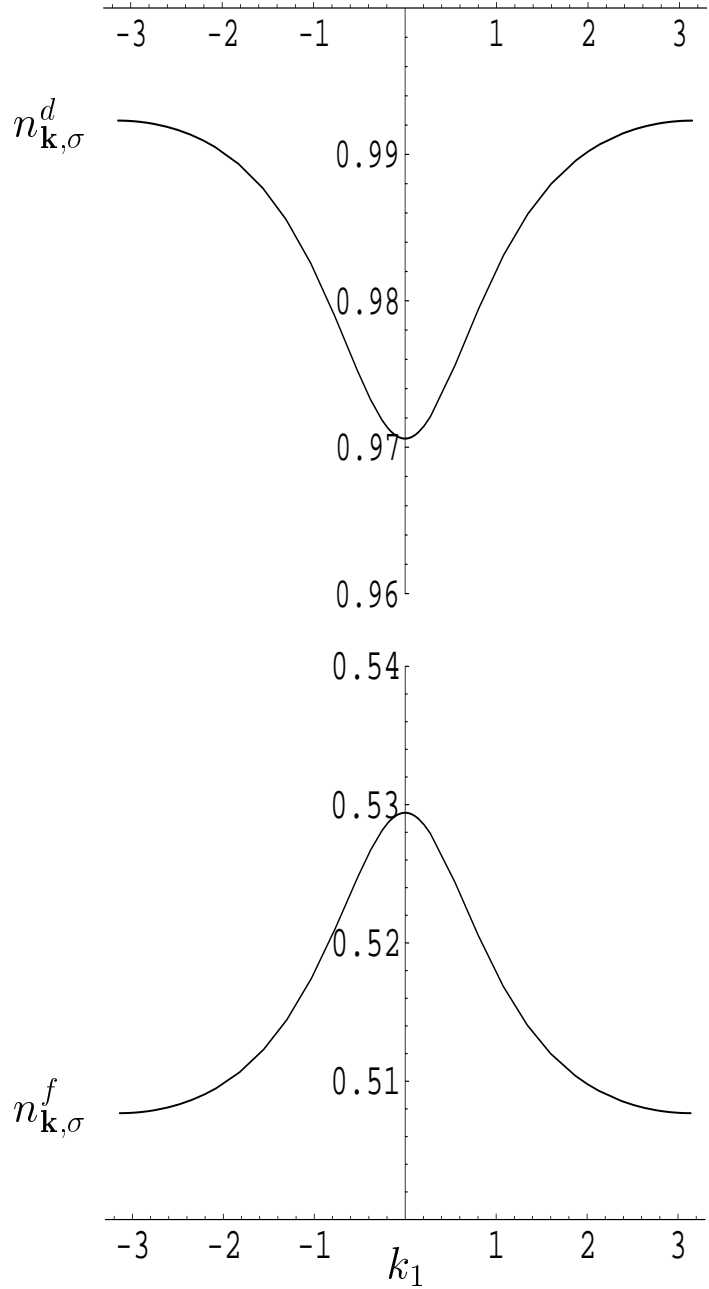}} \caption{Momentum
distribution functions $n^d_{\mathbf{k},\protect\sigma}$,
$n^f_{\mathbf{k},\protect\sigma}$ for the conducting solution
discussed in Section IV.B, for $|t^d_1|/|t^d_2| = 5.0$, $%
|V_1|/|t^d_1|=0.5$, $t^d_1 > 0$, and $k_2=k_3=0$, where $k_{\protect\tau}=%
\mathbf{k} \mathbf{x}_{\protect\tau}$. The plot presents the
behavior in the first Brillouin zone for $k_1 \in
[-\protect\pi,\protect\pi]$. For other $\mathbf{k}$ directions a
similar behavior is found.} \label{fig6}
\end{figure}

\newpage

\begin{figure}[h]
\centerline{\epsfbox{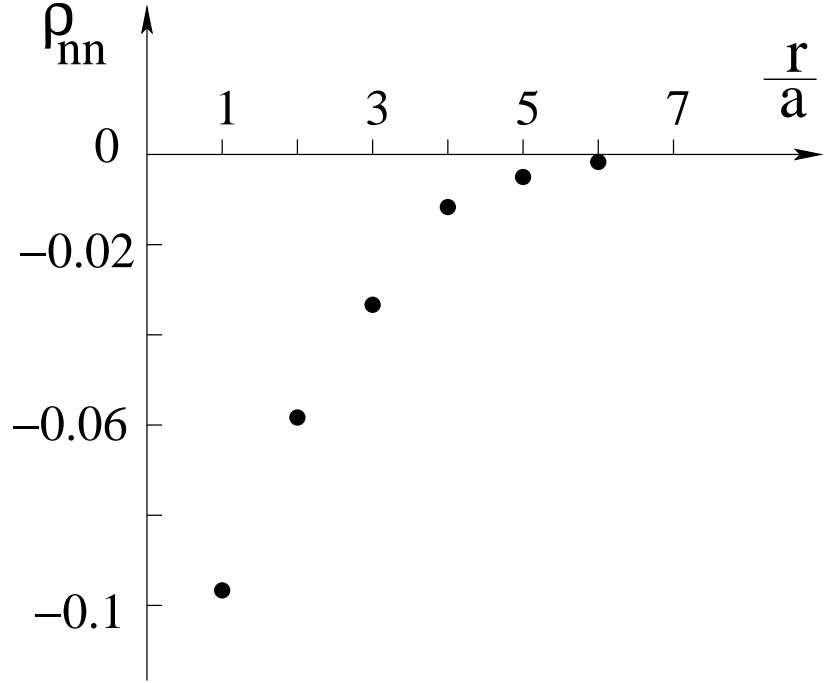}}
\caption{ Density-density correlation function for $|t^d_1/t^d_2|=4,$ $%
|V_1/t^d_1|=0.25$, $t^d_1 > 0$ for the itinerant case obtained
from (\protect\ref{Eq77})  in the thermodynamic limit. The
nine-dimensional integration was performed by a Monte Carlo method
using $69$ points. The distance $\mathbf{r}$ was taken in
$\protect\tau=1$ direction, and is expressed in units of the
lattice constant $a$. } \label{fig7}
\end{figure}

\newpage

\begin{figure}[h]
\centerline{\epsfbox{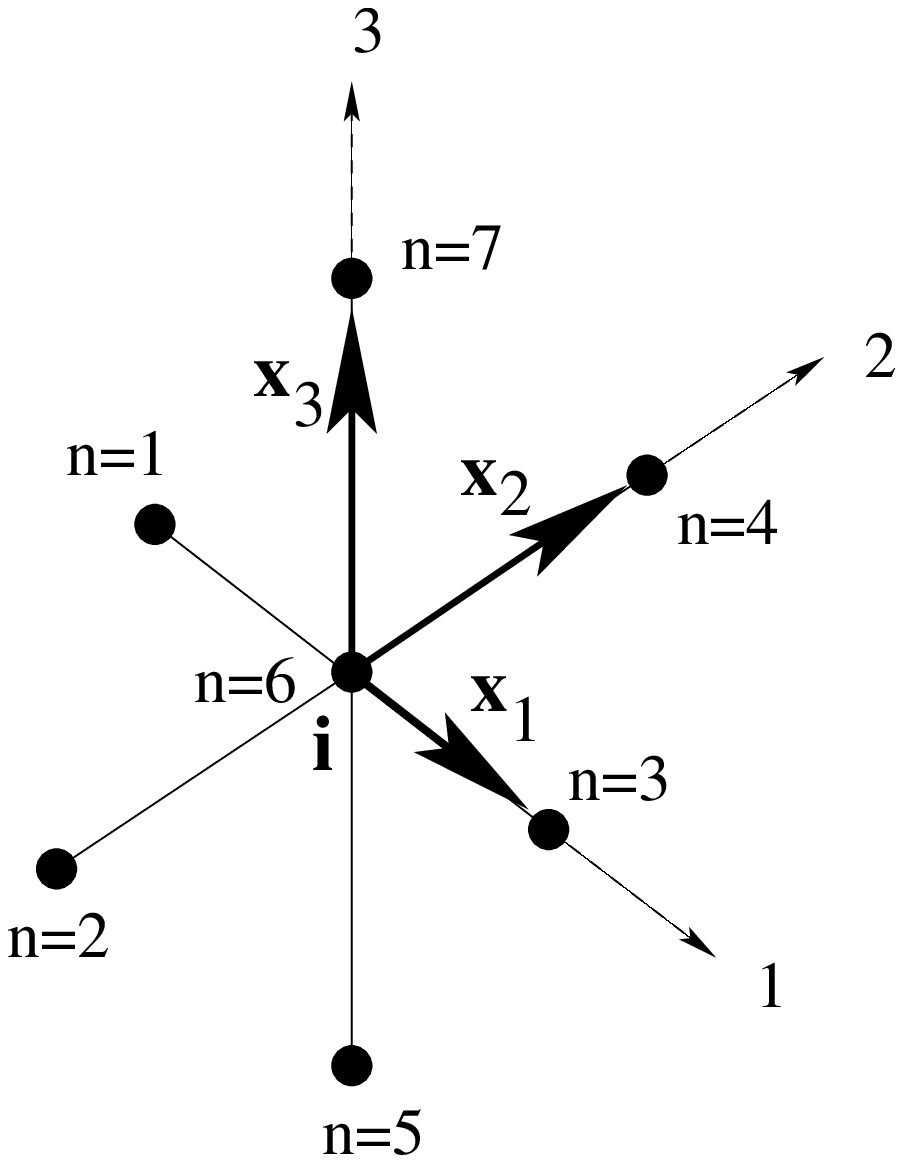}} \caption{Octahedral cell
defined at lattice site $\mathbf{i}$ as discussed in Section V.A.
The $\protect\tau = 1,2,3$ axes are represented by arrows with
broken lines, $\mathbf{x}_{\protect\tau}$ are indicated by thick
full line arrows, and $n$ represents the cell independent notation
of sites inside the octahedron.} \label{fig8}
\end{figure}

\newpage

\begin{figure}[h]
\centerline{\epsfbox{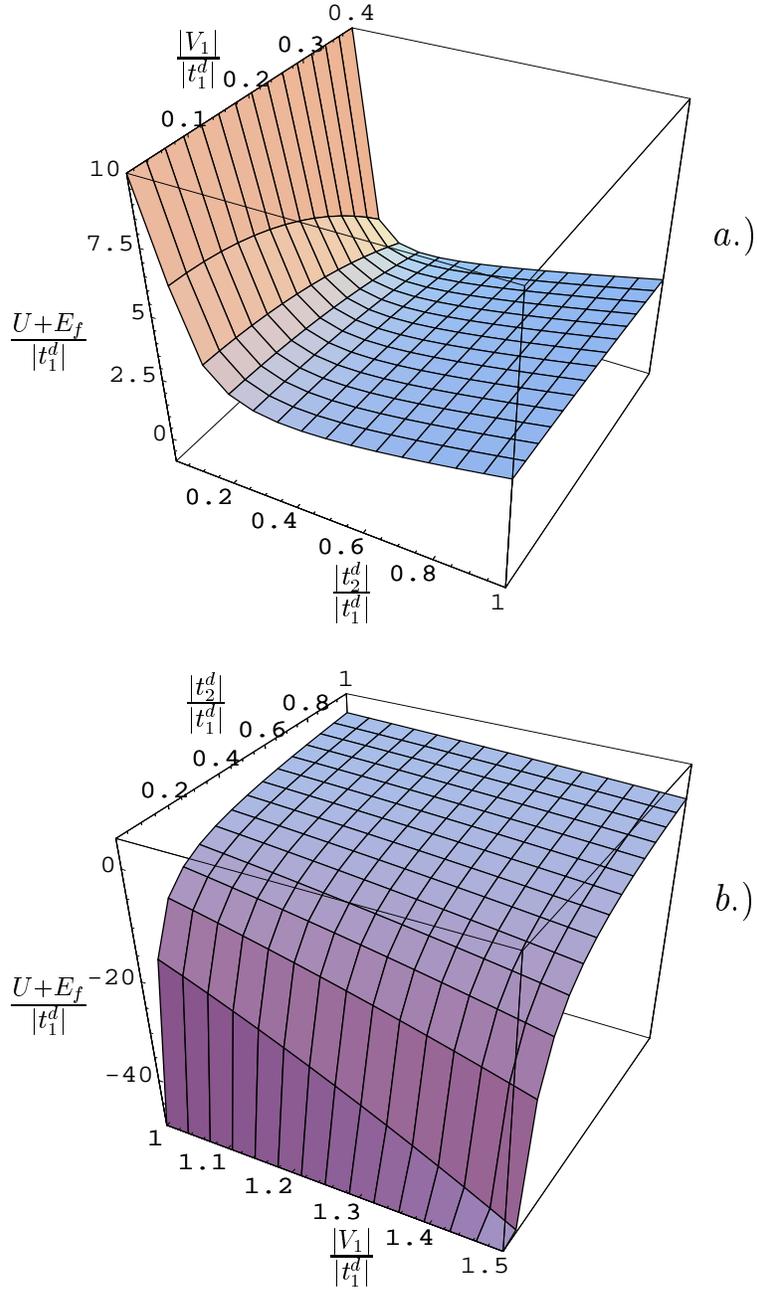}} \caption{Surfaces in
parameter space above which the conducting phase discussed in
Section V.A. is stable. a) $|V_1|/|t^d_1| < 0.5$;
$(U+E_f)/|t^d_1|$ is seen to diverge as $|t^d_2|/|t^d_1|$
approaches zero. b) $|V_1|/|t^d_1|> 0.5$. In contrast to Fig.5
where $V_1$ is imaginary, $V_1$ is real here.} \label{fig9}
\end{figure}

\newpage

\begin{figure}[h]
\centerline{\epsfbox{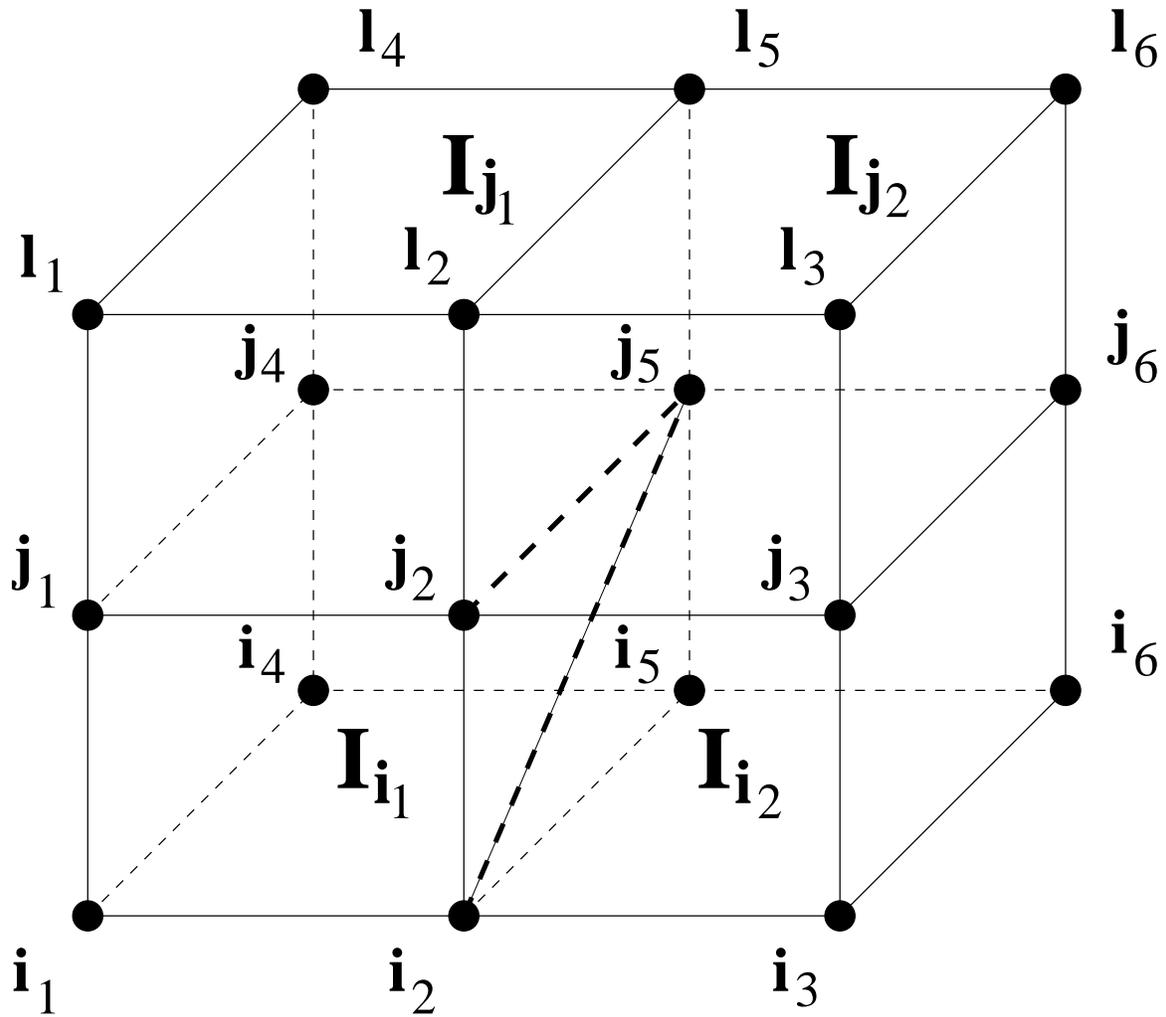}} \caption{Four neighboring unit
cells in $D=3$. Each cell (e.g., $I_{\mathbf{i}}$) is denoted by
one of the lattice sites at which it is located (e.g.,$\mathbf{
i}$); see Sec. II.B.1. For simplicity, orthorhombic unit cells are
used.} \label{fig10}
\end{figure}

\newpage


\begin{thebibliography}{99}
\bibitem{mi4} As will be discussed in Sec.V. this is not the only
possibility to cast the Hamiltonian into positive semi-definite
form.
\bibitem{aa1}
Zero total spectral weight is always possible when hopping and
nonlocal hybridization is  absent, and is not a peculiarity of the
model presented here (see, e.g., \cite{br}).
\bibitem{sspin}
The total spin operator $\hat {\mathbf{S}}$ is defined by
$\hat S^{\alpha} = \sum_{ \mathbf{i}=1}^{N_{\Lambda}} \hat S^{\alpha}_{%
\mathbf{i}}$, $\alpha=x,y,z$, where $\hat S^z_{\mathbf{i}} =
\frac{1}{2}(\hat n^d_{\mathbf{i},\uparrow} + \hat
n^f_{ \mathbf{i},\uparrow} - \hat n^d_{\mathbf{i},\downarrow} - \hat n^f_{%
\mathbf{i}, \downarrow})$, $\hat S^{+}_{\mathbf{i}} = (\hat S^{-}_{%
\mathbf{i}})^{\dagger} = \hat d^{\dagger}_{\mathbf{i},\uparrow} \hat d_{%
\mathbf{i},\downarrow} + \hat f^{\dagger}_{\mathbf{i},\uparrow} \hat f_{%
\mathbf{i},\downarrow}$, $\hat S^{\pm}_{\mathbf{i}} = \hat S^{x}_{%
\mathbf{i}} \pm i \hat S^{y}_{\mathbf{i}}$. By interchanging
$\mathbf{i}$ and $\mathbf{k}$, similar expressions are obtained in
${\mathbf{k}}$ space, where $\hat S^{\alpha} = \sum_{
\mathbf{k}=1}^{N_{\Lambda}} \hat S^{\alpha}_{\mathbf{k}}$.
\bibitem{commentx}
Eq.(\ref{Eq47}) has solutions for real $(V_0V_1)$ with $t^d_2 < 0$ and $(V_0V_1)/t^d_1 < 0$, arbitrary $%
|t^d_1|, |t^d_2|, V_1=|V_1| e^{i\phi_V}$, and $V_0 =
(t^d_1/t^d_2)V^{*}_1$ which have the form
$a_{2,d}=a_{4,d}=a_{5,d}$. In particular, one finds
$a_{1,d}=(|t^d_1|/\sqrt{|t^d_2|}) e^{i\phi}$,
$a_{2,d}=[-sign(t^d_1)] e^{i\phi} \sqrt{|t^d_2|}$,
$a_{1,f}=sign(t^d_1) [|V_1|/ \sqrt{|t^d_2|}] e^{-i\phi_V} e^{i
\phi}$, where $\phi$ is an arbitrary phase. For complex
$V_{\mathbf{k}}$ a solution  of the form $V_{\mathbf{k}}=V_0
+|V_1|e^{i \phi_V}
\sum_{\tau=1}^3 exp(ik_{\tau})$, emerges (see (\ref{Eq4})) where
$k_{\tau} = \mathbf{k} \mathbf{x}%
_{\tau}$. Since $t^d_{\mathbf{x}_{\tau}+\mathbf{x}_{
\tau^{\prime}}}=0$ while
$t^d_{\mathbf{x}_{\tau}-\mathbf{x}_{\tau^{\prime}}} \ne 0$, $\tau
> \tau^{\prime}$, this requires anisotropic hopping parameters.
\bibitem{commx}
For the operators $\hat C_{\delta,\mathbf{k},\sigma}$ to be
well-defined, $R_{\mathbf{k}}^{-1}$ has to be nonzero for all
$\mathbf{k}$. This holds for both itinerant solutions discussed in
Sec.IV (see (\ref{Eq50})). For example, for the solution obtained
in Sec.IV.B., one finds $|a_{\mathbf{k},f}|^2 = |a_f|^2 =
|V_1|^2/|t^d_2|$, $|a_{\mathbf{k},d}|^2 = |t^d_2| \{
[|t^d_1|/|t^d_2| - sign(t^d_1) ( \sum_{ \tau=1}^3 \cos k_{\tau} )
]^2 + [\sum_{\tau=1}^3 \sin k_{\tau}]^2 \}$.
\bibitem{newxx}
The ground state expectation value of the individual terms from
$\hat H_{itin}$ can also be calculated, e.g., $\sum_{\sigma}
\langle \hat d^{\dagger}_{\mathbf{i},\sigma}
\hat d_{\mathbf{j},\sigma} \rangle = \sum_{\mathbf{k}}[(2|a_{\mathbf{k},d}|^2%
+|a_{\mathbf{k},f}|^2)/(|a_{\mathbf{k},d}|^2+|a_{\mathbf{k},f}|^2)]exp[i%
\mathbf{k}(\mathbf{i}-\mathbf{j})]$.
\bibitem{xxy}
Using $\hat A_{I_{\mathbf{i}},\sigma} = \sum_{\mathbf{k}} e^{i
\mathbf{k} \mathbf{i}} \hat A^{ \dagger}_{\mathbf{k}\sigma}$ and
$(\hat A^{\dagger}_{\mathbf{k}\sigma})^2 = 0$ one finds
$\prod_{\mathbf{i}}^{N_{ \Lambda}} \hat A^{\dagger}_{I_{%
\mathbf{i}},\sigma} = Z \prod_{\mathbf{k}}^{N_{\Lambda}} \hat A^{\dagger}_{%
\mathbf{k}\sigma}$, where $Z=\sum_P (-1)^{P_p} exp [i(\mathbf{i}_1 \mathbf{k%
}_{i_1} + \mathbf{i}_2 \mathbf{k}_{i_2} + ... +
\mathbf{i}_{N_{\Lambda}} \mathbf{k}_{ i_{N_{\Lambda}}})]$ is a sum
over all
permutations $P$ of $(1,2,...,N_{\Lambda})$ to $(i_1,i_2,...,i_{N_{%
\Lambda}}) $ with $P_p$ as the parity of $P$.
\bibitem{xxy1}
For example, we can choose ${\cal D}_{\uparrow}$ to be defined by
the set
$(\mathbf{x}_1,%
\mathbf{x}_2, 2 \mathbf{x}_3)$ of primitive vectors, and $\mathbf{R}=\mathbf{%
x}_3$.
\bibitem{commz}
There exist special cases (see, e.g., Sec.IV.B for
$|t^d_1/t^d_2|<3$ and footnote (\cite{commx})) where for some
vector $\mathbf{k}^{*}$ one has $a_{\mathbf{k}^{*},d}=0$ and hence
$X_{\mathbf{k}=\mathbf{k}^{*}}=0$. Nevertheless, as long as
$R^{-1}_{\mathbf{k}} \ne 0$ for all $\mathbf{k}$ the operators
$\hat C_{\delta,\mathbf{k},\sigma}$ are well-defined, and
itinerant solutions are still possible. In this case (see the
second equality in (E9)), the momentum distribution function will
have a discontinuity at a single point, i.e., $\langle \hat %
C^{\dagger}_{2,\mathbf{k},\sigma} \hat C_{2,\mathbf{k},\sigma} \rangle$
is 1/2 for all ${\bf k}$, except for ${\bf k}^{*}$ where it is
zero.
\bibitem{mi7} Another possibility to define a cell operator is discussed in
Ref.[\cite{tiot1}] for 2D. In this case a localized phase is
obtained without third neighbor contributions of the hopping and
hybridization amplitudes.
\bibitem{xxx0}
Eq.(\ref{Eq88}) has solutions only for real $V_{\mathbf{x}_{\tau}}$, $%
t^d_{2\mathbf{x}_{\tau}} < 0$, with $sign(V_{\mathbf{x}_{\tau}})
sign(Re(V_0)) = - sign(t^d_{\mathbf{x}_{\tau}})$, real or complex
values of $V_0 = |V_0|e^{i\phi_{V_0}}$, with $Re(V_0) = - t^d_1
V_1/|t^d_2|$ and arbitrary $Im(V_0)$. In the isotropic case when
$t^d_1 = t^d_{\mathbf{x}_1} = t^d_{ \mathbf{x}_2} =
t^d_{\mathbf{x}_3}$, $t^d_2= t^d_{\mathbf{x}_{\tau}
\pm \mathbf{x}_{ \tau^{\prime}}}$, $V_1 = V_{\mathbf{x}_1} = V_{\mathbf{x}%
_2} = V_{\mathbf{x}_3}$, $t^{d^{\prime}}_2 = t^d_2/2 =
t^d_{2\mathbf{x}_1} =
t^d_{2\mathbf{x}_2} = t^d_{2\mathbf{x}_3}$, one finds $%
a_{1,d}=a_{2,d}=a_{3,d}=a_{4,d}=a_{5,d}=a_{7,d}$.
\bibitem{xxx}
The $a_{\mathbf{k},b}$ parameters presented in (\ref{Eq50}), are
now
defined by the Fourier transform of $\hat B^{\dagger}_{B_{\mathbf{i}%
},\sigma}$ which have the form $\hat
B^{\dagger}_{\mathbf{k},\sigma} = \sum_{b=d,f}
a^{*}_{\mathbf{k},\sigma} \hat b^{\dagger}_{\mathbf{k},\sigma}$, where $a_{%
\mathbf{k},f} = a_{6,f}$, and $a^{*}_{\mathbf{k},d} = a^{*}_{6,d} +%
a^{*}_{1,d} e^{- i \mathbf{k}\mathbf{x}_1} + a^{*}_{2,d} e^{- i \mathbf{k}%
\mathbf{x}_2} + a^{*}_{3,d} e^{+ i \mathbf{k}\mathbf{x}_1} +
a^{*}_{4,d} e^{+ i \mathbf{k}\mathbf{x}_2} + a^{*}_{5,d} e^{- i
\mathbf{k}\mathbf{x}_3} + a^{*}_{7,d} e^{+ i
\mathbf{k}\mathbf{x}_3}$. For the solution defined below
(\ref{Eq90}), one finds $a_{\mathbf{k},d} = \sqrt{2 |t^d_2|} e^{i
\phi} [ (\sum_{\tau=1}^3 \cos
k_{\tau}) + (|V_0|/(2 V_1)) e^{-i \phi_{V_0}} ]$, $a_{\mathbf{k}%
,f} = - V_1 \sqrt{2/|t^d_2|} e^{i \phi}$ . The parameters
$R^{-1}_{\mathbf{k}}$, $|a_{\mathbf{k},f}|$ are nonzero for
all $\mathbf{k}$; (see (\ref{Eq51})), so the operators $\hat C_{\delta,%
\mathbf{k}, \sigma}$ are well-defined. For
real $V_0$ one finds $|a_{\mathbf{k},d}| \ne 0$ for $|t^d_1/t^d_2|> 6$, while for $%
Im(V_0) \ne 0$ one always has $|a_{\mathbf{k},d}| > 0$. We note that $a_{%
\mathbf{k},d} \ne 0$ implies $X_{\mathbf{k}} \ne 0$, where
$X_{\mathbf{k}}$ has been defined in (\ref{Eq68}).
\bibitem{compop}
%
%
Complementary anti-commuting operators, but with additional
constraints on lattice site, have previously been employed in the
construction of exact ground states corresponding to extended
spiral structures of the 1D Hubbard model\cite{complop1}, and for
ferromagnetism on decorated lattices \cite{complop2}.
\bibitem{itinq}
The proof of the finite conductivity starts from the observation
that the ground state (\ref{mag4}) can be generalized beyond
quarter filling via (\ref{mag6}),
, see
Sect.V.B.2. Using $E'_g$ from (58), one finds $\mu^{+}=E'_g(N)-E'_g(N-1)=%
K_d$ ,  $\mu^{-}=E'_g(N-1)-E'_g(N-2)=K_d$, i.e.
$\mu^{+}-\mu^{-}=0$. Consequently, the described state is
conducting \cite{cond}.
\end{thebibliography}
\end{document}